\documentclass[11pt]{article}

\usepackage{graphicx}
\usepackage{float}
\usepackage[top=1in,bottom=1in,left=1in,right=1in]{geometry}
\usepackage{amsthm,amsmath,amsfonts,amscd,amssymb,bm,epsfig,epsf,bbm,amssymb,algorithm,algpseudocode,prettyref,tikz,mathtools,pgfplots,dsfont,makecell,caption,subcaption,enumerate} 

\usepackage[toc,page]{appendix}
\usepackage{setspace}
\usepackage[T1]{fontenc}
\usepackage[utf8]{inputenc}
\usepackage{authblk}
\usepackage{xcolor}
\usepackage{cite}
\usepackage{natbib}
\usepackage{longtable}
\usepackage{arydshln}

\definecolor{darkred}{RGB}{100,0,0}
\definecolor{darkgreen}{RGB}{0,100,0}
\definecolor{darkblue}{RGB}{0,0,150}
\definecolor{red}{RGB}{255,0,0}
\usepackage{hyperref}
\hypersetup{colorlinks=true, linkcolor=darkred, citecolor=darkgreen, urlcolor=darkblue}
\usepackage{url}
\usepackage[capitalize]{cleveref}

\theoremstyle{remark}

\newcommand{\beginsupplement}{
  \setcounter{table}{0}  
  \renewcommand{\thetable}{S\arabic{table}} 
  \setcounter{figure}{0} 
  \renewcommand{\thefigure}{S\arabic{figure}}
}

\newcommand{\var}{\textrm{var}}

\renewcommand{\vec}[1]{{\boldsymbol{#1}}}

\newcommand{\diag}{\operatorname{diag}}

\newcommand{\dprime}{\prime\prime}

\newcommand{\Datau}{D^{a,\tau}}

\newcommand{\Dxtau}{D^{x,\tau}}

\newcommand{\Dxf}{D^{x}(\omega)}
\newcommand{\Dphitau}{D^{\phi,\tau}}

\newcommand{\Dphif}{D^{\phi}(\omega)}
\newcommand{\tbin}{t_{\text{bin}}}
\newcommand{\phiprime}{\left<\phi^{\prime}\right>}
\newcommand{\phiprimei}{\left<\phi^{\prime}_i\right>}

\newcommand{\Dlinf}{D_{\text{lin}}(\omega)}

\newcommand{\Cxt}{C^{x}(\tau)}
\newcommand{\Cxtau}{C^{x,\tau}}
\newcommand{\Cxf}{C^{x}(\omega)}

\newcommand{\Cphit}{C^{\phi}(\tau)}
\newcommand{\Cphitau}{C^{\phi,\tau}}
\newcommand{\Cphif}{C^{\phi}(\omega)}

\newcommand{\Cetaf}{C^{\eta}(\omega)}

\newcommand{\Czero}{c_0}

\newcommand{\Caf}{C^a(\omega)}
\newcommand{\Czetaf}{C^{\zeta}(\omega)}
\newcommand{\Czetaomega}{C^{\zeta,\omega}}

\newcommand{\Pxtau}{p^{x,\tau}(\lambda)}

\newcommand{\Pxf}{p^{x}(\lambda;\omega)}
\newcommand{\Pphitau}{p^{\phi,\tau}(\lambda)}

\newcommand{\Pphif}{p^{\phi}(\lambda;\omega)}

\newcommand{\psiphif}{\psi^{\phi}(\omega)}
\newcommand{\psixf}{\psi^{x}(\omega)}
\newcommand{\psiphitf}{\psi^{\phi}(\vec{\omega})}
\newcommand{\psixtf}{\psi^{x}(\vec{\omega})}
\newcommand{\mean}[1]{\left<#1\right>}

\newcommand{\lrsq}[1]{\left[#1 \right]}
\newcommand{\sigeff}{\sigma_{\text{eff}}}
\newcommand{\lrbk}[1]{\left(#1 \right)}
\newcommand{\lrabs}[1]{\left|#1 \right|}
\newcommand{\free}{\text{free}}

\newcommand{\MCxf}{\mathbf{C}^{x}\left(\omega\right)}

\newcommand{\MCxomega}{\mathbf{C}^{x,\omega}}
\newcommand{\MCphif}{\mathbf{C}^{\phi}\left(\omega\right)}
\newcommand{\MCphiomega}{\mathbf{C}^{\phi,\omega}}

\newcommand{\MCatau}{\mathbf{C}^{a,\tau}}
\newcommand{\MCxtau}{\mathbf{C}^{x,\tau}}

\newcommand{\MCphitau}{\mathbf{C}^{\phi,\tau}}

\newcommand{\MCphixif}{\mathbf{C}^{\phi\xi}\left(\omega\right)}

\newcommand{\MCxxif}{\mathbf{C}^{x\xi}\left(\omega\right)}

\newcommand{\MJ}{\mathbf{J}}
\newcommand{\MJeff}{\mathbf{J}_{\text{e}}}
\newcommand{\MI}{\mathbf{I}}
\newcommand{\MC}{\mathbf{C}}

\newcommand{\MB}{\mathbf{B}}

\newcommand{\MClinf}{\mathbf{C}_{\text{lin}}(\omega)}

\newcommand{\MPhiPrime}{\mathbf{D}_{\left<\phi^{\prime}\right>}}
\newcommand{\MP}{\mathbf{P}}
\newcommand{\MCzetaf}{\mathbf{C}^{\zeta}(\omega)}

\newcommand{\MA}{\mathbf{A}}

\newcommand{\M}[1]{\mathbf{#1}}

\newcommand{\MCx}{\mathbf{C}^{x}}
\newcommand{\geff}{g_{\text{eff}}}

\numberwithin{equation}{section}

\newcommand{\keywords}[1]{\vspace{2mm}\noindent\textbf{Keywords:} #1}

\pgfplotsset{compat=1.17}

\title{Covariance Spectrum in Nonlinear Recurrent Neural Networks: A Near-Critical Effective Connection Strength Throughout Chaos}
\author[1]{Xuanyu Shen}
\author[1,2,3]{Yu Hu\thanks{To whom correspondence should be addressed: huyu@ion.ac.cn}}
\affil[1]{Institute of Neuroscience, Center for Excellence in Brain Science and Intelligence Technology, State Key Laboratory of Brain Cognition and Brain-inspired Intelligence Technology, Chinese Academy of Sciences, Shanghai 200031, China}
\affil[2]{Department of Mathematics, The Hong Kong University of Science and Technology, Hong Kong SAR, China}
\affil[3]{Division of Life Science, The Hong Kong University of Science and Technology, Hong Kong SAR, China}
\date{}

\begin{document}

\maketitle

\begin{abstract}
Advances in simultaneous recordings of large numbers of neurons have driven significant interest in the structure of neural population activity, such as dimensionality, and in how it arises mechanistically from network connectivity. The covariance eigenvalue distribution, or spectrum, which can be derived analytically in random recurrent networks, was proposed as a robust description of population geometry beyond dimensionality. The theoretical spectrum has since matched data across brain areas and species, serving as a computational biomarker linking neural population activity to network connectivity. However, this empirical success highlights a gap in theory, as the model used to derive the spectrum was minimal with linear neurons. Here we close this gap by studying the covariance spectrum in networks with nonlinear neurons and across broader dynamical regimes, including chaos. Surprisingly, the spectrum is precisely described by equations analogous to the linear theory, substituted with an effective recurrent connection strength, which reflects both the connection weights and the dynamic gain of neurons. Across dynamical regimes, this effective parameter gives a unified account of the spectrum and dimensionality changes, and stays near its critical value throughout the chaotic regime, suggesting a mechanism for brain criticality without fine-tuning and a new interpretation for previously observed near-critical covariance spectra. The key step of our theory is proposing a covariance matrix ansatz under nonlinear dynamics, which explicitly links neuronal dynamics, recurrent connectivity, and all pairwise correlations in the network. The ansatz is supported by extensive numerical and theoretical validations, including extensions to other network models such as partially symmetric connectivity. These results further our understanding of nonlinear neural population dynamics and provide new theoretical support and analysis method, including identifying dynamical regimes, for applying covariance spectrum analysis in biological circuits.
\end{abstract}


\section{Introduction}
\label{sec:introduction}
Understanding the relationship between neural population dynamics and the underlying network connectivity is a central question in theoretical neuroscience and is also of common interest to other fields studying complex and extensively interacting systems.
In recent years, there have been significant advances in large-scale simultaneous recordings of neural activity \citep{Ahrens2012,James2017,Semedo2019,Stringer2019Science}, which not only increase the quantity of data but also allow for new ways of analyzing population activity structures at the multi-dimensional level, rather than averaging across neurons. Examples of such analyses include principal component analysis (PCA) \citep{Stringer2019Science}, dimensionality \citep{Rajan2011,Dahmen2026,Fusi2016}, neural manifolds \citep{Gallego2017}, and communication subspaces \citep{Semedo2019}, among others. This, in turn, has driven intense interest in theoretical studies aimed at understanding how these properties of population dynamics arise mechanistically and relate to the network connectivity structure in neural circuit models \citep{Ostojic2018,Recanatesi2019,Clark2023,Hu2022,Gozel2024}.

The covariance spectrum \citep{Hu2022} is a curve representing the distribution of covariance matrix eigenvalues that can be computed from simultaneously recorded neural activity.
The covariance spectrum includes the widely used dimensionality information but is more robust than scalar measures and further describes the geometry or shape of neural population activity—for example, it helps reveal a brain-wide scale-invariant coding property \cite{Wang2025}. Hu and Sompolinsky developed a theory \citep{Hu2022} predicting the covariance spectrum in recurrent neural circuits, using simple network models with random connections \cite{Sompolinsky1988}, second-order motifs \cite{Zhao2011}, and linear neurons. This theoretical spectrum has been found to match well with experimental data despite having only a single parameter $g$ \citep{Hu2022,Morales2023,Calvo2024}, including calcium imaging in zebrafish, multi-electrode recordings in mice, and human magnetoencephalography (MEG) data. The fitting to the theoretical spectrum also estimates a recurrent connection strength $\hat{g}$, serving as a computational biomarker that interprets neural activity features, such as MEG recordings of Parkinson's disease patients \citep{Calvo2024}, in terms of network connectivity mechanisms.

However, this empirical success of the model covariance spectrum highlights a gap in theory. Although the linear dynamics assumed in \cite{Hu2022} may be an acceptable approximation for spontaneous brain activity, biological neural networks can also be in strongly nonlinear dynamical regimes such as chaos \citep{Sompolinsky1988,Kadmon2015,Schuecker2018}, which has been proposed to be a state potentially advantageous for functions such as generating rich temporal patterns and working memory \citep{Maass2002,Sussillo2009,Toyoizumi2011}. 
Therefore, it is an important question to generalize the covariance spectrum theory to nonlinear dynamics, which is also crucial for properly interpreting the biological meaning of the fitted connectivity parameter $\hat{g}$.

Describing the covariance spectrum in nonlinear recurrent networks is technically challenging. For example, one needs to consider the collective effect of small correlations between neurons (vanishing in a large network with $N$ neurons as $O(1/\sqrt{N})$) that were ignored in the classic mean-field theory. 
A recent breakthrough by Clark et al. \citep{Clark2023} developed a two-site cavity method to characterize the dimensionality, which depends on the first two orders of statistics of correlations; whereas to determine the spectrum, one needs to consider all orders.

In this study, we close the aforementioned theoretical gap by determining the covariance spectrum in nonlinear randomly connected recurrent neural networks. 
Our theory builds on the results of \cite{Hu2022} and \cite{Clark2023}, and a key step is to propose a new ansatz for the covariance matrix,
which combines single-neuron mean-field theory \cite{Sompolinsky1988,Schuecker2018} with network connectivity in a closed form (\cref{eq:ansatz_of_Cov_phi_omega}).
This covariance matrix ansatz describes all pairwise correlations in the network and is valid across all dynamical regimes including chaos, as supported both theoretically and numerically.  
From the covariance matrix ansatz, one can then apply random matrix theory to compute the (frequency-dependent) covariance spectra for both firing rates and currents.

As a corollary of the theory, the changes in the covariance spectrum shape and dimension can be understood in a unified manner in terms of an effective recurrent connection strength parameter $\geff$ (\cref{eq:g_eff_definition}), which serves as a counterpart to the connection strength $g$ in networks with linear dynamics \citep{Hu2022}. 
This provides a precise interpretation for the inferred parameter $\hat{g}$ when fitting the (linear) theoretical spectrum to data \citep{Hu2022,Morales2023,Calvo2024}, as it incorporates the effect of the dynamic gain of neurons.
Intriguingly, $\geff$ stays close to the critical value of 1 throughout the chaotic regime (\cref{sec:effective_g_and_transition_to_chaos}). This predicts an approximate power-law covariance spectrum (\cref{eq:spectrum_power_law_tail}), supplementing the previous result of population activity being low-dimensional in this regime \citep{Clark2023} with geometric information. The results suggest a new mechanism for attaining approximate brain criticality without network parameter fine-tuning and may help explain the near-critical spectra observed in many experimental datasets \cite{Morales2023}.
To apply the theory to real data, we also provide a method to infer the network parameters and the dynamical regimes (\cref{sec:infer_parameter_main_text}).

For simplicity, we mainly focus on an archetypical recurrent neural network model \citep{Sompolinsky1988,Schuecker2018}, but the core theory does not depend on model details and can be generalized to other network models, as we demonstrate with other neuronal firing-rate functions, sparse E-I networks, and networks with partially symmetric connections (\cref{sec:extension_of_theory}). 

The rest of the paper is organized as follows. In \cref{sec:network_model}, we introduce the recurrent network model and define the covariance matrix, participation ratio dimension, and covariance spectrum. In \cref{sec:moment_to_ansatz}, we propose the ansatz for the covariance matrix and discuss its numerical and theoretical validations. 
In \cref{sec:eigenvalue_spectrum_of_covariance_matrices}, we use the ansatz and its corollary to derive the covariance spectrum for firing rates and currents. 
In \cref{sec:effective_g_and_transition_to_chaos}, we study the behavior of the effective recurrent connection strength $\geff$ across dynamical regimes and investigate the relationship between the dimension gap and the transition to chaos, as well as the method to infer the dynamical regime from neural recordings.
In \cref{sec:extension_of_theory}, we demonstrate several extensions of our theory to network models with other neuronal firing-rate functions and beyond Gaussian random connections.

\section{Neural network model}
\label{sec:network_model}
We study the dynamics of a recurrent network of $N$ neurons, where the total current of neuron $i$, $x_i(t)$, evolves according to 
\begin{equation}
    \tau \frac{dx_i (t)}{dt} = 
    -x_i+\sum_{j=1}^NJ_{ij}\phi(x_j)+\xi_i(t),
    \quad i=1,2,\cdots, N.
    \label{eq:network_model}
\end{equation}
The neurons are coupled by a random network (fixed during the neural dynamics) \citep{Sompolinsky1988,Schuecker2018,Hu2022,Clark2023}, specified by the connectivity matrix $\mathbf{J}=(J_{ij})_{1\le i,j\le N}$, whose entries are drawn i.i.d. from a Gaussian distribution with zero mean and variance $g^2/N$.
In particular, such a network is balanced, on average, in excitation and inhibition, and the parameter $g$ controls the recurrent connection strength.
For simplicity, we set the neuronal time constant $\tau=1$ in the rest of the paper.
The nonlinear firing-rate function $\phi(x)=\tanh{(x)}$ converts the neuron's current to its firing rate. For ease of notation, we write $\phi_i(t) =\phi(x_i(t))$.
The neurons are also driven by external inputs $\xi_i(t)$ that are independent across neurons $i$ and are modeled as Gaussian white noise with strength $\sigma$, such that $\langle \xi_i(t)\xi_j(t+\tau)\rangle=\sigma^2\delta_{ij} \delta(\tau)$. 
We use the notation $\langle \cdot \rangle := \mathbb{E}_t[\cdot]$ to denote the temporal average over the stochastic input.

We are interested in the population dynamics of the network described by the covariance matrix and related quantities. 
In particular, the time-lagged covariance, or correlation function, between neurons $i$ and $j$ is $C_{ij}^x(\tau)=\langle \Delta x_i(t)\Delta x_j(t+\tau)\rangle$, 
$C_{ij}^{\phi}(\tau)=\langle \Delta \phi_i(t)\Delta \phi_j(t+\tau)\rangle$, 
where $\Delta x_i=x_i-\langle x_i\rangle$, $\Delta\phi_i=\phi_i-\langle \phi_i\rangle$.
Due to symmetry, $\langle x_i\rangle=\langle \phi_i\rangle=0$, so we drop the $\Delta$ notation when writing covariances.
Note that we have two covariance matrices $\mathbf{C}^{a}(\tau) =(C_{ij}^a(\tau))$, $a\in\{x, \phi\}$, for the currents and firing rates, respectively. More generally, we use notations such as $\mathbf{C}^{x\phi}(\tau)=(C^{x\phi}_{ij}(\tau))=(\langle x_i(t)\phi_j(t+\tau)\rangle)$ to denote the cross-covariance.
We assume the dynamics is stationary, so the covariance depends only on the time delay $\tau$. 
The covariance matrices can also be equivalently described in the frequency domain by performing an entry-wise Fourier transform $C^{a}_{ij}(\omega)=\mathcal{F}\left[C^a_{ij}(\tau)\right]:=\int_{-\infty}^{\infty} e^{-i\omega \tau}C^a_{ij}(\tau)\,d\tau$ to obtain the frequency-dependent covariance matrices $\mathbf{C}^{a}(\omega)=(C^{a}_{ij}(\omega))$.
For ease of notation, we write the zero-timelag covariance matrix as $\mathbf{C}^a_{\tau=0}=\mathbf{C}^a(\tau=0)$ and the zero-frequency covariance matrix as $\mathbf{C}^a_{\omega=0}=\mathbf{C}^a(\omega=0)$. Similar notations are used for $\phi$, $\eta$, and other related quantities, including dimensions.

The structure of neuron population activity or dynamics can then be characterized in terms of dimension and shape.  
Specifically, the (relative) participation ratio dimension \citep{Rajan2011,Litwin-Kumar2017,Recanatesi2019} can be calculated from the covariance matrix as
$D(\MC)=\frac{1}{N}\frac{\left(\sum_{i=1}^N\lambda_i\right)^2}{\sum_{i=1}^N\lambda_i^2}$, where $\lambda_i$ are eigenvalues of the covariance matrix $\MC$. Here $D$ is normalized by the number of neurons and thus lies within $[0,1]$.
The covariance spectrum additionally describes the shape, besides the dimension, of the population activity \citep{Hu2022, Wang2025}; it is the probability density function of the covariance eigenvalues, whose empirical distribution $p(\lambda)=\frac{1}{N}\sum_{i=1}^N\delta(\lambda-\lambda_i)$ may converge to a continuous function in the large-network limit $N\to \infty$ that can be determined from the model parameters, such as the recurrent connection strength $g$ \citep{Hu2022}.

With nonlinear neurons, the network can be in different dynamical regimes, including chaos, depending on the model parameters $(g,\sigma)$ (\cref{fig:Eigen_spectrum_total}A, \citep{Schuecker2018}).
For a fixed $\sigma>0$, as we increase the recurrent connection strength $g$ to a value $g_{c_1}$, the network switches from stable fluctuations around a fixed point (linearly stable regime) to a regime where the Jacobian matrix characterizing the linear stability has eigenvalues with positive real parts (linearly unstable regime) \citep{Schuecker2018}.
Despite the linear instability, the dynamics is not yet chaotic, which is characterized by a positive leading Lyapunov exponent and occurs at a larger $g=g_{c_2}$. Importantly, due to the presence of fluctuating input $\xi_i(t)$, a larger $g_{c_2}$ is required than in the autonomous or noiseless case of $\sigma=0$, where $g_{c_2}=1$ (and without the linear unstable regime) \citep{Sompolinsky1988,Schuecker2018}.
The critical values $g_{c_1}$ and $g_{c_2}$ can be determined using a mean-field theory focusing on a single, typical neuron \citep{Schuecker2018,Sompolinsky1988}. 
For example, $g_{c_1}=1/\sqrt{\mean{\lrsq{\phi^{\prime}(x)}^2}}$ \citep{Schuecker2018}.
Note that we dropped the neuron index in the above expression and in other places, such as $C^x(\tau)$ for the autocorrelation function instead of $C^x_{ii}(\tau)$. This is justified because neurons become homogeneous or equivalent to each other, in the large-network limit $N\rightarrow \infty$.
The transition value $g_{c_2}$ for chaos, and statistics such as $C^a(\tau)$ and $\phiprime = \mean{\phi^\prime(x_i(t))}$, can be determined from a mean-field equation for a single neuron's dynamics
\citep{Sompolinsky1988,Schuecker2018}, 
\begin{equation}
    (1+\omega^2)\Cxf = g^2\Cphif + \sigma^2,
\label{eq:single_neuron_MFT}
\end{equation}
which stems from analyzing the statistics of the total recurrent input to a neuron $\eta_i(t):=\sum_{j=1}^{N}J_{ij}\phi_j(t)$ over the ensemble of the random connectivity $\MJ$.

\section{Covariance matrix ansatz}
\label{sec:moment_to_ansatz}
To achieve our goal of characterizing the shape of neural population activity using the covariance spectrum, we need a description of the covariance matrix. In particular, the lower-order statistics of $C^a_{ij}$, although they can be used to derive the dimension $D$ \citep{Clark2023,Recanatesi2019}, are not sufficient for the spectrum.
Here we propose a guess or ansatz for the covariance matrix $\MCphif$ based on a crucial observation that the frequency-dependent dimension $D^{\phi}(\omega):=D(\MCphif)$ has a simple expression (\cref{eq:D_phi_omega}) closely related to the linear dynamics counterpart. 
Although, as we elaborate below, $D^{\phi}(\omega)$ can be readily derived from \cite{Clark2023}, and the linear dimension and spectrum theory have been empirically applied to potentially strongly nonlinear networks \citep{Hu2022,Morales2023,Calvo2024}, this connection between the nonlinear and linear network dynamics has not been explicitly made, to the best of our knowledge.

As a slight extension of the result on dimension in \cite{Clark2023}, which focuses on the case without external input $\sigma=0$, we derive the frequency-dependent dimension $\Dphif$ for $\sigma\geq0$ valid for all values of $g$, by computing the four-point function $\psiphif:=N\left.\mean{ C^{\phi}_{ij}(\omega)C^{\phi}_{ij}(-\omega)} _{\MJ}\right|_{i\neq j}$ 
following a similar two-site cavity mean-field theory, as in \cite{Clark2023} (\cref{app:cavity_unit_four_point_function}). 
Importantly, a factor of the single-neuron dynamics quantity $\lrsq{\Cphif}^2$ appearing in both the first and second moments of covariance eigenvalues cancels when computing the dimension,
\begin{equation}
    \Dphif=\left(1-\frac{g^2\phiprime^2}{1+\omega^2}\right)^2.
\label{eq:D_phi_omega}
\end{equation}
This simple expression takes an analogous form to the dimension of the linear network \citep{Hu2022}, $\Dlinf=\left(1-\frac{g^2}{1+\omega^2}\right)^2$. In particular, if we replace $g$ with an \emph{effective recurrent connection strength} (\cite{Schuecker2018,Clark2023}; see further discussions in \cref{sec:effective_g_and_transition_to_chaos}, including that it is always less than 1),
\begin{equation}
\geff=g\phiprime,
\label{eq:g_eff_definition}
\end{equation}
then the nonlinear dynamics dimension $\Dphif$ matches $\Dlinf$ across all frequencies $\omega$.
This motivates us to propose an ansatz for the firing rate covariance matrix under nonlinear dynamics,
\begin{equation}
\text{Ansatz:} \quad
    \MCphif=\frac{\Cphif(1-g^2\phiprime^2+\omega^2)}{1+\omega^2}
    \left(\MI-\frac{\phiprime}{1+i\omega}\MJ\right)^{-1}\left(\MI-\frac{\phiprime}{1-i\omega}\MJ^T\right)^{-1}.
\label{eq:ansatz_of_Cov_phi_omega}
\end{equation}
Here $\MI$ is the identity matrix. 
The ansatz is based on the counterpart in a linear dynamics network driven by independent white noise: $\MClinf=\frac{\sigma^2}{1+\omega^2}\left(\MI-\frac{1}{1+i\omega}\MJ\right)^{-1}\left(\MI-\frac{1}{1-i\omega}\MJ^T\right)^{-1}$ \citep{Hu2022,Trousdale2012,Gardiner2009}. The scalar factor in \cref{eq:ansatz_of_Cov_phi_omega} ensures the two sides match at the first moment (\cref{app:derivation_of_ansatz}).

\subsection{Covariance matrix of currents $\MCxf$}
\label{sec:cov_x}
As noted in \cite{Clark2023}, the firing rate and current covariances are different and, for example, they have different dimensions $D^{\phi}$ and $D^{x}$.  
To derive the covariance matrix for currents $\MCxf$, we start by taking the Fourier transform of the neural dynamics equation (\cref{eq:network_model}): $\vec{x}(\omega) = \frac{1}{\sqrt{T}}\int_{0}^T e^{-i\omega t} \vec{x}(t) dt$, where $\vec{x} = (x_i)_{i=1}^N$ and multiply it by its conjugate transpose to obtain $\MCxf=\lim_{T\rightarrow \infty}\mean{\vec{x}(\omega)\vec{x}^{\dagger}(\omega)}$ (Wiener-Khinchin theorem),
\begin{equation}
(1+\omega^2)\MCxf=\MJ\MCphif\MJ^T+\MJ\MCphixif+\M{C}^{\xi\phi}(\omega)\MJ^T+\sigma^2\MI.
\label{eq:Cx_dynamic_eq}
\end{equation}
Here $\vec{x}^{\dagger}$ represents the conjugate transpose of the vector $\vec{x}$.
Without external inputs $\sigma=0$, this combined with the $\MCphif$ ansatz (\cref{eq:ansatz_of_Cov_phi_omega}) immediately gives an expression for $\MCxf$.

For the case with external inputs, we further assume the following identity for $\MCphixif$:
\begin{equation}
    \MCphixif = \phiprime\MCxxif,
\label{eq:Cphi_xi_1}
\end{equation}
which can be justified by Stein's lemma if $x_i(t)$ and $\xi_j(t^\prime)$ are jointly Gaussian distributed:
$\mean{\phi(x_i(t))\xi_j(t+\tau)}$ $=\phiprimei\mean{x_i(t)\xi_j(t+\tau)}$, and, by using the homogeneity, $\phiprimei\to\phiprime$ as $N\to\infty$. Note that $x_i(t)$ and $\xi_j(t^\prime)$ are indeed \textit{marginally} Gaussian, as $x_i(t)$ becomes a Gaussian process in the large-network limit \citep{Sompolinsky1988}.

To proceed, substituting $\vec{x}$ in $\MCxxif$ using the dynamics equation \cref{eq:network_model} gives
\begin{equation}
\MCxxif=\frac{1}{1+i\omega}\lrsq{\MJ\MCphixif+\sigma^2\MI}.
\label{eq:Cphi_xi_2}
\end{equation}
We can now solve for $\MCphixif$ from \cref{eq:Cphi_xi_1,eq:Cphi_xi_2} and plug it into \cref{eq:Cx_dynamic_eq} to get (\cref{app:second_moment_and_dimension}),
\begin{equation}
\begin{split}
\MCxf &=
\frac{\Cphif-\phiprime^2 \Cxf}{1+\omega^2}
 \MJ \left(\MI-\frac{\phiprime}{1+i\omega}\MJ\right)^{-1}\left(\MI-\frac{\phiprime}{1-i\omega}\MJ^T\right)^{-1} \MJ^T\\
&+\frac{\sigma^2}{1+\omega^2} 
\left(\MI-\frac{\phiprime}{1+i\omega}\MJ\right)^{-1}\left(\MI-\frac{\phiprime}{1-i\omega}\MJ^T\right)^{-1}.
\end{split}
\label{eq:Cov_x_omega_full}
\end{equation}

Comparing with linear dynamics theory \citep{Hu2022}, the above expression intuitively bridges the theory of linear and chaotic dynamics (see also \cref{eq:Cx_with_zeta}). 
The first term in \cref{eq:Cov_x_omega_full} represents the contribution from nonlinear dynamics such as chaos (see \cref{eq:Cx_with_zeta}) and causes $\MCx$ to differ from $\MC^{\phi}$. This term is zero when the neurons are linear, $\phi(x) = x$, in that case, only the second term corresponding to external noise input remains.

\subsection{Interpretation and numerical supports}
\label{sec:effective_dynamics}
\begin{figure}[!htbp]
    \centering
    \includegraphics[trim=0cm 4cm 0cm 0cm, clip, width=\linewidth]{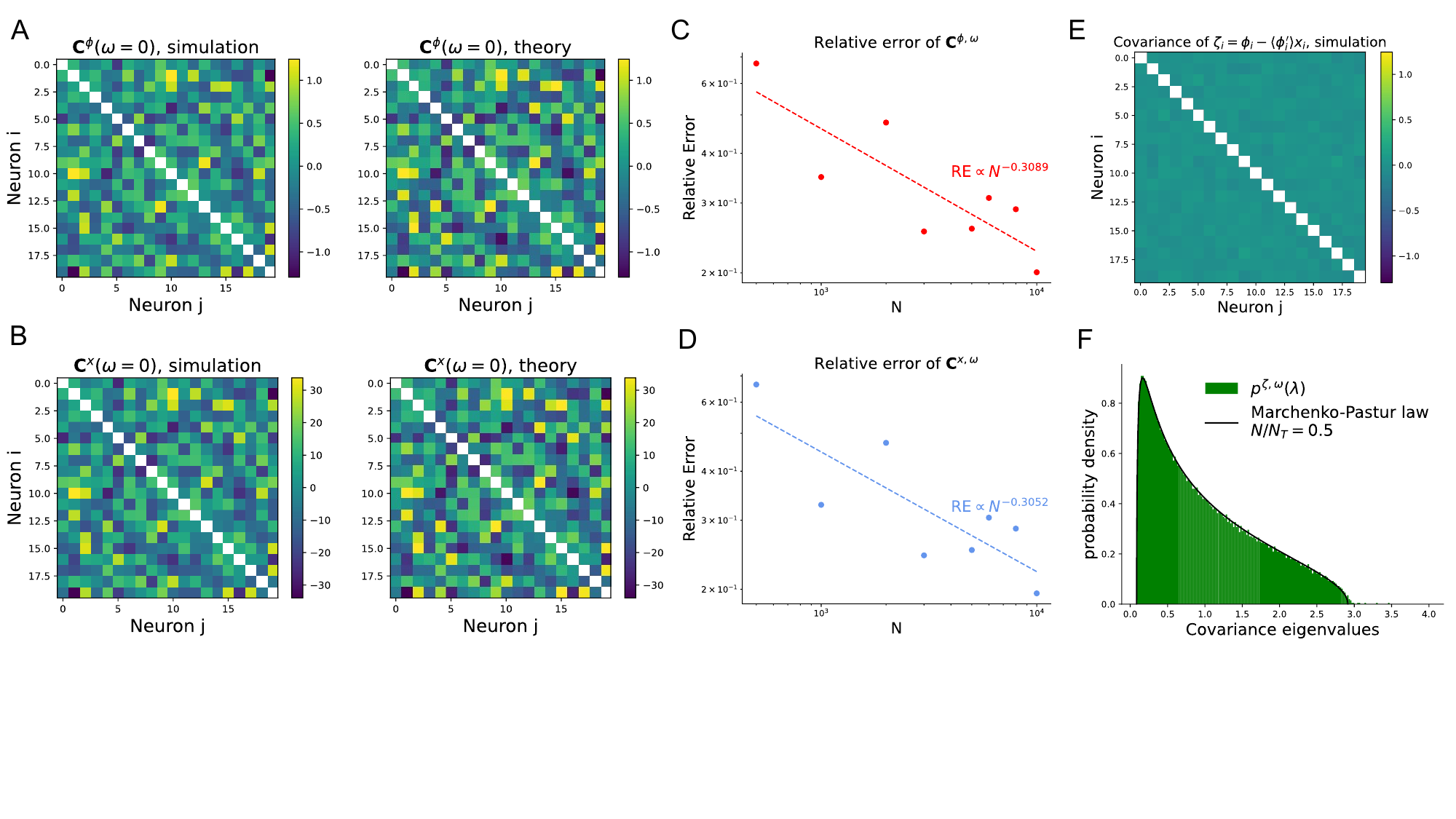}
    \caption{
    (\textbf{A}) Comparing the zero-frequency covariance matrix of firing rates $\MCphiomega$ from network simulations (\cref{eq:network_model}, left) with the ansatz (\cref{eq:ansatz_of_Cov_phi_omega}, right) for a subset of 20 neurons in a network with $N=10000$ (under a single realization of connectivity $\MJ$). Diagonals are not shown to better visualize the off-diagonal entries. The network is in the chaotic regime with $g=5$, $\sigma=0.495$ (\cref{fig:Eigen_spectrum_total}).
    (\textbf{B}) Same as (A), but comparing the covariance matrix for currents $\MCxomega$ from simulation and theory (\cref{eq:Cov_x_omega_full}, right).
    (\textbf{C, D}) Relative error between simulation and theory for $\MCphiomega$ and $\MCxomega$ (in Frobenius norm) decreases with the network size $N$.
    (\textbf{E}) Covariance matrix $\MC^{\zeta,\omega}$ estimated from simulation (for the same 20 neurons as in (A) and (B)). 
    The relative magnitude of its off-diagonal entries is quantified by
    $\overline{\vert \rho\vert} = \frac{1}{N(N-1)/2}\sum_{i<j}|C_{ij}|/\sqrt{C_{ii}C_{jj}}$, which is 0.0057.
    The matrix is scaled such that it has the same average diagonal values as $\MCphiomega$ in (A) and is shown under the same color map to allow visual comparison of the $\zeta$ and $\phi$ off-diagonal entries (the latter has a much larger $\overline{\vert \rho\vert}=0.069$).    
    (\textbf{F}) The spectrum of the sample covariance matrix $\MC^{\zeta,\omega}$ (computed from $N_T=2N$ time samples (\cref{sec:finite-sample_theory}) matches closely with the Marchenko-Pastur law expected for i.i.d. $\zeta_i$'s.
    }
\label{fig:example_covariance_and_relative_error}
\end{figure}

Here we provide some intuitive explanations of the covariance matrix ansatz, along with numerical verifications.
We want to emphasize that the covariance matrix expressions (\cref{eq:ansatz_of_Cov_phi_omega,eq:Cov_x_omega_full}) hold for each individual realization of the network connectivity $\MJ$, rather than only describing the covariance matrices on a statistical level over the distributions of the random $\MJ$. This is verified by direct simulations of the nonlinear network (\cref{fig:example_covariance_and_relative_error}A,B). For a fixed $\MJ$, the theoretical covariance matrices computed using \cref{eq:ansatz_of_Cov_phi_omega,eq:Cov_x_omega_full} and the mean-field quantities match closely with those computed from simulated neural activity, and the relative error between the theory and simulation decreases with the network size $N$ (\cref{fig:example_covariance_and_relative_error}C,D).

The covariance matrix ansatz (\cref{eq:ansatz_of_Cov_phi_omega}) has a heuristic explanation by introducing an effective noise. For simplicity, we first explain this for the case without external input $\sigma=0$.
Define the \emph{effective recurrent noise} as 
\begin{equation}
    \zeta_i:=\phi_i-\mean{\phi_i^{\prime}} x_i.
\label{eq:zeta_define}
\end{equation}
We can compute the variance of $\zeta_i$ in the frequency domain.
According to the classic mean-field theory, $\{x_i(t)\}_t$ is a Gaussian process \citep{Sompolinsky1988}, therefore, $\mean{x_i(\omega)\phi_i^{\ast}(\omega)}=\phiprimei\mean{x_i(\omega)x_i^{\ast}(\omega)}$ by Stein's lemma.
Then, using \cref{eq:zeta_define}, 
$\mean{\zeta_i(\omega)\zeta_i^{\ast}(\omega)}$ $=\mean{\phi_i\phi_i^{\ast}}$ $-\phiprimei\lrsq{\mean{x_i\phi_i^{\ast}} +\mean{\phi_ix_i^{\ast}}}$ $+\phiprimei^2\mean{x_ix_i^{\ast}}$ $= \mean{\phi_i\phi_i^{\ast}}$ $-\phiprimei^2\mean{x_ix_i^{\ast}}$. Note that the neuron index $i$ can be dropped due to the homogeneity of neurons as $N\rightarrow \infty$. Then the Fourier transform of the autocorrelation function of $\zeta$, or $\Czetaf$, is
\begin{equation}
\Czetaf = \Cphif-\phiprime^2\Cxf
= \left(1-\frac{g^2\phiprime^2}{1+\omega^2}\right)\Cphif.
\label{eq:Czeta_diag}
\end{equation}
Here we used the mean-field relationship \cref{eq:single_neuron_MFT} (with $\sigma=0$) to obtain the second equality.

We can use $\vec{\zeta}=(\zeta_i)$ to express $\vec{\phi}$. 
Plugging the Fourier transform of the dynamics equation (\cref{eq:network_model}) into \cref{eq:zeta_define} gives 
$
\zeta_i(\omega)=\phi_i(\omega)-\frac{\mean{\phi_i^{\prime}}}{1+i\omega} \sum_{j=1}^{N} J_{ij} \phi_j(\omega)
$. 
In matrix form, we can solve for $\vec{\phi}$,
\begin{equation}
    \vec{\phi}(\omega) = \lrsq{\MI-\frac{1}{1+i\omega}\MPhiPrime\MJ}^{-1} \vec{\zeta}(\omega).
\label{eq:effective_dynamics_of_phi}
\end{equation}
Here $\MPhiPrime = \diag(\mean{\phi_1^\prime}, \mean{\phi_2^\prime},\ldots, \mean{\phi_N^\prime})$ is a diagonal matrix. The covariance matrix of $\vec{\phi}$ is then (again invoking the homogeneity of neurons as $N\rightarrow \infty$),
\begin{equation}
\MCphif=\lrsq{\MI-\MJeff}^{-1}\MCzetaf \lrsq{\MI-\MJeff^\dagger}^{-1}, \quad \text{where } \MJeff := \frac{\phiprime}{1+i\omega} \MJ.
\label{eq:Cov_phi_omega_from_zeta}
\end{equation}
Comparing this with \cref{eq:ansatz_of_Cov_phi_omega}, we see that our previous ansatz for $\MCphif$ can be justified if $\MCzetaf$ is a scalar matrix,
\begin{equation}
\text{Ansatz:} \quad \MCzetaf = C^{\zeta}(\omega) \cdot \MI \;\;\text{ as } N\to\infty.
\label{eq:ansatz_Cov_zeta}
\end{equation}
This means that $\zeta_i$ are uncorrelated with each other and homogeneous across neurons $i$, and the scalar value of $ C^{\zeta}(\omega)$ is given by \cref{eq:Czeta_diag}.

In the general case with external inputs, the total effective noise to neuron $i$ will be
$\tilde{\zeta}_i = \zeta_i + \frac{\phiprimei}{1+i\omega}\xi_i$, which will replace $\zeta_i$ in \cref{eq:effective_dynamics_of_phi}.
The rest of the argument can be generalized accordingly (\cref{app:extend_zeta}).
We further conjecture that in the large-network limit $N\rightarrow \infty$, $\zeta_i$ are uncorrelated with the external noise $\vec{\xi}$.
Interestingly, we can use this property $\MC^{\zeta\xi}(\omega)=0$ as an alternative to \cref{eq:Cphi_xi_1} to derive the current covariance matrix $\MCxf$ (\cref{eq:Cov_x_omega_full}), which can be succinctly written using $\Czetaf$, 
\begin{equation}
\MCxf =
\frac{\Czetaf}{\phiprime^2}
 \MJeff \left(\MI-\MJeff\right)^{-1}\left(\MI-\MJeff^\dagger\right)^{-1} \MJeff^\dagger\\
 +\frac{\sigma^2}{1+\omega^2} 
\left(\MI-\MJeff\right)^{-1}\left(\MI-\MJeff^\dagger\right)^{-1}.
\label{eq:Cx_with_zeta}
\end{equation}

According to the ansatz (\cref{eq:ansatz_Cov_zeta}), the off-diagonal entries of $\MCzetaf$ should vanish in the large-network limit.
This is consistent with numerical simulations of the nonlinear network, where we find that the off-diagonal entries of the sample estimated $\MC^{\zeta}$ are much smaller relative to its diagonal (\cref{fig:example_covariance_and_relative_error}E). In other words, the average magnitude of correlations is small ($\overline{\vert \rho\vert} \approx 0.0057$). Moreover, this is drastically smaller than the $\overline{\vert \rho\vert}\approx 0.069$ of $\MC^{\phi}$ (\cref{fig:example_covariance_and_relative_error}A,E).
To further check the collective effect of $\MC^{\zeta}$ off-diagonal entries, we compute its eigenvalues. If the $\zeta_i$ are indeed uncorrelated across neurons, or sufficiently close to, these sample covariance matrix eigenvalues would be described by the Marchenko-Pastur law. This is indeed the case in our simulations (\cref{fig:example_covariance_and_relative_error}F).

\subsection{Theoretical validation based on dimension}
\label{sec:theoretical_validation}
As a validation of our covariance matrix expression for $\MCxf$ (\cref{eq:Cov_x_omega_full}), we use it to provide an alternative derivation of the dimension $\Dxf$ previously obtained in \cite{Clark2023}. It is important to note that we did not use $\Dxf$ and related moments in deriving \cref{eq:Cov_x_omega_full} based on the $\MCphif$ ansatz. Therefore, the derivation of dimension is indeed a validation rather than a tautology.

Note that the moments of the eigenvalues can be computed using the trace of the covariance matrix powers. In particular,  
define $\varphi [\cdot]\coloneqq\lim_{N\to\infty}\frac{1}{N}\mathrm{Tr}\mean{\cdot}_{\MJ}$, where $\mean{ \cdot }_{\MJ}$ denotes averaging over the realizations of the random connectivity $\MJ$.
Then the $n$-th moment of the eigenvalues of $\MA$ is $\varphi_n[\MA]\coloneqq\varphi[\MA^n]$.
First, if we compute the first moment $\varphi[\cdot]$ on both sides of \cref{eq:Cov_x_omega_full}, we recover the mean-field theory for the diagonal entry or autocorrelation $\Cxf$ (\cref{eq:single_neuron_MFT}; derivations in \cref{app:second_moment_and_dimension}).
Similarly, further computing the second moment $\varphi_2[\cdot]$ of \cref{eq:Cov_x_omega_full} gives the (frequency-dependent) dimension of $x$,
\begin{equation}
    \Dxf =\frac{\left(\varphi_1\lrsq{\MCxf}\right)^2}{\varphi_2\lrsq{\MCxf}}= \frac{\left(1+\omega^2-g^2\phiprime^2 \right)^2}{\left(1+\omega^2\right)^2+\left(1+\omega^2-\frac{\sigma^2}{\Cxf}\right)^2-g^4\phiprime^4}.
\label{eq:dim_x_w_sigma}
\end{equation}
Interestingly, when setting $\sigma=0$, \cref{eq:dim_x_w_sigma} readily reduces to exactly the same result previously derived in \cite{Clark2023}, albeit here we derive it using random matrix theory (\cref{app:second_moment_and_dimension}), which is a quite different approach from the two-site cavity method used in \cite{Clark2023} (\cref{app:cavity_unit_four_point_function}).

In sum, supported by multifaceted theoretical and numerical evidence (\cref{fig:example_covariance_and_relative_error,sec:effective_dynamics,sec:theoretical_validation}), we hypothesize that the proposed covariance matrix expressions \cref{eq:ansatz_of_Cov_phi_omega,eq:Cov_x_omega_full} are asymptotically exact in the large-network limit of $N\to \infty$, across dynamical regimes including chaos.

\section{Covariance spectrum}
\label{sec:eigenvalue_spectrum_of_covariance_matrices}
We are now ready to study the spectrum of the covariance matrices $\MCphif$ and $\MCxf$, which describe additional information about the shape of neural population activity beyond the dimension.
The ansatz for $\MCphif$ (\cref{eq:ansatz_of_Cov_phi_omega}) directly indicates that we can describe its (frequency-dependent) spectrum $\Pphif$ using the expression for the spectrum under linear dynamics $p_{\text{lin}}(\lambda;\omega)$ (Eqs.~5 and 20 in \cite{Hu2022}) by substituting $g$ with $\geff$ (\cref{eq:g_eff_definition}),
\begin{equation}
    p^{\phi}(\lambda;\omega, g, \sigma)=p_{\text{lin}}(\lambda; \omega, \geff, \sigeff(\omega)),
\label{eq:Pphif_using_effective_Plin}
\end{equation}
where $\geff=g\phiprime$ (\cref{eq:g_eff_definition}) and $\sigeff^2(\omega)=(1-g^2\phiprime^2+\omega^2)\Cphif = (1+\omega^2) C^{\tilde{\zeta}}(\omega)$ (see \cref{sec:effective_dynamics} and \cref{app:extend_zeta}).

For the covariance matrix of currents $\MCxf$, \cref{eq:Cov_x_omega_full} also allows us to determine its spectrum in terms of network parameters. In particular, once $\phiprime$ and $\Cphif$ are determined by the mean-field theory for given parameters $g$ and $\sigma$ (\cite{Schuecker2018}, \cref{eq:single_neuron_MFT}), determining the spectrum of $\MCxf$ becomes a random matrix problem, albeit an involved one. 
Here, instead of deriving an analytical expression for the spectrum $\Pxf$, we simply apply the Monte Carlo method based on \cref{eq:Cov_x_omega_full} with randomly generated $\MJ$
matrices, and $\Pxf$ can then be estimated by averaging the resulting eigenvalues across many $\MJ$ realizations. Note that the scalar quantities such as $\phiprime$ and $\Cphif$ are still computed from the mean-field theory, and this Monte Carlo method remains vastly more efficient than direct simulations of nonlinear networks (\cref{eq:network_model}).

\begin{figure}[!htbp]
    \centering
    \includegraphics[width=\linewidth]{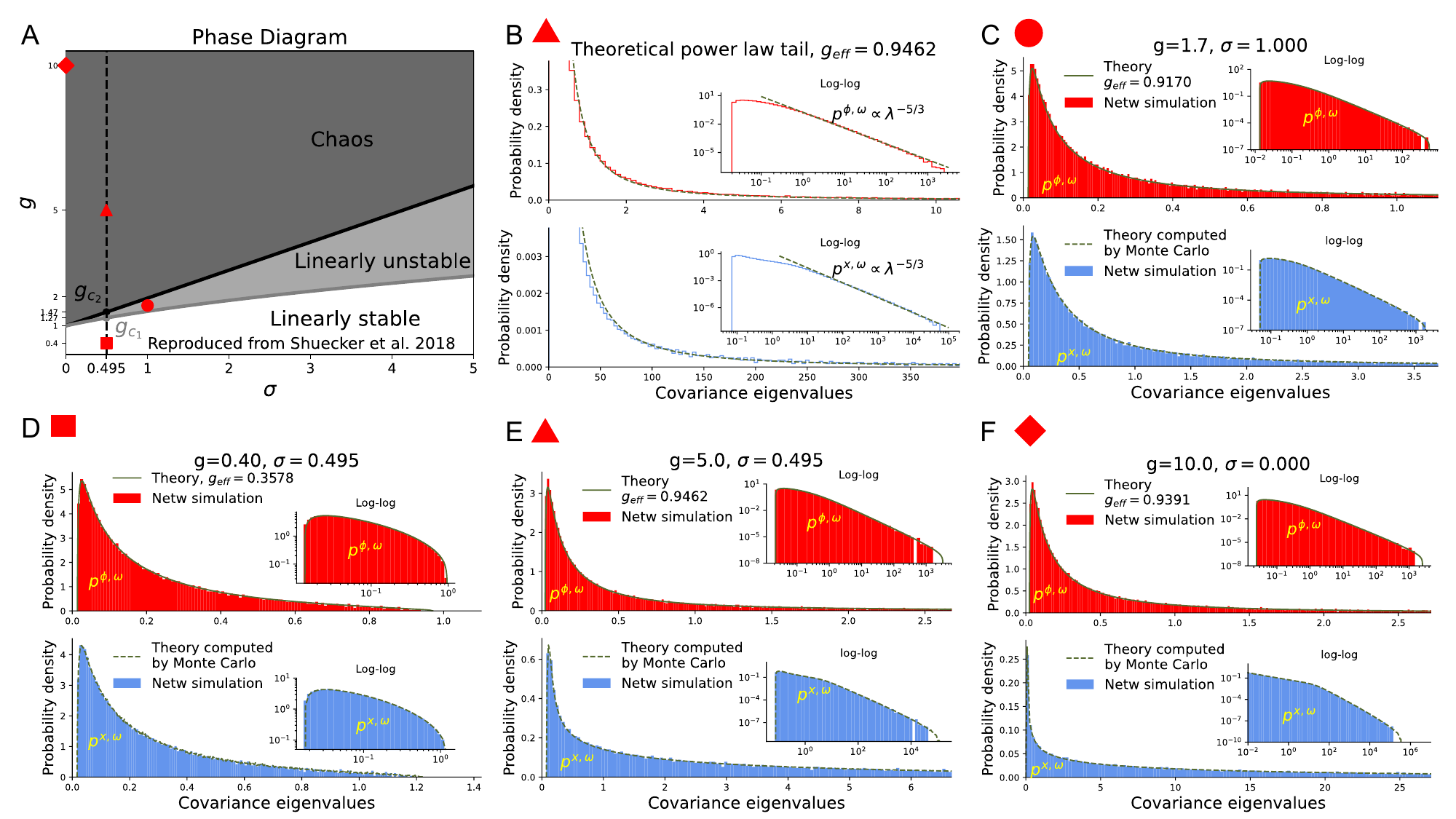}
    \caption{(\textbf{A}) Phase diagram of dynamical regimes for the nonlinear randomly connected RNN (\cref{eq:network_model}), reproduced from \cite{Schuecker2018}. For a fixed input level $\sigma>0$, there are two critical values $g_{c_1}$ and $g_{c_2}$ of the connection strength $g$ at the regime boundaries, where the larger $g_{c_2}$ separates the chaotic regime.
    (\textbf{B}) Power-law approximations (see text) of the tail of covariance spectra from network simulations in the chaotic regime (triangle marker in (A)).
    (\textbf{C-F}) Comparison of the zero-frequency covariance spectra from simulation (a single network and $\MJ$, histograms) and theory (curves, see text) with corrections for finite time samples (\cite{Hu2022} and \cref{sec:finite-sample_theory}) across dynamical regimes including chaos (noise-driven and deterministic).
    For the $x$ theoretical spectra, these are computed using the Monte Carlo method based on \cref{eq:Cov_x_omega_full} (see text).
    The $(g,\sigma)$ values correspond to the markers in (A). $N=4000$ for (C, E, F) and $N=3000$ for (D).
    }
    \label{fig:Eigen_spectrum_total}
\end{figure}

Our theoretical spectra are confirmed by numerical simulations, which show close matching of the probability density curves and the eigenvalue histograms computed from neural activity (\cref{fig:Eigen_spectrum_total}C-F). In particular, the matching is accurate across different dynamical regimes, which is also evidenced by comparing the estimate $\geff$ from simulations with that from the mean-field theory (\cref{fig:g_eff_phase_diagram}B).
As the connection strength $g$ increases (while fixing $\sigma$), both the $\phi$ and $x$ spectra develop longer tails of large eigenvalues (see also \cref{eq:spectrum_power_law_tail} below) corresponding to lower dimension, analogous to the phenomenon seen under linear dynamics \citep{Hu2022}.
However, this trend will eventually saturate as $g\rightarrow \infty$, where the spectra converge to a limiting shape and dimension \citep{Clark2023} (see also \cref{sec:effective_g_and_transition_to_chaos}).
The spectrum of the currents $\Pxf$ differs from that of the firing rates $\Pphif$ at small eigenvalues, with a higher peak density near 0 (\cref{fig:Eigen_spectrum_total}D-F, see also the log-log scale insets). In fact, we can use \cref{eq:Cov_x_omega_full} to prove that when $\sigma=0$ (for all $g$), the left edge of the support of $\Pxf$ is exactly 0 (\cref{app:left_support_of_x_spectrum}, \cref{fig:Eigen_spectrum_total}F), which is in contrast to the positive left edge of $\Pphif$. These results are consistent with the observation that the dimension of the currents $x$ is lower than that of the firing rates $\phi$, previously noted in \cite{Clark2023} (see also \cref{sec:transition_to_chaos_and_dimension_gap,app:non-negative_dimension_gap}).

Although we do not yet have a closed-form expression for the current covariance spectrum $\Pxf$, we can analytically characterize the large eigenvalues or tail of the spectrum $\Pxf$ when $\geff\to 1^-$ (its supremum is 1, see \cref{sec:proof_geff_bound}). This is based on the method from \cite{Hu2022}, which relates the orders of moments $\mathbb{E}\lrsq{\lambda^n}$ in terms of $(1-\geff^2)$ and the power-law exponent. We can then use the analytical expressions of the first two moments of $\Pxf$  (\cref{eq:first_moment_Cov_x_omega} and \cref{eq:second_moment_Cov_x_nonzero_frequency_sigma>0}) 
to show that (\cref{app:power_law_tail}),
\begin{equation}
    \Pphif \propto \lambda^{-\frac{5}{3}},\quad \Pxf\propto\lambda^{-\frac{5}{3}},\quad \text{when}\quad\geff\to1^-,\;\;\lambda\to\infty. 
\label{eq:spectrum_power_law_tail}
\end{equation}
Here we also include the result for $\Pphif$ for comparison, which directly follows from the expression in \cref{eq:Pphif_using_effective_Plin} based on results from \cite{Hu2022}.
Interestingly, despite clear differences for small eigenvalues (\cref{fig:Eigen_spectrum_total}D-F), both spectra $\Pphif$ and $\Pxf$ follow power laws with an identical exponent of $-5/3$ 
, which is the same as that in the linear dynamics \citep{Hu2022}.

As we show later, for the case with noise input $\sigma>0$, the effective connection strength $\geff$ is finitely bounded away from the critical value 1 (\cref{fig:g_eff_phase_diagram}D and \cref{sec:proof_geff_bound}). This is different from the situation when $\sigma=0$, where $\geff$ approaches 1 as $g\rightarrow 1^+$. This means that the power law in \cref{eq:spectrum_power_law_tail} is generally an approximation applicable when $\geff$ is close to 1, and can be accurate even for $\geff\approx 0.7$ (\cite{Hu2022} and \cref{fig:different_combination_geff=0.7}).
Moreover, such results are important and broadly applicable for describing the covariance spectrum under chaos. This is because $\geff$ stays close to 1 in the chaotic regime (described later in \cref{sec:effective_g_and_transition_to_chaos}), so the power-law theory (\cref{eq:spectrum_power_law_tail}) can be highly accurate (\cref{fig:Eigen_spectrum_total}B).

We want to note that our frequency-dependent covariance spectra contain temporal information about the network dynamics \citep{Hu2022,Calvo2024}. In particular, the zero-frequency $\omega=0$ spectrum describes correlations over a long time window and, for example, is suitable for analyzing calcium imaging data \citep{Hu2022}.
Activity and correlations at fast time scales are described by the covariance matrix and spectrum with nonzero $\omega$, where $\geff$ is replaced by 
\begin{equation}
 \geff(\omega) = \frac{\geff}{\sqrt{1+\omega^2}}=\frac{g\phiprime}{\sqrt{1+\omega^2}},
\label{eq:geff_w}
\end{equation}
according to \cref{eq:ansatz_of_Cov_phi_omega,eq:Cov_x_omega_full} (\cref{fig:special_spectrum_autocorrelation}A). 
In particular, the power-law approximations (\cref{eq:spectrum_power_law_tail}) still apply when $\geff(\omega)$ is close to 1 (\cref{fig:special_spectrum_autocorrelation}A).
We can also obtain the time-lagged covariance matrix $\MC^{a}(\tau)$ and the zero-timelag covariance spectrum $p^a(\lambda;\tau=0)$ (note that the spectrum is only meaningfully defined for $\tau=0$ where $\MCphitau$ is symmetric positive semidefinite) by numerically taking an entry-wise inverse Fourier transform of the frequency-dependent covariance matrix expressions (\cref{eq:ansatz_of_Cov_phi_omega,eq:Cov_x_omega_full}, see \cref{sec:time-lag_dimension}). The zero-timelag covariance spectra computed this way match closely with direct nonlinear network simulations (\cref{fig:special_spectrum_autocorrelation}BC).

\section{Effective recurrent connection strength $\geff$ and transition to chaos} 
\label{sec:effective_g_and_transition_to_chaos}
The effective recurrent connection strength $\geff$ (\cref{eq:g_eff_definition}) provides a unified description of how both the dimension and covariance spectrum change across dynamical regimes. While $\geff$ has previously been linked to network memory capacity, neuronal autocorrelation time constants \cite{Schuecker2018}, and dimension \cite{Clark2023}, we show here that it also determines the shape of the firing-rate covariance spectrum (\cref{eq:Pphif_using_effective_Plin}). This can be seen clearly from \cref{eq:ansatz_of_Cov_phi_omega}: if we introduce a scaled connectivity matrix $\MJ_0=\MJ/g$, then the matrix $\MCphif$ is, up to a scalar and the shape of its spectrum, dependent only on $\geff$ and $\MJ_0$ (and not on $g$). A similar result also holds for the currents $x$ when there is no noise (i.e., for the first term in \cref{eq:Cov_x_omega_full} and for the dimension, \cref{eq:dim_x_w_sigma0} in \cref{app:second_moment_of_Cov_x_omega}). When there is nonzero noise $\sigma>0$, the covariance matrix and spectrum of $x$ further depend on the relative magnitude among $\sigma^2$, $\Cxf$, and 
$\Cphif$, as governed by the mean-field theory \cref{eq:single_neuron_MFT}, which can also be seen in the dimension expression \cref{eq:dim_x_w_sigma}.

\begin{figure}[!htbp]
    \centering
    \includegraphics[trim=0cm 1.5cm 0cm 0cm, clip, width=\linewidth]{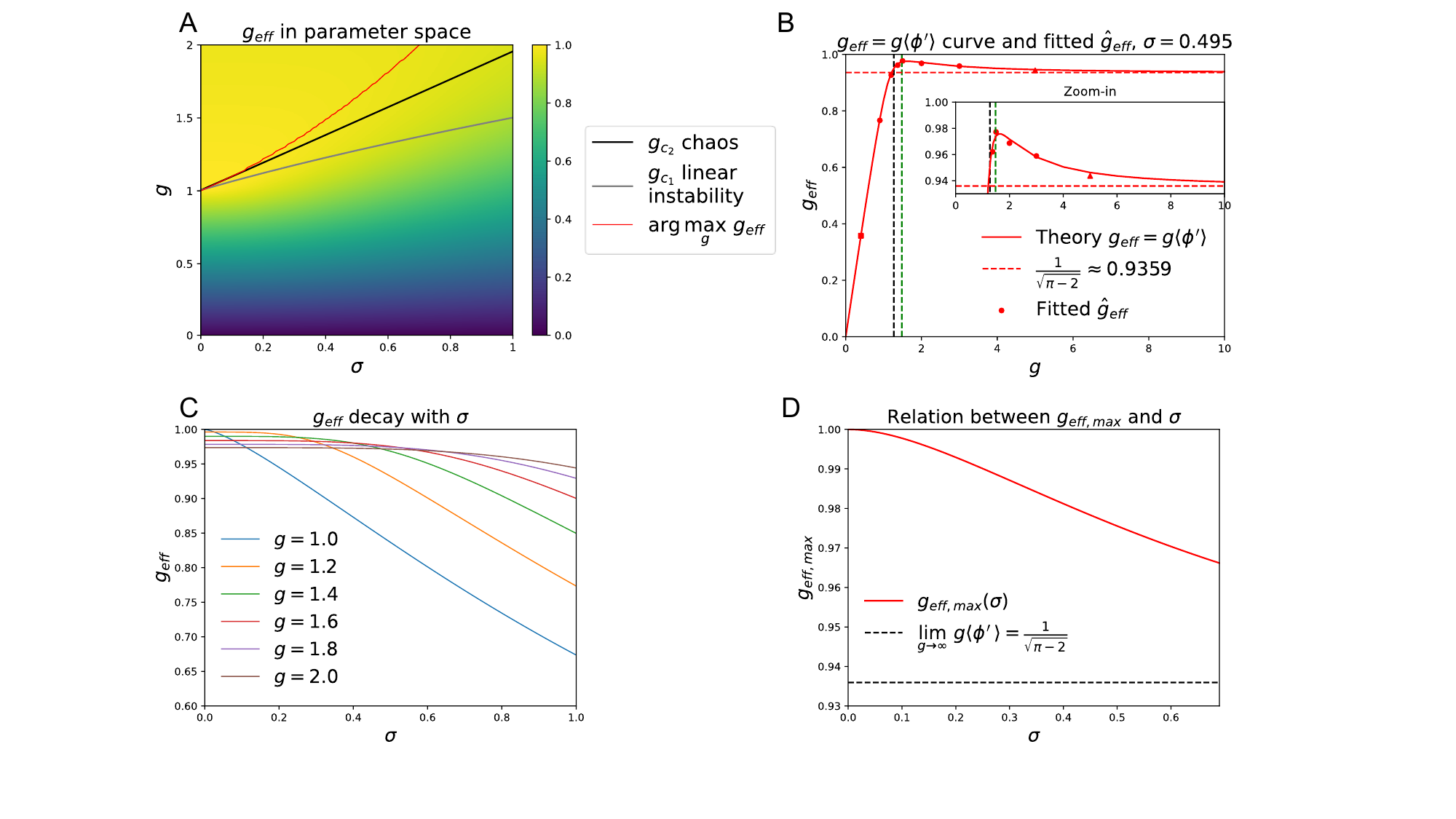}
    \caption{(\textbf{A}) The effective connection strength $\geff$ (color) as a function of $(g,\sigma)$ in the network model \cref{eq:network_model}. The black and gray curves are boundaries for the chaos and linear instability regions \citep{Schuecker2018}, respectively, and are the same as in \cref{fig:Eigen_spectrum_total}A.
    The red curve shows the location of the maximum of $\geff$ for each fixed $\sigma$.
    (\textbf{B}) The effective connection strength $\geff=g\phiprime$ (\cref{eq:g_eff_definition}) solved from the mean-field theory (curve) compared with the estimates $\hat{g}_{\text{eff}}$ (dots) obtained by fitting the theoretical covariance spectrum (\cref{eq:Pphif_using_effective_Plin}) to network simulations.
    (\textbf{C}) $\geff$ as a function of $\sigma$ for different values of $g$.
    (\textbf{D}) The relationship between the maximum of $\geff$ over $g$ (while fixing $\sigma$, e.g., in (B)) and $\sigma$.
    }
    \label{fig:g_eff_phase_diagram}
\end{figure}

We can then focus on how $\geff$ changes with network parameters $g,\sigma$ and across dynamical regimes, especially around the transition to chaos (\cref{fig:g_eff_phase_diagram}).
First, one can prove that $\geff = g\phiprime < 1$ for all $g, \sigma>0$ as long as the dynamics is not a fixed point (\cref{sec:proof_geff_bound}).
For a fixed $\sigma>0$, in the linear regime (both linearly stable and unstable, \cref{fig:Eigen_spectrum_total}A), $\geff$ increases quickly with $g$ from 0 until the transition to chaos ($g<g_{c_2}$). 
In the chaotic regime, $\geff$ stays close to $1$ and eventually converges to $1/\sqrt{\pi-2}\approx 0.9359$ as $g\rightarrow \infty$ (\cref{fig:g_eff_phase_diagram}B,  \cite{Clark2023,Crisanti2018}). 
This asymmetric behavior of $\geff$, and consequently of the spectrum (\cref{fig:Eigen_spectrum_total}C-F) and dimension (\cref{fig:Dimension_gap}CD), on the two sides of the chaos boundary is reminiscent of previous results about memory properties of such networks \citep{Toyoizumi2011}.
The near-critical $\geff$ in the chaotic regime without fine tuning is intriguing. One implication is that we can use a simple power-law approximation (\cref{eq:spectrum_power_law_tail}) to describe the covariance spectrum. The near-critical $\geff$ may also be related to findings in experimental data, where the fitted $\hat{g}_{\text{eff}}$'s in various brain regions \citep{Morales2023} and frequency bands \citep{Calvo2024} are often close to the critical value of 1.

When fixing $g$, $\geff$ decreases monotonically with $\sigma$ (\cref{fig:g_eff_phase_diagram}C). 
The existence of nonzero noise also prevents the correlation time $\tau_{1/2}$ of the neuron dynamics (defined by $ C^{a}(\tau_{1/2})=\frac{1}{2}C^a(0)$, $a\in\{x,\phi\}$) from diverging to infinity.
Moreover, the maximum of $\geff$ across $g$ is bounded away from 1 for any fixed $\sigma>0$ (\cref{fig:g_eff_phase_diagram} and \cref{sec:proof_geff_bound}), and the only possibility of achieving $\geff\to 1^{-}$ is when $\sigma \rightarrow 0$ (and $g\to 1$).

\subsection{Dimension gap near the transition to chaos}
\label{sec:transition_to_chaos_and_dimension_gap}
\begin{figure}[!htb]
    \centering
    \includegraphics[trim=0cm 2.5cm 0cm 1cm, clip, width=\linewidth]{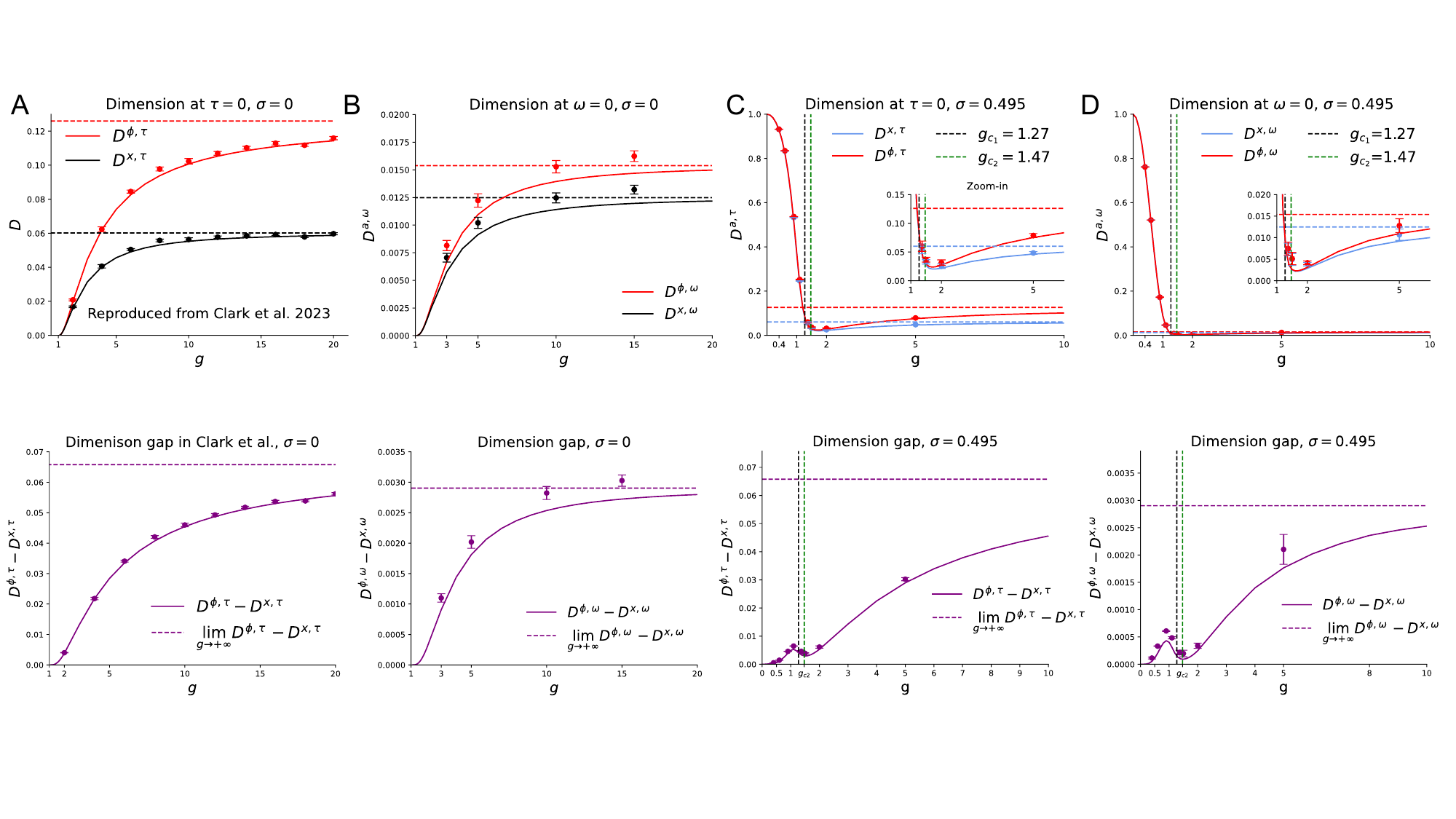}
    \caption{(\textbf{A}) Zero-timelag dimensions $D^{a,\tau}(\tau=0)$, $a\in\{x,\phi\}$ for currents and firing rates in the deterministic case with $\sigma=0$ (top, reproduced from \cite{Clark2023}) and the dimension gap $\Delta D^{\tau}$ (bottom). The solid curves are theory (\cite{Clark2023}, \cref{sec:time-lag_dimension}) and dots are averages from simulations of \cref{eq:network_model} ($N=4000$ with 5 realizations of $\MJ$; error bars are standard errors across network realizations). Dashed lines are the asymptotic values as $g\rightarrow \infty$.
    (\textbf{B}) Same as (A), but for the zero-frequency dimensions $D^{a}(\omega=0)$ based on the frequency-dependent theory \cref{eq:D_phi_omega,eq:dim_x_w_sigma} (see also \cref{fig:zero-frequency_dim_change_with_N} for a discussion of deviations from simulations). 
    (\textbf{C}) Same as (A), but with noise input $\sigma=0.495$.
    (\textbf{D}) Same as (C), but for zero-frequency dimensions.
    }
    \label{fig:Dimension_gap}
\end{figure}

It was noted in \cite{Clark2023} that currents and firing rates have different dimensions $D^{\phi,\tau} >D^{x,\tau}$ (\cref{fig:Dimension_gap}A, $D^{a,\tau}$ is the zero-timelag dimension for $\MC^{a}(\tau=0)$), and the \emph{dimension gap} 
\begin{equation}
    \Delta D:=D^\phi - D^x,
\label{eq:dimension_gap_zero_timelag}
\end{equation}
decreases to 0 when $g\to 1^+$ approaching its critical value for chaos (\cite{Clark2023}, \cref{fig:Dimension_gap}A, bottom). However, in this deterministic model considered in \cite{Clark2023}, one is limited to the range of $g>1$ (the network remains at a fixed point for $g<1$). In contrast, our noise-driven model \cref{eq:network_model} allows us to examine the full picture of the dimension gap during the transition from linear to chaotic regimes (\cref{fig:Eigen_spectrum_total}A). It is then natural to ask for this general noise-driven system: Does the dimension gap still vanish at the transition to chaos? Are the rate and current dimensions exactly equal within the linear regimes?

We give conclusive answers to these questions using the theoretical expressions for the dimension (\cref{eq:D_phi_omega,eq:dim_x_w_sigma}), which are also supported by extensive direct simulations of the network. 
We find that the dimension gap $\Delta D$ is always non-negative (see \cref{app:non-negative_dimension_gap} for a proof for the frequency-dependent dimensions) and has a non-monotonic dependence on $g$ for $\sigma>0$. It has a local maximum in the linear stable regime, and a local minimum before increasing with $g$ in the chaotic regime (\cref{fig:Dimension_gap}CD). 
The nontrivial local minimum of $\Delta D$ is analogous to the vanishing of $\Delta D$ at $g\to 1^+$ in the $\sigma=0$ case, but the local minimum is located near but not coinciding with the chaos boundary (\cref{fig:dimension_gap_and_derivative}).

The above phenomena of dimension gap hold for both zero-timelag (\cref{fig:Dimension_gap}A,C) and frequency-dependent dimensions (\cref{fig:Dimension_gap}B,D).
For the zero-timelag dimensions, $\Dphitau$ and $\Dxtau$ converge to their theoretical asymptotic values of 0.126 and 0.0602, respectively (\cref{sec:time-lag_dimension}, \cite{Clark2023}).
These asymptotic values are identical with or without noise input (\cref{fig:Dimension_gap}A,C, see \cref{sec:time-lag_dimension}) and match previous results on the noiseless case \citep{Clark2023}.
For the frequency-dependent dimensions, their asymptotic values can be obtained from \cref{eq:dim_x_w_sigma,eq:D_phi_omega} using the limit $\geff = g\phiprime\to 1/\sqrt{\pi-2}$ \citep{Clark2023,Crisanti2018},
\begin{equation}
\lim_{g\to\infty}\Dphif=\left[\frac{(1+\omega^2)(\pi-2)-1}{(1+\omega^2)(\pi-2)}\right]^2, \quad\lim_{g\to\infty}\Dxf=\frac{\lrsq{(1+\omega^2)(\pi-2)-1}^2}{2(1+\omega^2)^2(\pi-2)^2-1}.
\label{eq:asymptotic_frequency_dimension}
\end{equation}
Again there is no dependence on the magnitude of noise input $\sigma$ (\cref{fig:Dimension_gap}B,D). At the zero-frequency $\omega=0$, the limiting dimensions are approximately 0.0154 and 0.0125, respectively.

We note that since the dimension gap, and even the dimensions themselves, are very small in magnitude around the chaotic regime (see y-scale in \cref{fig:Dimension_gap}C,D, and the zoomed-in insets), it is difficult to determine the above properties of $\Delta D$ ``by eye'' using only simulations of the network. Therefore, the accurate and asymptotically exact ($N\rightarrow \infty$) theory (\cref{eq:D_phi_omega,eq:dim_x_w_sigma}) is crucial for reaching our conclusions delineating the relationship between dimension gap and transition to chaos.

\subsection{Inferring dynamical regimes and network parameters}
\label{sec:infer_parameter_main_text}
\begin{figure}[H]
    \centering
    \includegraphics[trim=0cm 11cm 0cm 0cm, clip, width=1\linewidth]{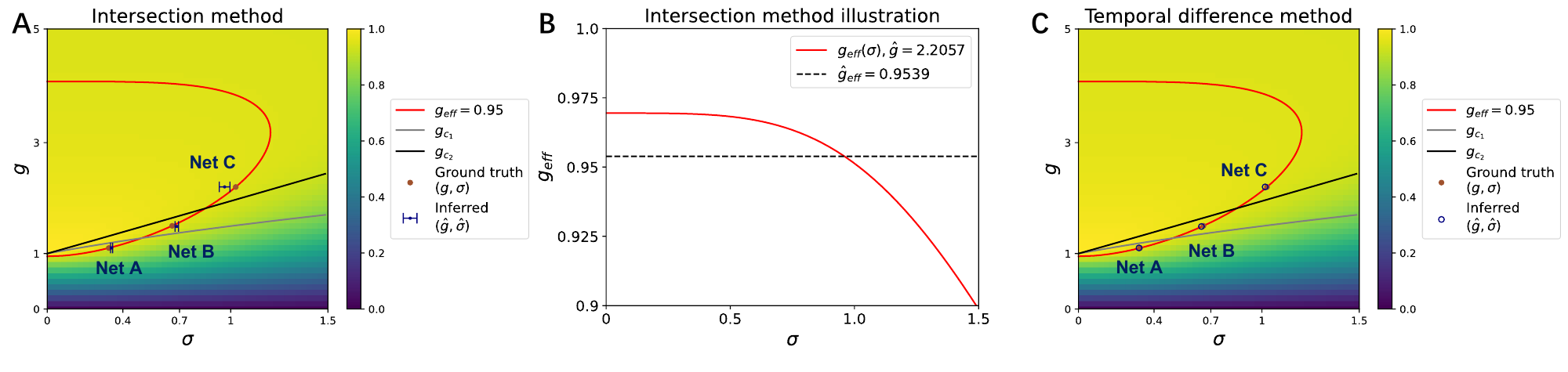}
    \caption{Estimating network parameters from three different dynamical regimes. The $(g,\sigma)$ parameters for network A, B, C are (1.1, 0.3324), (1.5, 0.6742), and (2.2, 1.0173), respectively.
    (\textbf{A}) Estimated parameters using the intersection method (see text) plotted on the phase diagram similar to \cref{fig:Eigen_spectrum_total} with color representing the $\geff$ value. 
    The red curve is the contour for $\geff=0.95$. 
     The inferred $(\hat{g},\hat{\sigma})$ (mean $\pm$ SD across 10 bootstrap resamples of the large time bins, \cref{sec:methods}) are: Net A: $(1.1015\pm0.0036, 0.3475\pm0.0055)$, Net B: $(1.4885\pm0.0040, 0.6993\pm0.0095)$, Net C: $(2.2017\pm0.0038, 0.9573\pm0.0285)$; the error bar shows the standard deviation of $\hat{\sigma}$.
    (\textbf{B}) Illustration of the intersection method used to find $\hat{\sigma}$ in (A).
    (\textbf{C}) Same as (A), but using the temporal difference method (\cref{app:data_analysis_method}) to obtain more accurate $\sigma$ estimates (mean $\pm$ SD; error bars are omitted because they are too small to be visible): Net A: $0.3266\pm0.0004$, Net B: $0.6635\pm0.0005$, Net C: $1.0073\pm0.0008$.
}
    \label{fig:contour_geff0.95}
\end{figure}

As discussed above, our theory (\cref{eq:Pphif_using_effective_Plin}) shows that the shape of the firing-rate covariance spectrum is determined solely by the effective recurrent connection strength $\geff$. However, for a given $\geff$ there can be many combinations of network parameters and from different dynamical regimes lead to the same $\geff$ value (\cref{fig:contour_geff0.95}A). This means that one may not be able to identify the dynamical regime by fitting the covariance spectrum to the firing rate data alone. Here, we introduce a method to address this limitation by inferring the dynamical regimes and network parameters (i.e., $g$ and $\sigma$).

First, it estimates the firing-rate function $\phi(x)$ following the approach in \cite{Lim2015}, which assumes that the currents $\{x_i(t)\}$ are Gaussian distributed. This assumption is justified in large randomly connected neural networks by mean-field theory \citep{Sompolinsky1988}.
We can then estimate the currents $\{x_i(t)\}$ via $\hat{\phi}^{-1}$ and compute the dynamic gain $\phiprime$ within $\geff$ (\cref{eq:g_eff_definition}). Lastly, to estimate the external input magnitude $\sigma$, recall that $\geff$ decays monotonically as $\sigma$ increases for a fixed $g$ (\cref{fig:g_eff_phase_diagram}C, \cref{sec:par_geff_par_sigma}). After obtaining $\hat{g}$ and computing its corresponding $\geff(\sigma)$ curve, one can draw a horizontal line at $\hat{\geff}$, and the intersection gives an estimate $\hat{\sigma}$ (\cref{fig:contour_geff0.95}B).

To demonstrate the effectiveness of the method, we simulated three networks with different combinations of parameters $g,\sigma$.
As shown in \cref{fig:contour_geff0.95}A, the method accurately estimated $g, \sigma$ in the simulated networks, which allows one to correctly identify the different dynamical regimes these networks are in, despite all the networks having the same $\geff$ and the same shape of firing rate covariance spectrum. Additional details and an improved method applicable to high temporal resolution data are described in \cref{app:data_analysis_method}.

\section{Extensions to other network models}
\label{sec:extension_of_theory}
Up to now, we have focused on a simple yet fundamental network model (\cref{eq:network_model}, \cite{Sompolinsky1988,Schuecker2018,Hu2022,Clark2023}). It is important to ask whether our theory of the covariance matrix and spectrum (\cref{eq:ansatz_of_Cov_phi_omega,eq:Cov_x_omega_full,eq:Pphif_using_effective_Plin}) can be applied to broader models and what modifications may be needed. In this section, we make some initial steps toward this question by considering a number of model variations in the neuronal firing-rate function and network connectivity. The numerical and theoretical results below strongly support the notion that the core theory, our effective linearized dynamics ansatz, can be generalized to a broad range of network models. This also leads us to propose a generalized form of the covariance matrix ansatz (\cref{eq:ansatz_of_Cov_phi_omega_symmetric}) and effective connection strength (\cref{eq:g_c_eff_in_part_symm}).

\subsection{Different nonlinear firing-rate function}
\label{app:different_nonlinearity}
We have been focusing on a specific firing-rate function $\phi(x)=\tanh{(x)}$. However, we believe that the validity of our ansatz and spectrum theory does not rely on the details of $\phi(x)$. To illustrate, we consider a piecewise linear firing-rate function (\cref{fig:summary_Plin}A).
\begin{equation}
    \phi(x)=\left\{\begin{split}
        & -1, & x<-1\\
        & x, & x\in[-1,1]\\
        & 1, & x>1
    \end{split}  \right. .
    \label{eq:def_Plin}
\end{equation}
Under this $\phi(x)$, the covariance matrix ansatz and theoretical spectra (\cref{eq:ansatz_of_Cov_phi_omega,eq:Cov_x_omega_full,eq:Pphif_using_effective_Plin}) continue to apply without modification (\cref{fig:summary_Plin}C-F). The equation $\geff=g\phiprime$ still gives the correct effective recurrent connection strength when computing the dynamic gain using mean-field theory under this new $\phi(x)$ \citep{Schuecker2018}. The dependence of $\geff$ on $g$ (\cref{fig:summary_Plin}B) is qualitatively similar to that of the $\phi(x)=\tanh(x)$ case and shares the same asymptotic value as $g\rightarrow \infty$ (\cref{fig:g_eff_phase_diagram}B).

\subsection{Sparse and E-I networks}
\label{app:sparse_connection_and_EI}
The Gaussian random connectivity (\cref{sec:network_model}) we have considered thus far is a minimal model for recurrent networks, but it is all-to-all fully connected with probability 1, and the outgoing connections of a neuron can have both positive and negative weights. In contrast, biological neural networks often have sparse connections and different neuron types, such as excitatory and inhibitory. It is therefore important to check whether the results can apply to these more biologically realistic connectivity models. 
To this end, we consider two network models, described as follows. First is a sparse random connectivity matrix, whose entries are drawn i.i.d. from a spiked Gaussian distribution: $J_{ij}=0$ with probability $1-p$, and $J_{ij}\neq0$ and is drawn from $\mathcal{N}\lrbk{0,\frac{g^2}{pN}}$ with probability $p$ (\cref{fig:summary_sparse_J}A). Such networks have a connection probability equal to $p$ and still have the same statistics of zero mean and $g^2/N$ variance as our original model. 

Another network model we consider here is an E-I network with excitatory (E) and inhibitory (I) neurons obeying Dale's law; that is, the outgoing connections are all positive and all negative, respectively.
In this simple E-I network, the same as in \cite{Hu2022}, there are $N/2$ E and $N/2$ I neurons. The connections are formed independently with probability $K/N$, while the connection weight is $w_e=w_0/\sqrt{K\lrbk{1-K/N}}$ or $w_i=-w_e$, depending on the neuron type from which it originates.
Importantly, the mean connection weight is zero and the variance is $\var\lrsq{J_{ij}}=w_0^2/N$, regardless of the cell types of neurons $i,j$, and 
this E-I network is expected to have similar dynamics and covariance spectrum as a Gaussian random network with parameter $g=w_0$ \citep{Hu2022,Kadmon2015}. 

For both the sparse and E-I networks, our theory of covariance matrix ansatzes (\cref{eq:ansatz_of_Cov_phi_omega,eq:Cov_x_omega_full}, \cref{fig:summary_sparse_J}CE, \cref{fig:general_theory_main_text}E) and spectra (\cref{fig:summary_sparse_J}DF, \cref{fig:general_theory_main_text}F), and the effective connection strength $\geff$ (\cref{eq:g_eff_definition}, \cref{fig:summary_sparse_J}B, \cref{fig:general_theory_main_text}D) remain closely matching with network simulations.

\subsection{Partially symmetric connections and generalized ansatz}
\label{app:symmetric_connection}
\begin{figure}[H]
    \centering
    \includegraphics[trim=0cm 3cm 0cm 0cm, clip, width=\linewidth]{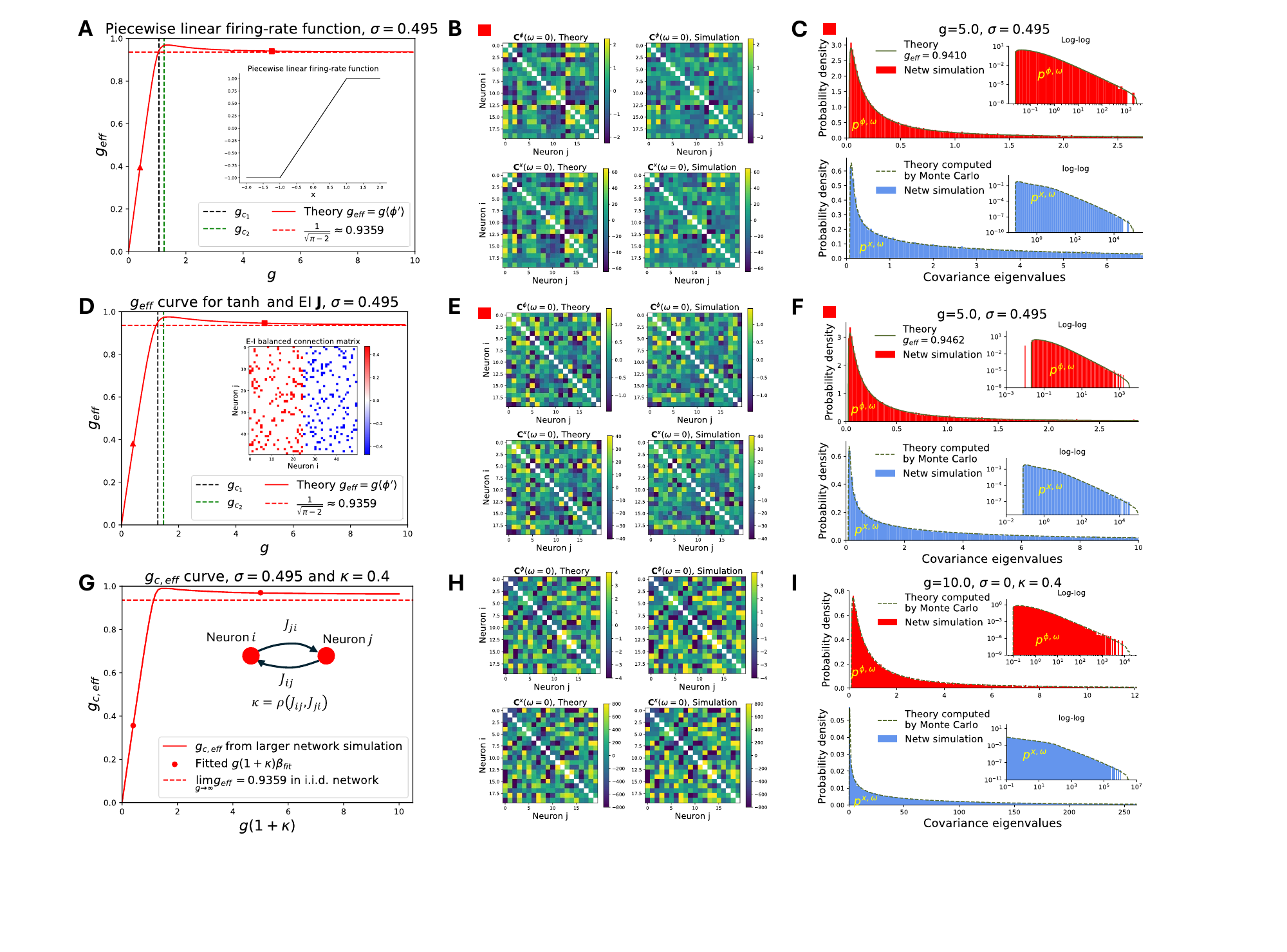}
    \caption{(\textbf{A-C}) Result summary for the piecewise linear firing-rate function (\cref{app:different_nonlinearity}). The corresponding $\geff$ curve and fitted $\hat{g}_{\text{eff}}$ (A); the inset of (A) is the piecewise linear firing-rate function and different markers represent different simulated networks. Example covariance matrices (B) and corresponding covariance spectra of the simulated network in the chaotic regime. (\textbf{D,E,F}) Similar to (A-C), but for the E-I network (\cref{app:sparse_connection_and_EI}), the inset in (A) shows an example of E-I network connections. (\textbf{G-I}) Similar to (A-C) but for the partially symmetric network (\cref{app:symmetric_connection}). Note that (H,I) are from networks without input and therefore do not correspond to the example networks in (G), but correspond to networks in \cref{fig:summary_symm_J}BC.
    }
    \label{fig:general_theory_main_text}
\end{figure}

Here we consider another way of generalizing the network connectivity model, where the connections are no longer independent but are correlated,
\begin{equation}
    \kappa=\rho(J_{ij},J_{ji}),\quad -1< \kappa < 1.
\label{eq:kappa_def}
\end{equation}
Such correlations are also related to \emph{reciprocal motifs} (\cref{fig:general_theory_main_text}G inset), which are widely observed in brain networks \citep{Song2005}, where the probability of having a pair of bidirectional connections differs significantly from that of a random shuffled network.
In particular, we consider a partially symmetric (or anti-symmetric when $\kappa<0$) random connectivity model satisfying \cref{eq:kappa_def} with zero-mean Gaussian distributed connections, as in \cite{Sompolinsky1987,Marti2018,Hu2022}.
The effect of partially symmetric connections on nonlinear network dynamics in terms of dimension has been studied in \cite{Clark2023}. Here we want to study its impact on the level of covariance spectrum and examine the validity of our covariance matrix ansatz (\cref{sec:moment_to_ansatz}). For simplicity, we focus on the autonomous case without external input ($\sigma=0$) unless stated otherwise.

Intriguingly, a direct application of the ansatz \cref{eq:ansatz_of_Cov_phi_omega,eq:Cov_x_omega_full} does not match network simulations (\cref{fig:part_symm_spectrum_supplement}A), indicating that the effect of partial symmetry $\kappa$ cannot be captured solely by plugging the partially symmetric $\MJ$ into the ansatz. 
We then explore whether a generalized ansatz of similar form can capture the covariance spectrum for partially symmetric connectivity. We replace the dynamic gain $\phiprime$ in \cref{eq:ansatz_of_Cov_phi_omega,eq:Cov_x_omega_full} by an undetermined positive number $\beta$ and check whether any choice of $\beta$ can fit the shape of the simulated spectra well (ignoring the scale factor for now, which we will return to later). 
This is indeed the case (\cref{fig:part_symm_spectrum_supplement}BC). 
Importantly, this matching between the ansatz and simulations is nontrivial, as the optimal $\beta$ values achieving the best fit to both the $\phi$ and $x$ spectra are very close (\cref{fig:summary_symm_J}AD, yellow curves). 
In contrast, if one uses the i.i.d. connectivity spectra (or equivalently plugs an i.i.d. random $\MJ$ into the ansatz), the optimal $\beta$ values are very different for fitting to the simulated $x$ and $\phi$ spectra, respectively (\cref{fig:summary_symm_J}AD, black curves).

Motivated by these numerical results, we heuristically derived a frequency-dependent expression for the generalized dynamic gain $\beta$ (\cref{app:dynamic_gain_part_symm}),
\begin{equation}
    \beta(\omega) = S^{\phi}(\omega) / S^x(\omega), 
\label{eq:beta_equation}
\end{equation}
where $S^a (\tau) = \mean{\delta a_i(t)/\delta\eta_i(t-\tau)}_{t}$ is the response function \citep{Clark2023}. 
\cref{eq:beta_equation} is derived by considering the properties of $\zeta_i(\omega) := \phi_i(\omega)- \beta(\omega) x_i(\omega)$ in the frequency domain, analogous to the arguments in \cref{sec:effective_dynamics}.
This theoretical $\beta(\omega=0)$ (vertical red dashed lines in \cref{fig:summary_symm_J}AD) not only matches closely with the optimal numerical $\beta$ obtained by fitting to the simulated spectra but also determines an expression for the scalar factor in the covariance matrix ansatz: 
$C^{\zeta}(\omega)=\Cphif-\beta(\omega) C^{x\phi}(\omega)-\beta^{\ast}(\omega)C^{\phi x}(\omega)+\lrabs{\beta(\omega)}^2C^x(\omega)$ (\cref{app:dynamic_gain_part_symm,app:estimate_beta_and_C_zeta}).
In sum, we arrived at a generalized covariance matrix ansatz: 
\begin{equation}
\begin{aligned}
    \MCphif =& \Czetaf\lrsq{\MI - \frac{\beta(\omega)}{1+i\omega}\MJ}^{-1}\lrsq{\MI - \frac{\beta(\omega)}{1+i\omega}\MJ}^{-\dagger},\\
    \MCxf =& \frac{\Czetaf}{1+\omega^2}\MJ\lrsq{\MI - \frac{\beta(\omega)}{1+i\omega}\MJ}^{-1}\lrsq{\MI - \frac{\beta(\omega)}{1+i\omega}\MJ}^{-\dagger}\MJ^T.
\end{aligned}
\label{eq:ansatz_of_Cov_phi_omega_symmetric}
\end{equation}
Note that when $\MJ$ is i.i.d. random (i.e., $\kappa=0$), $S^{\phi}(\omega) = \frac{\phiprime}{1+i\omega}$ and $S^x(\omega) = \frac{1}{1+i\omega}$ \citep{Clark2023}; then $\beta(\omega) = \phiprime$ and \cref{eq:ansatz_of_Cov_phi_omega_symmetric} reduces to the case of the original ansatz \cref{eq:ansatz_of_Cov_phi_omega,eq:Cov_x_omega_full}.

The generalized covariance matrix ansatz (\cref{eq:ansatz_of_Cov_phi_omega_symmetric}) indicates that the shape of the firing rate $\phi$ spectrum is the same as that of a linear network with partially symmetric random connectivity, whose variance and correlation parameters are $g\beta$ and $\kappa$ (focusing on zero-frequency $\omega=0$ for simplicity). Considering the critical transition in linear networks \citep{Hu2022}, we can define a $g_{c,\text{eff}}$ that would be on a similar scale to $\geff$ for the i.i.d. connectivity case and allows for intuitive comparisons:
\begin{equation}
    g_{c,\text{eff}} := g(1+\kappa) \beta.
\label{eq:g_c_eff_in_part_symm}
\end{equation}
\cref{fig:g_eff_curve_for_diff_kappa}AC shows example curves of how $g_{c,\text{eff}}$ depends on the parameter $g$ while fixing $\sigma$ and $\kappa$, along with verifications from network simulations (with the extended $\beta(\omega)$ expression for the case with external input $\sigma\ge 0$ in \cref{app:estimate_beta_and_C_zeta}). 
Importantly, $g_{c,\text{eff}}$ has the same qualitative shape as the $\geff$ curve in the i.i.d. random networks (\cref{fig:g_eff_phase_diagram}B, \cref{fig:summary_Plin}B, \cref{fig:summary_sparse_J}B, \cref{fig:summary_EI_J}B) and stays close to 1 when $g$ is large.

At the covariance matrix level, we numerically verified the generalized covariance matrix ansatz for the partially symmetric $\MJ$ (\cref{eq:ansatz_of_Cov_phi_omega_symmetric}) with network simulations and showed a close match (\cref{fig:summary_symm_J}BCEF). 
Furthermore, we provide theoretical support for the ansatz by computing the first two moments of the generalized covariance matrix $\MCphif$ (\cref{eq:ansatz_first_moment_part_symm,eq:ansatz_second_moment_part_symm}) using random matrix theory and showing that the moments, and consequently the frequency-dependent dimension, are exactly equal to the results derived using the two-site cavity methods in \cite{Clark2023} (\cref{app:first_second_moment_part_symm}).
Taken together, these results support the validity of the generalized covariance matrix ansatz and spectra for networks with motifs and non-i.i.d. connections.

\section{Discussion}
\label{sec:discussion}
In this study, we derived a theory for the covariance spectrum in randomly connected recurrent networks with nonlinear neuronal dynamics. The results are exact in the large-network limit and are valid across different dynamical regimes, including chaos. These results on the covariance spectrum provide a precise description and important insights into neural population dynamics, particularly regarding their shape beyond dimensionality.
The covariance matrix and spectrum are expressed in a frequency-dependent form (\cref{eq:ansatz_of_Cov_phi_omega,eq:Cov_x_omega_full}); they describe the neural activity in each frequency band and contain information about the temporal dynamics. In particular, the time-domain quantities, such as the zero-timelag covariance spectrum and dimension, can be obtained through an inverse Fourier transform (\cref{fig:special_spectrum_autocorrelation}BC).

The key step in deriving these results is a covariance matrix ansatz (\cref{eq:ansatz_of_Cov_phi_omega,eq:Cov_x_omega_full}), which provides an explicit link between neuronal dynamics, recurrent network connectivity, and all pairwise correlations within the network. 
This ansatz is confirmed through extensive numerical simulations (\cref{fig:example_covariance_and_relative_error,fig:Eigen_spectrum_total}), as well as by cross-checking with previous theoretical results on dimensionality.
The fact that the same population dynamics quantity, such as dimensionality, can be derived via two distinct approaches reveals a new connection between two lines of methods for analyzing the nonlinear dynamics of random networks: dynamical mean-field theory \citep{Sompolinsky1988,Kadmon2015,Schuecker2018,Clark2023} and random matrix theory \citep{Aljadeff2015b,Hu2022,Shao2023}.
These findings significantly extend previous theory on the covariance spectrum derived under linear dynamics \citep{Hu2022} and therefore provide further theoretical support for applying the covariance spectrum analysis to data. In particular, our results offer new insights for interpreting the fitted recurrent connection strength $\hat{g}$ as an effective parameter $\geff$, reflecting both anatomical connection weights and the dynamic gain of nonlinear neurons.

Our theory on the covariance matrix and spectrum highlights the unified role of the effective recurrent connection strength $\geff$, whose importance now extends beyond determining dimensionality \citep{Clark2023}. This suggests a simple, conceptual framework for understanding nonlinear network dynamics: while $\geff$ can be derived from classical mean-field theory focusing on a single neuron's dynamics \citep{Schuecker2018}, the effects of the recurrent connections are then summarized in our closed-form covariance matrix expressions (\cref{eq:ansatz_of_Cov_phi_omega,eq:Cov_x_omega_full}), which contain interaction terms of all orders, such as  $\MJ\MJ^T$, $\MJ^2(\MJ^{T})^2$, and $\MJ\MJ^T \MJ\MJ^T \MJ\MJ^T$.

We studied how $\geff$ changes across different dynamical regimes and through the transition to chaos (\cref{fig:g_eff_phase_diagram}B). 
Importantly, this revealed asymmetric trends on the two sides of the edge of chaos \citep{Toyoizumi2011}: a rapid increase in the linear regimes, while $\geff$ remains consistently close to the critical value of 1 in the chaotic regime without fine-tuning. This finding is potentially related to the observation that fitting experimental neural activity data often results in near-critical $\hat{g}_{\text{eff}}$ values \citep{Morales2023,Calvo2024}.
Furthermore, we characterized a more general version of the phenomenon of a diminishing dimension gap near the transition to chaos (\cref{fig:Dimension_gap}), previously noted in \cite{Clark2023}, thus revealing a full picture across various dynamical regimes.
Based on the theoretical insights, in particular the two factors in the $\geff$ expression (\cref{eq:g_eff_definition}), we developed an algorithm to infer the network parameters and therefore the dynamical regime the system is in from firing-rate data (\cref{sec:infer_parameter_main_text}), which may not be unambiguously determined by covariance spectrum fitting alone.

This work serves as a basis for further exploring the geometry of population dynamics in nonlinear systems, particularly in the chaotic regime, and many interesting questions remain open for future investigation.
For example, it would be interesting to explore more general types of recurrent network connectivity, such as those containing motifs \citep{Zhao2011,Hu2013}, spatially dependent connectivity \citep{Huang2019}, and connectivity with heterogeneous degrees \citep{Hu2014}; our results on extensions of the network model (\cref{sec:extension_of_theory}) provide encouraging evidence on the generality of the theory.
On the technical side, it is important to develop methods that directly derive the covariance matrix ansatz (\cref{eq:ansatz_of_Cov_phi_omega}), for example, by proving the uncorrelated property of the effective noise $\zeta$ (\cref{eq:zeta_define,eq:ansatz_Cov_zeta,app:extend_zeta}).

\section*{Acknowledgment}
This work was partially supported by an Early Career Scheme (Project No. 26303921) and a General Research Fund (Project No. 16103725) from the Research Grants Council of Hong Kong, and by a start-up fund from the Hong Kong University of Science and Technology.

\bibliographystyle{apalike} 
\bibliography{ref}

\newpage
\section*{Appendix}
\begin{appendix}

\section{Supplementary figures}
\label{app:supplementary_figure}
\beginsupplement
\begin{figure}[H]
    \centering
    \includegraphics[trim=0cm 10.3cm 0cm 0cm, clip, width=\linewidth]{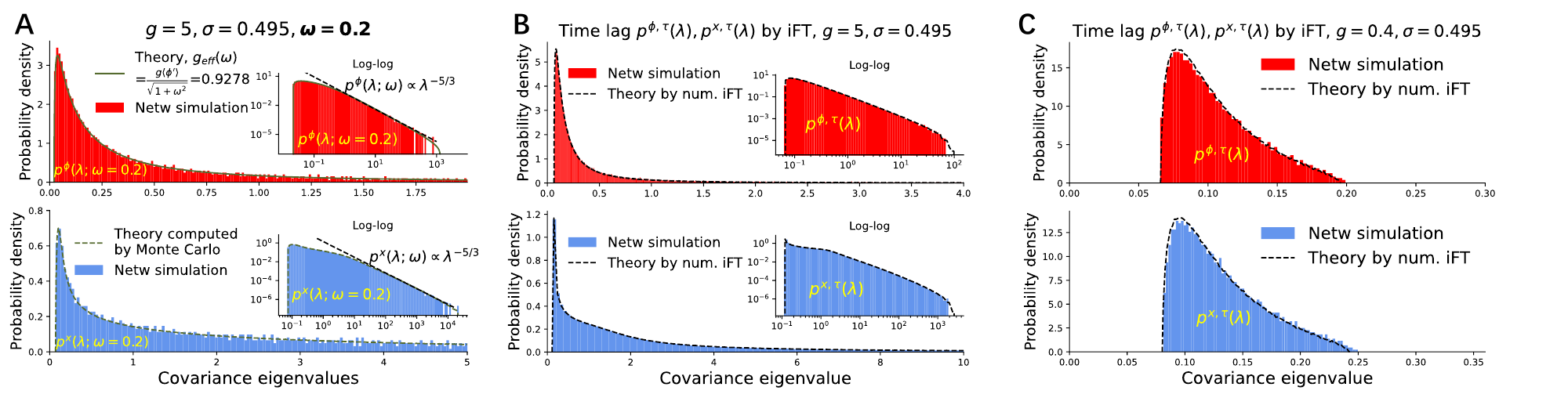}
    \caption{(\textbf{A}) Nonzero-frequency covariance spectrum $p^{\phi}(\lambda;\omega=0.2)$ and $p^{x}(\lambda;\omega=0.2)$ of a network at $g=5$, $\sigma=0.495$ (same as \cref{fig:Eigen_spectrum_total}E in the chaotic regime) but $N=2000$. The frequency-dependent effective connection strength is $\geff(\omega)=g\phiprime/\sqrt{1+\omega^2}$ (\cref{eq:geff_w}).
    (\textbf{B}) Zero-timelag covariance spectrum $\Pphitau$ and $\Pxtau$ of a network at $g=5$, $\sigma=0.495$ (same as (A) and \cref{fig:Eigen_spectrum_total}E) and $N=10000$. The theory curve is computed by an entry-wise inverse Fourier transform of the covariance matrices (\cref{eq:ansatz_of_Cov_phi_omega,eq:Cov_x_omega_full}) and averaged over 10 realizations of $\MJ$ with $N=10000$ (see \cref{sec:time-lag_dimension}).
    (\textbf{C}) Same as (B), but for a network in the linearly stable regime (\cref{fig:Eigen_spectrum_total}A, $g=0.4$, $\sigma=0.495$, same as \cref{fig:Eigen_spectrum_total}D but with $N=2000$.). 
    }
\label{fig:special_spectrum_autocorrelation}
\end{figure}

\begin{figure}[H]
    \centering
    \includegraphics[trim=0cm 5cm 0cm 3.5cm, clip, width=1\linewidth]{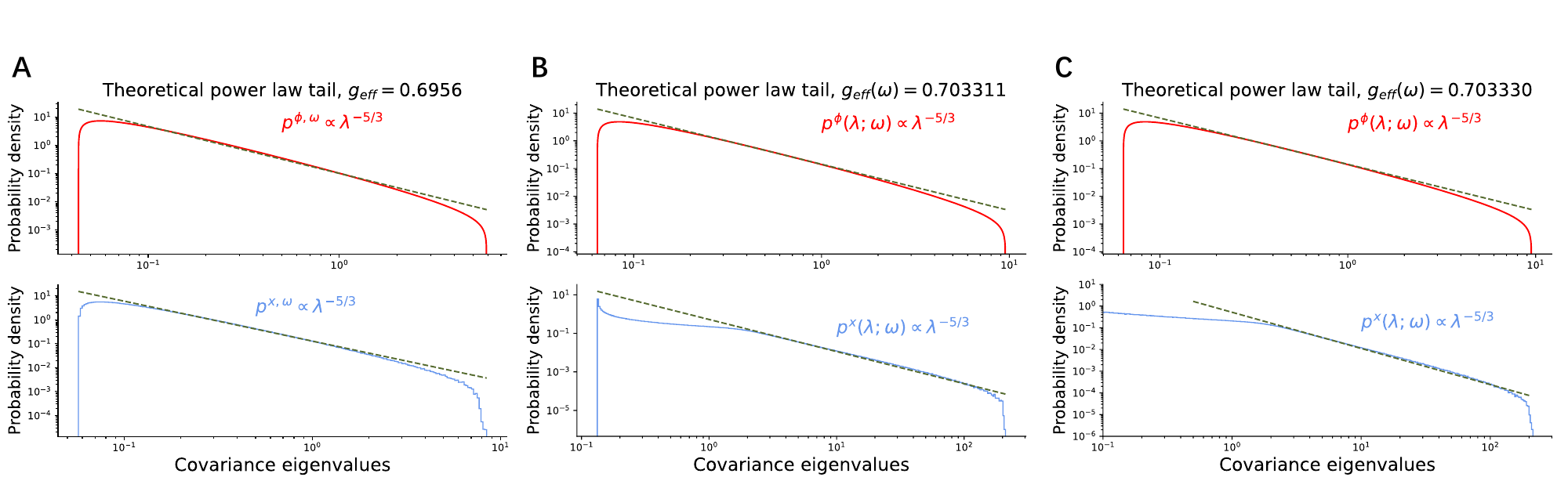}
    \caption{Power-law approximations to the theoretical covariance spectra $\Pphif$ and $\Pxf$ (\cref{eq:spectrum_power_law_tail}) at  $\geff(\omega)=g\phiprime/\sqrt{1+\omega^2} \approx0.7$.
    Note that $\geff(\omega)$ depends on $g$, $\sigma$, and the frequency $\omega$. Therefore, we show here three different combinations of parameters corresponding to $\geff(\omega)\approx0.7$. In all cases, the tail of the theoretical spectra (\cref{sec:eigenvalue_spectrum_of_covariance_matrices}) are well approximated by the power law given in \cref{eq:spectrum_power_law_tail}.
     (\textbf{A}) $g=0.8$, $\sigma = 0.495$, $\omega=0$, and    $\geff=0.6956$.
      (\textbf{B}) $g=5$, $\sigma = 0.495$, $\omega=0.9$, and $\geff(\omega)=0.703311$.   
      (\textbf{C}) $g=5$, $\sigma = 0$, $\omega=0.9$, and $\geff(\omega)=0.703330$.
    }
    \label{fig:different_combination_geff=0.7}
\end{figure}

\begin{figure}[H]
    \centering
    \includegraphics[trim=0cm 4.2cm 0cm 0cm, clip, width=\linewidth]{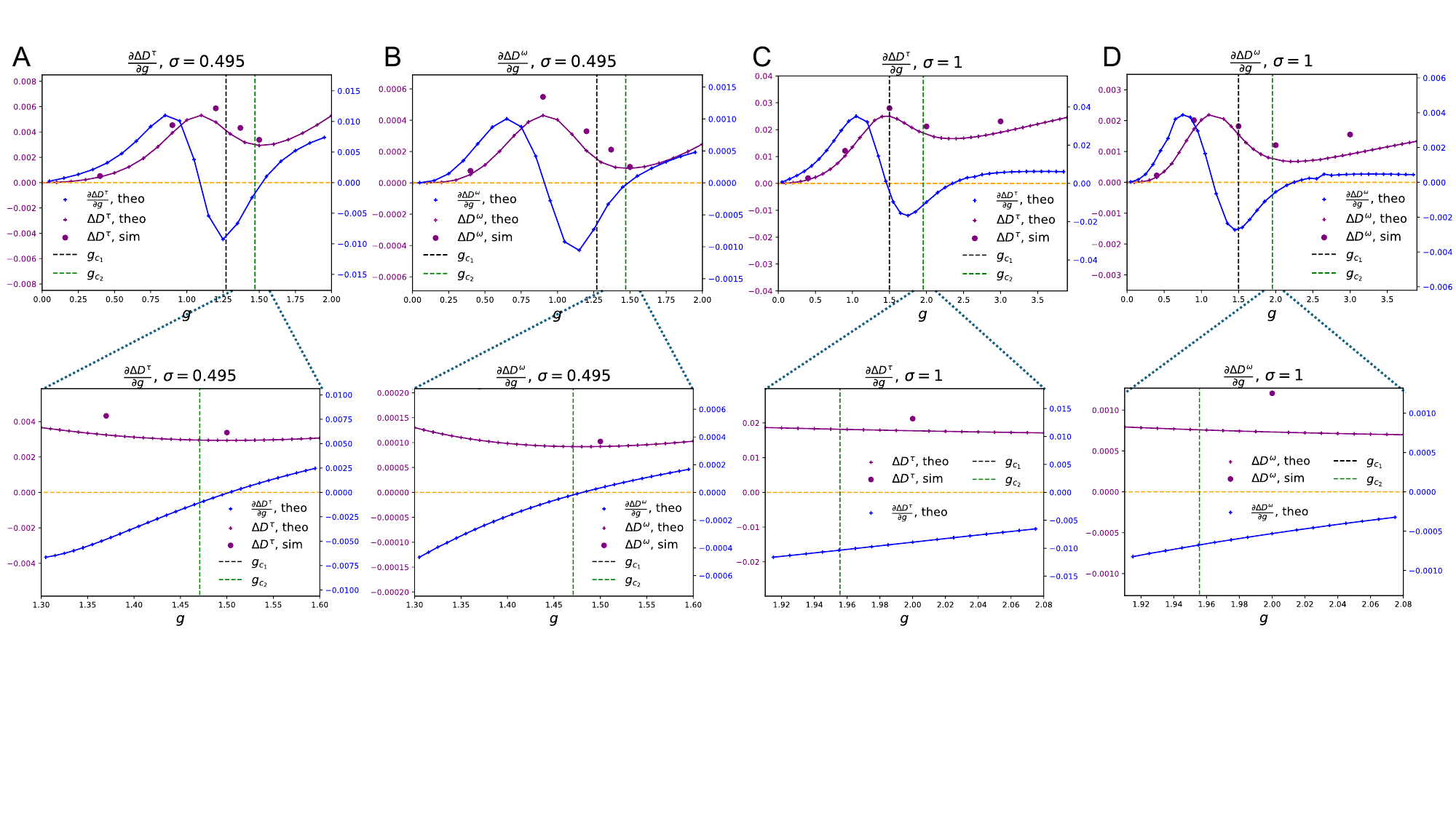}
    \caption{
    (\textbf{A}) The theoretical zero-timelag dimension gap $\Delta D^{\tau} = D^{\phi,\tau} -D^{x,\tau}$ (purple cross and line, \cref{sec:time-lag_dimension}) and its derivative (blue cross and line) for $\sigma=0.495$. Direct network simulations (large purple dots) confirm the trend predicted by the theory. The bottom panel shows a zoom-in near the transition to chaos ($g_{c_2}=1.47$). 
    (\textbf{B}) Same as (A), but for the zero-frequency dimension gap $\Delta D^{\omega}=D^{\phi,\omega} -D^{x,\omega}$. Note the closeness of $\partial\Delta D^{\omega}/\partial g=0$ and $g_{c_2}$ is a coincidence (see (D) for a different $\sigma$).
    (\textbf{C,D}) Same as (A,B), but for $\sigma=1$. It is clear that the local minimum of the dimension gap $\Delta D$ lies within the chaotic regime. 
    }
    \label{fig:dimension_gap_and_derivative}
\end{figure}

\begin{figure}[H]
    \centering
    \includegraphics[trim=1cm 0cm 4cm 0cm, clip, width=\linewidth]{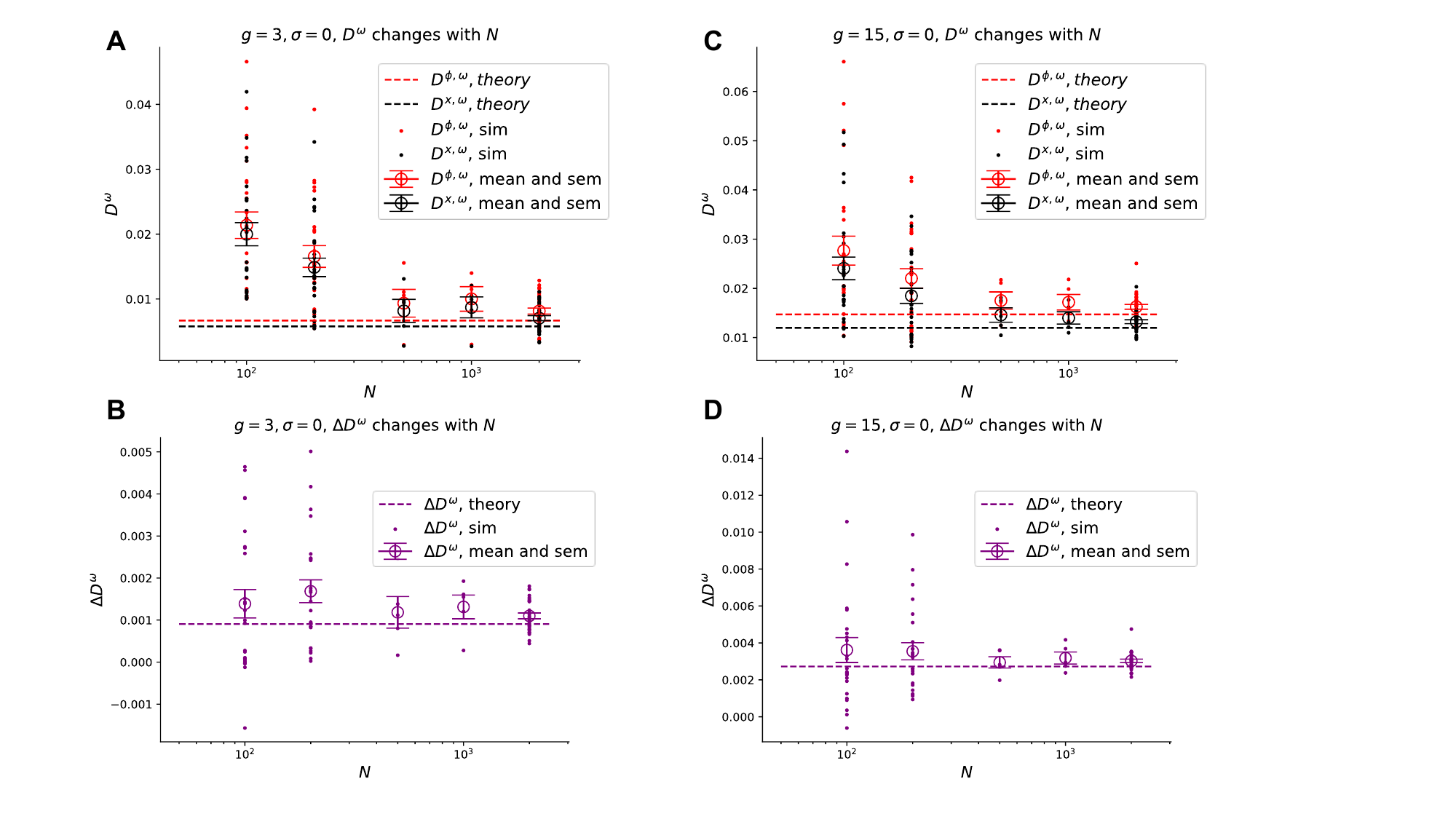}
    \caption{The zero-frequency dimensions (\textbf{A,C}) and the dimension gap (\textbf{B,D}) converge to the mean-field theory values as $N\to\infty$. The dots are computed from network simulations (\cref{eq:first_order_finite_difference_simulation}), and circles and error bars denote the mean and standard error of the mean (SEM) across network realizations, respectively. The dashed lines correspond to theoretical values (\cref{eq:D_phi_omega,eq:dim_x_w_sigma}).
    The networks are deterministic with $\sigma=0$ corresponding to \cref{fig:Dimension_gap}B, and the parameter values of $g=3$ (A,B) and $g=15$ (C,D) are representative of the plotted range of $g$. These figures therefore support that the deviations seen in \cref{fig:Dimension_gap}B are likely finite-size effects.
    }
    \label{fig:zero-frequency_dim_change_with_N}
\end{figure}

\begin{figure}[H]
    \centering
    \includegraphics[trim=0cm 0cm 0cm 0cm, clip, width=0.8\linewidth]{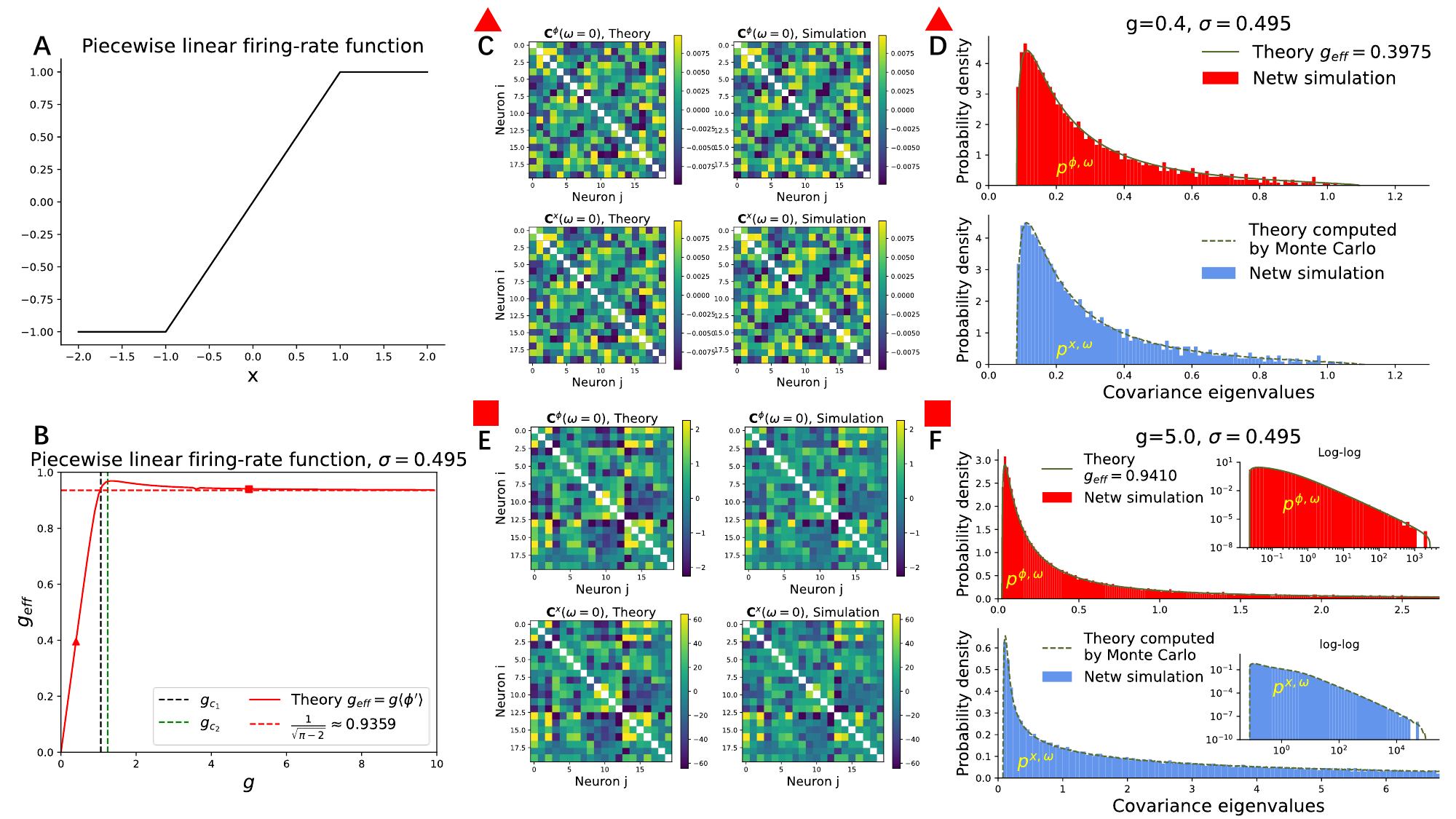}
    \caption{Networks with a piecewise linear firing-rate function $\phi(x)$ (\cref{eq:def_Plin}). 
    (\textbf{A}) Piecewise linear firing-rate function $\phi(x)$ (\cref{eq:def_Plin}).
    (\textbf{B}) $\geff$ curve similar to \cref{fig:g_eff_phase_diagram}B. The solid line is the mean-field theory (\cref{sec:effective_g_and_transition_to_chaos}) and $\hat{g}_{\text{eff}}$ fitted to network simulations (different shaped markers), this curve is qualitatively similar to \cref{fig:g_eff_phase_diagram}B.
    (\textbf{C,D}) Covariance matrix and spectrum comparisons between ansatz and network simulations, similar to \cref{fig:example_covariance_and_relative_error}AB and \cref{fig:Eigen_spectrum_total}D, at $g=0.4$, $\sigma=0.495$ with $N=1000$.
    (\textbf{E,F}) Same as (CD) but for $g=5$, $\sigma=0.495$ with $N=4000$.}
    \label{fig:summary_Plin}
\end{figure}

\begin{figure}[H]
    \centering
    \includegraphics[trim=0cm 0cm 0cm 0cm, clip, width=0.8\linewidth]{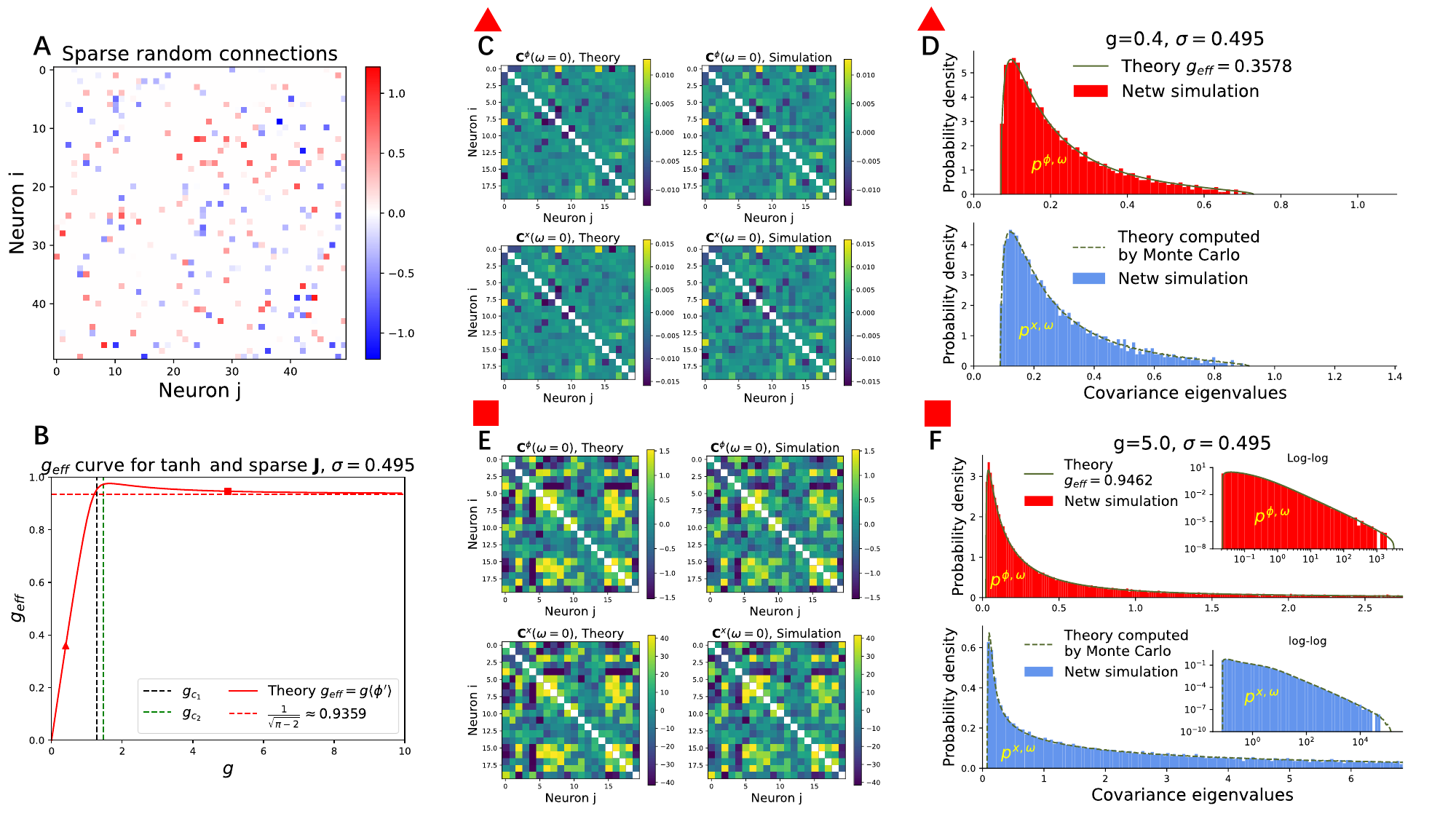}
    \caption{Networks with sparse connectivity. (\textbf{A})Illustration of an example of connection matrix with sparsity $p=0.1$. 
    (\textbf{B}) $\geff$ curve from MFT and fitted $\hat{g}_{\text{eff}}$ (dots). 
    (\textbf{C,D}) $g=0.4$, $\sigma=0.495$, $N=1000$. Comparing theoretical covariance matrix (using MFT) and simulated ones (C), spectrum comparison (D). (\textbf{E,F}) similar to (CD) but for $g=5$, $\sigma=0.495$, $N=4000$.}
    \label{fig:summary_sparse_J}
\end{figure}

\begin{figure}[H]
    \centering
    \includegraphics[trim=0cm 0cm 0cm 0cm, clip, width=0.8\linewidth]{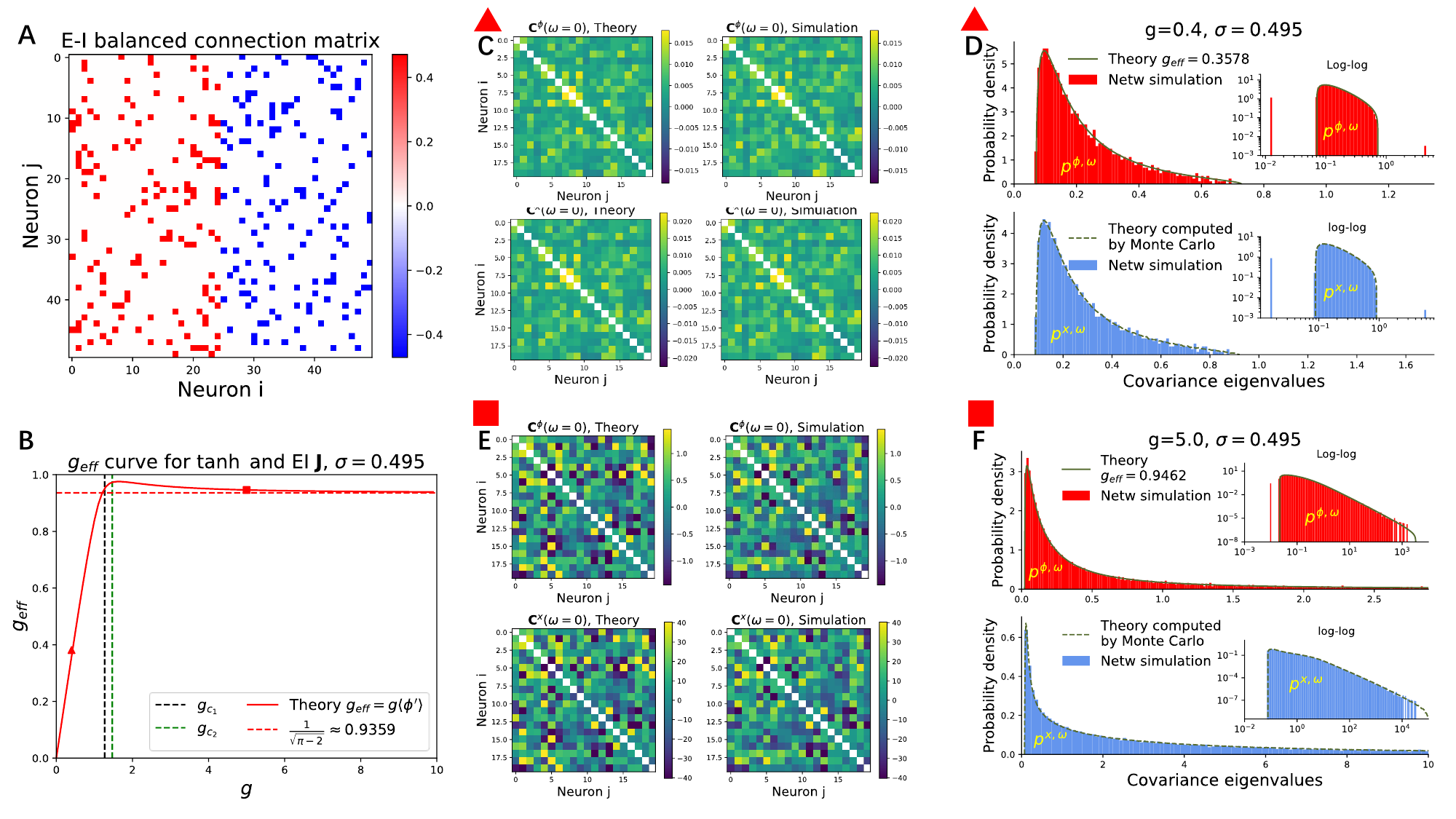}
    \caption{E-I networks obeying Dale's law. 
    (\textbf{A}) Illustration of the sparse E-I connection matrix obeying Dales' law, with $N/2=25$ E and I neurons and an average degree of $K=5$ (i.e., connection probability $K/N=0.1$ as the network in (CD)).
    (\textbf{B}) $\geff$ curve from the MFT (\cref{eq:g_eff_definition}) comparing with the estimated $\hat{g}_{\text{eff}}$ by fitting the spectrum to network simulations (dots). 
    (\textbf{C,D}) Comparing the covariance matrix ansatz (\cref{eq:ansatz_of_Cov_phi_omega,eq:Cov_x_omega_full}) and spectrum (\cref{eq:Pphif_using_effective_Plin}) with network simulations, where $g=w_0=0.4$, $\sigma=0.495$, $N=1000$, $K=100$.
    (\textbf{E,F}) Same as (CD) but for $g=w_0=5$, $K=32$ and $N=4000$.
    }
    \label{fig:summary_EI_J}
\end{figure}

\begin{figure}[H]
    \centering
    \includegraphics[trim=0cm 2cm 5cm 0cm, clip, width=1\linewidth]{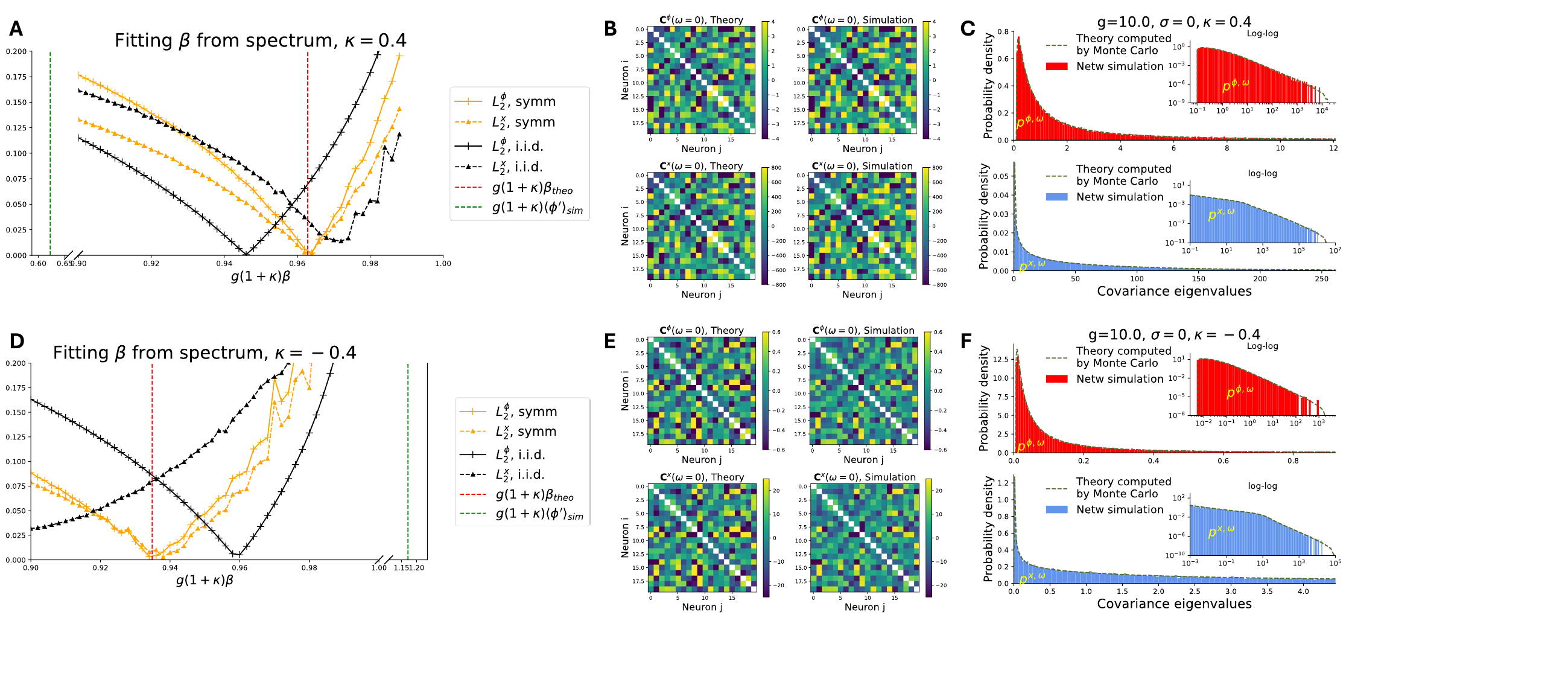}
    \caption{
    Generalizing covariance matrix ansatz for partially symmetric networks.
    (\textbf{A}) The error between the simulated spectra and the ones from i.i.d. network theory (black) and partially symmetric theory (orange) across different values of $\beta$ (scaled as $g(1+\kappa)\cdot \beta$ to an intuitive range).
    The solid and dashed lines are fitting errors of the $\phi$ and $x$ spectrum, respectively. Note that the minima of the partially symmetric error curves co-localize (see text). 
    The vertical dashed lines mark the theoretical $\beta$ values from i.i.d. connectivity (green) and partially symmetric connectivity (red). 
    The error is quantified by Cramer-von-Mises statistics and we normalized the spectrum to have $\mathbb{E}[\lambda]=1$ to compare the shape only.
    (\textbf{B,C})
    Comparisons of theoretical covariance matrices (B, \cref{eq:ansatz_of_Cov_phi_omega_symmetric,eq:Cov_matrix_estimation}), spectra (C) and simulated ones. Using theoretical $\beta$ and $\Czetaf$ (\cref{eq:beta_estimation,eq:C_zeta_estimation} in \cref{app:estimate_beta_and_C_zeta}).
    (\textbf{D,E,F}) Same as (A,B,C) ($\kappa=0.4$) but for $\kappa=-0.4$.
    In panels DEF, $N=4000$, $g=10$, $\sigma=0$.
    }
    \label{fig:summary_symm_J}
\end{figure}

\begin{figure}[H]
    \centering
    \includegraphics[trim=0cm 3cm 0cm 0cm,clip, width=1\linewidth]{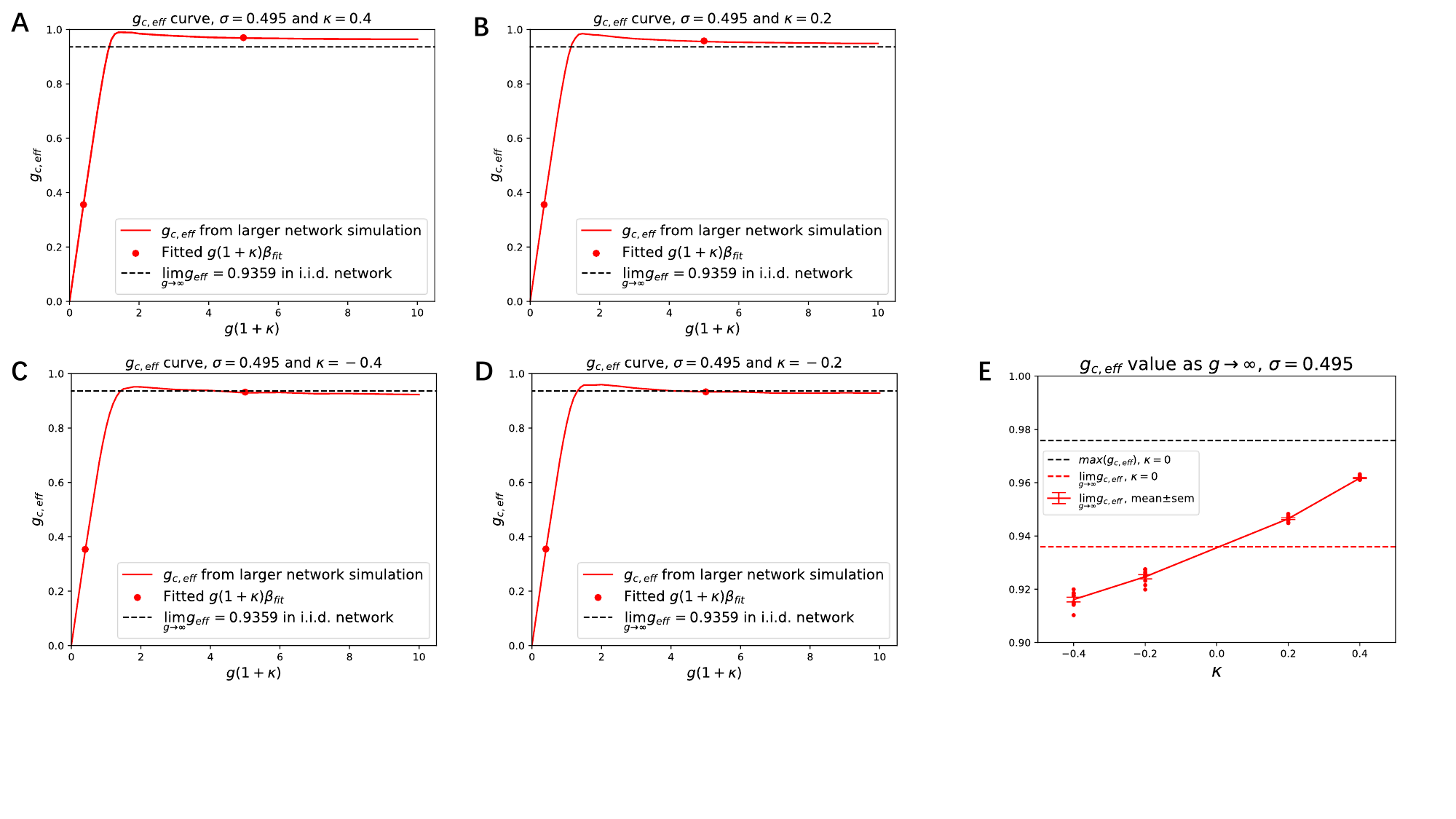}
    \caption{
    (\textbf{A-D})  $g_{c,\text{eff}}$ curve corresponds to different $\kappa$ values. The curves are computed from the mean across 5 realizations of each network (15 realizations for $\kappa=-0.4$, more details can be found in \cref{tab:table_of_parameters}).
    The horizontal dashed line is the asymptotic value of $\geff$ in the i.i.d. random network as $g\to\infty$ for reference.
    (\textbf{E}) The estimated asymptotic value of $g_{c,\text{eff}}$ (approximated as the theory curve when $g(1+\kappa)=100$) across different $\kappa$ values, which shows an increasing relationship with  $\kappa$. The two horizontal dashed lines are $\geff$ values from the i.i.d. random network for reference: the red is the asymptotic value as $g\to\infty$ and black is the maximum value across all $g$ (under the same $\sigma=0.495$).
    }
    \label{fig:g_eff_curve_for_diff_kappa}
\end{figure}

\begin{figure}[H]
    \centering
    \includegraphics[trim=0cm 10cm 0cm 0cm, clip,width=1\linewidth]{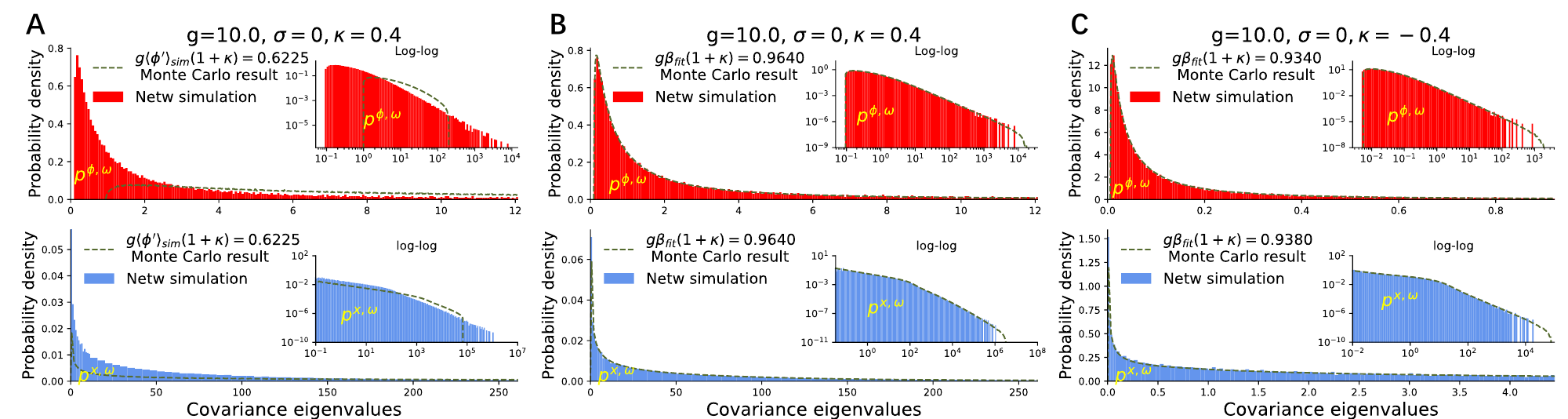}
    \caption{
    (\textbf{A}) For $\kappa=0.4$, comparing the shape of the covariance spectrum between network simulations and the spectrum from $\lrsq{\MI-\phiprime_{\text{sim}}\MJ}^{-1}\lrsq{\MI-\phiprime_{\text{sim}}\MJ}^{-T}$ (for $\phi$) and $\MJ\lrsq{\MI-\phiprime_{\text{sim}}\MJ}^{-1}\lrsq{\MI-\phiprime_{\text{sim}}\MJ}^{-T}\MJ^T$ (for $x$) using partially symmetric $\MJ$ with the same $\kappa$ using Monte Carlo method, here $\phiprime_{\text{sim}}$ is estimated from the network simulation. The Monte Carlo estimated spectra are rescaled such that $\mathbb{E}\lrsq{\lambda^a_{sim}}=\mathbb{E}\lrsq{\lambda^a_{MC}}$, for $a\in\{x,\phi\}$. 
    Note that for $\kappa=-0.4$, $g\phiprime_{\text{sim}}(1+\kappa)>1$, invalidating the linear network ansatz, we thus omit the plot.
    (\textbf{B, C}) Comparing the shape of the covariance spectra from network simulations ($\kappa=\pm 0.4$ for (B) and (C), respectively) and the ones based on the generalize covariance matrix ansatz (\cref{eq:ansatz_of_Cov_phi_omega_symmetric}) using the optimal $\beta_{\text{fit}}$ minimizing the Cramer-von-Mises statistic for the $\phi$ spectrum and $x$ spectrum, respectively (the minima of yellow curves in \cref{fig:summary_symm_J}AD). 
    Both simulated and theoretical spectra are scaled the same as in (A).
    }
    \label{fig:part_symm_spectrum_supplement}
\end{figure}

\section{Four-point function $\psi^{a}(\omega_1,\omega_2)$ and dimension when $\sigma\ge 0$ using the cavity method}
\label{app:cavity_unit_four_point_function}
Here we derive the frequency-dependent four-point function $\psiphitf$ and $\psixtf$, $\vec{\omega}=(\omega_1,\omega_2)$ in the case with white noise input, by following and slightly extending the two-site cavity method in \cite{Clark2023} from the case without external input $\sigma=0$. 
While the majority of steps are the same as in \cite{Clark2023} and we include here for completeness, several steps around \cref{eq:two-point_function_of_a} are different from the $\sigma=0$ case (Eq.(B3) in \cite{Clark2023}).

We consider two \textit{cavity neurons} with indices $0,0'$, and the \textit{reservoir neurons} are index by $i\in\{1,\cdots,N\}$.
The cavity neurons are driven by three components: (i) input from the reservoir, (ii) input from mutual- and self-connections, and (iii) from external noise input. 
We have
\begin{equation}
\left\{
\begin{aligned}
    (1+\partial_t)x_{i}^{\free}(t)&=\sum_{j=1}^NJ_{ij}\phi_j^{\free}(t)+\xi_i(t)=\eta_i(t)+\xi_i(t),\\
    (1+\partial_t)x_{\mu}^{\free}(t)&=\sum_{i=1}^{N}J_{\mu i}\phi_{i}^{\free}(t)+\xi_{\mu}(t)=\eta_{\mu}^{\free}(t)+\xi_{\mu}(t).
\end{aligned}
\right.
\label{eq:unperturbed_activity}
\end{equation}
Here $\phi_{\mu}^{\free}(t)=\phi(x_{\mu}^{\free}(t))$. 
As in the main text, we denote $a\in\{\phi, x\}$ instead of writing multiple similar equations.
The superscript ``free'' denotes the value when there is no feedback from the cavity neurons to the reservoir, that is, the reservoir is not perturbed by the cavity neurons by setting $J_{i\mu}=0$, $J_{\mu\nu}=0$, but $J_{\mu i}\neq0$. 
Note that $\left(\eta^{\free}_{0}(t), \eta^{\free}_{0'}(t')\right)$ are jointly Gaussian distributed but $\lrbk{\eta_i(t), \eta_j(t')}$ are not \citep{Clark2023}. Moreover, denote $\tilde{\eta}=\eta+\xi$ and $\tilde{\eta}^{\free}=\eta^{\free}+\xi$ to be the overall Gaussian input. 

We expand the cavity neuron activities (both $\phi_{\mu}$ and $x_{\mu}$) perturbed by cavity-to-reservoir connection, mutual- and self-connections up to leading orders in $1/\sqrt{N}$. First, introducing the perturbation through cavity-to-reservoir connection $J_{j\nu}$, the $\tilde{\eta}_{j}(t)$ is perturbed by $\delta\tilde{\eta}_j(t)=\sum_{\nu\in\{0,0'\}}J_{j\nu}\phi_{\nu}^{\free}(t)$, and
\[
\delta\phi_i(t') = \int^{t'}dt^{\dprime}\sum_{j=1}^NS^{\phi}_{ij}\lrbk{t',t^{\dprime}}\delta\tilde{\eta}_{j}(t')dt' = \int^{t'}dt^{\dprime}\sum_{j=1}^NS^{\phi}_{ij}\lrbk{t',t^{\dprime}}\sum_{\nu\in\{0,0'\}}J_{j\nu}\phi_{\nu}^{\free}(t^{\dprime}).
\]
Here, $S^{a}_{ij}(t,t')$ is linear response function for $a\in\{x,\phi\}$, which is similarly defined as in \cite{Clark2023} with the modification of $\tilde{\eta}_{j}=\eta_{j}+\xi_{j}$, 
\begin{equation*}
    S^{a}_{ij}(t,t') = \delta a_{i}(t)/\delta\tilde{\eta}_{j}(t').
\end{equation*}
Here $\delta$ is the functional derivative. The time-averaged linear response only depends on the time lag $\tau=t-t'$, so we write $S^{a}_{ij}(\tau)=\mean{S^{a}_{ij}(t,t-\tau)}_t$ and $S^{a}(\tau)=S^{a}_{ii}(\tau)$, we dropped the subscript because neurons become homogeneous as $N\to\infty$. Next, introduce the mutual- and self-connections, the perturbation of overall input $\tilde{\eta}$ is
\begin{equation*}
\delta \tilde{\eta}_\mu(t') = \sum_{\nu\in\{0,0'\}}J_{\mu\nu}\phi_\nu^{\mathrm{free}}(t') + \sum_{i=1}^N J_{\mu i}\delta\phi_i(t')=\frac{1}{\sqrt{N}}\int^{t'}dt^{\dprime}\sum_{\nu\in\{0,0'\}}F_{\mu\nu}(t',t^{\dprime})\phi_{\nu}^{\free}(t^{\dprime}).
\end{equation*}
Where 
\begin{equation*}
    F_{\mu\nu}(t,t')=\sqrt{N}\left[\sum_{i,j=1}^NJ_{\mu i}S^{\phi}_{ij}\left(t,t'\right)J_{j\nu}+J_{\mu\nu}\delta(t-t')\right],
\end{equation*}
it describes the overall influence of $\nu$ on $\mu$, and we denote the time-average version $F_{\mu\nu}(\tau)=\mean{F_{\mu\nu}(t,t-\tau)}$ with $\tau=t-t'$. 
Then use the expression of $\delta \tilde{\eta}_\mu(t')$ to derive the perturbed activity $a_{\mu}(t)=a_{\mu}^{\free}(t)+\int^tdt'S^{a}_{\mu\mu}(t,t')\delta\tilde{\eta}_{\mu}(t')+O\lrbk{\frac{1}{N}}$, we get \citep{Clark2023}
\begin{equation}
    a_{\mu}(t)=a_{\mu}^{\free}(t) + \frac{1}{\sqrt{N}}\int^tdt'S^{a}_{\mu\mu}(t,t')\int^{t'}dt^{\dprime}\sum_{\nu\in\{0,0'\}}F_{\mu\nu}(t',t'')\phi^{\free}_{\nu}(t^{\dprime}) + O\left(\frac{1}{N}\right).
\label{eq:a_t_first_order_expansion}
\end{equation}

The four-point function of $a\in\{x,\phi\}$, defined as $\psi^{a}(\omega_1,\omega_2):=N\mean{C^{a}_{ij}(\omega_1)C^{a}_{ij}(\omega_2)}_{\MJ}$ is equivalent to $N\mean{C^{a}_{00'}(\omega_1)C^{a}_{00'}(\omega_2)}_{\MJ}$.
So, we compute $\mean{ a_0(t)a_{0'}(t+\tau) }_t$, which is $O\lrbk{\frac{1}{\sqrt{N}}}$, the time-lagged covariance (also named as two-point function), using \cref{eq:a_t_first_order_expansion} and keep terms up to $O\lrbk{\frac{1}{\sqrt{N}}}$,
\begin{equation}
    \begin{aligned}
        &\mean{ a_0(t)a_{0'}(t+\tau) } _t = \mean{ a_0^{\free}(t)a_{0'}^{\free}(t+\tau)} _t\\
        &+\frac{1}{\sqrt{N}}\mean{ a_0^{\free}(t)\int^tdt'S^{a}_{0'0'}(t+\tau,t')\int^{t'}dt^{\dprime}\sum_{\nu\in\{0,0'\}}F_{0'\nu}(t',t'')\phi^{\free}_{\nu}(t^{\dprime}) } _t\\
        &+\frac{1}{\sqrt{N}}\mean{ \int^tdt'S^{a}_{00}(t,t')\int^{t'}dt^{\dprime}\sum_{\nu\in\{0,0'\}}F_{0\nu}(t',t'')\phi^{\free}_{\nu}(t^{\dprime}) a_{0'}^{\free}(t+\tau)}_t + O\lrbk{\frac{1}{N}}.
    \end{aligned}
    \label{eq:a_free_t_average}
\end{equation}
The unperturbed activity $a^{\free}_{\mu}(t)$ can be rewritten as a functional of Gaussian fields $\tilde{\eta}^{\free}_{0}$ and $\tilde{\eta}^{\free}_{0'}$, and the first functional derivative of $a^{\free}_{\mu}(t)$ over $\tilde{\eta}^{\free}_{\mu}$ is $S^{a}_{\mu\mu}(t,s)$,
\[
\mean{ a_0^{\free}(t)a^{\free}_{0'}(t+\tau) }_t=\mean{ a_0^{\free}\lrsq{\tilde{\eta}^{\free}_0(s)}(t) a^{\free}_{0'}\lrsq{\tilde{\eta}^{\free}_{0'}(s')}(t+\tau) }_{t,s,s'}.
\]
Where $\mean{\cdot}_s$ represents the integral over $s$. We expand above equation in terms of $C^{\tilde{\eta}^{\free}}_{00'}(\tau) = O\lrbk{\frac{1}{\sqrt{N}}}$ and use the Price's theorem \citep{Price1958,Clark2023},
\begin{equation}
    \begin{aligned}
        &\mean{ a_0^{\free}(t)a^{\free}_{0'}(t+\tau) }_{t,s,s'}
        =\mean{\frac{\partial\mean{ a_0^{\free}(t)a^{\free}_{0'}(t+\tau) }_t}{\partial C^{\tilde{\eta}^{\free}}_{00'}(s'-s)}C^{\tilde{\eta}^{\free}}_{00'}(s'-s)}_{s,s'}+O\lrbk{\frac{1}{N}}\\
        &=\int ds \int ds' \mean{ \frac{\delta a^{\free}_{0}(t)}{\delta \tilde{\eta}^{\free}_0(s)}\frac{\delta a^{\free}_{0'}(t+\tau)}{\delta \tilde{\eta}^{\free}_{0'}(s')} }_tC^{\tilde{\eta}^{\free}}_{00'}(s'-s)+O\lrbk{\frac{1}{N}}\\
        &=\int ds \int ds' S^{a}(-s)S^{a}(\tau-s')C^{\eta^{\free}}_{00'}(s'-s)+O\lrbk{\frac{1}{N}}.
    \end{aligned}
    \label{eq:two-point_function_of_a}
\end{equation}
Note that in the last equal sign of the above, several terms appear due to $\sigma>0$, but they turn out to be all zero: $C^{\xi}_{00'}(s'-s)=C^{\eta^{\free}\xi}_{00'}(s'-s)=C^{\eta^{\free}\xi}_{0'0}(s'-s)=0$, so $C^{\tilde{\eta}^{\free}}_{00'}=C^{\eta^{\free}}_{00'}$. This is because in \cref{eq:unperturbed_activity} there is no feedback connections from cavity neurons to the reservoir and thus $\xi_{\mu}$ have no effect on reservoir $x_i$ and consequently no effect on $\eta^{\free}_{\nu}$.
The Fourier transform of $C^{a^{\free}}_{00'}(\tau)=\mean{ a_0^{\free}(t)a_{0'}^{\free}(t+\tau) } _t$ is $C^{a^{\free}}_{00'}(\omega)=|S^a(\omega)|^2C^{\eta^{\free}}_{00'}(\omega)$.

Next, for the second term and third term on the RHS of \cref{eq:a_free_t_average}, we need to compute time-lagged correlation $C^{\phi^{\free}a^{\free}}_{\mu\mu}(\tau)$. For $\MJ$ with i.i.d. connections, in \cref{eq:a_t_first_order_expansion}, the unperturbed activity $a^{\free}_{\mu}(t) = O(1)$, while the perturbation $\frac{1}{\sqrt{N}}\int S^{a}_{\mu\mu}\sum F_{\mu\nu}\phi_{\nu}^{\free} = O\lrbk{\frac{1}{\sqrt{N}}}$. Keeping only $O(1)$ term, $C^{\phi a}=C^{\phi^{\free}a^{\free}}+O\lrbk{\frac{1}{\sqrt{N}}}$.

Now we have the quantities needed to simplify \cref{eq:a_free_t_average}. Applying Fourier transform on \cref{eq:a_free_t_average} and keeping only $O\lrbk{\frac{1}{\sqrt{N}}}$ terms. Then, replace $C^{a^{\free}}_{00'}(\omega)$ by $|S^a(\omega)|^2C^{\eta^{\free}}_{00'}(\omega)$ and replace $C^{\phi^{\free}a^{\free}}(\omega)$ by $C^{\phi a}(\omega)$, \cref{eq:a_free_t_average} is now given by,
\begin{equation}
\begin{aligned}
    C^{a}_{00'}(\omega)&= C^{a^{\free}}_{00'}(\omega)+\frac{1}{\sqrt{N}}\left[ (F_{00'}(\omega)S^{a}(\omega))^*C^{\phi^{\free} a^{\free}}(\omega)+F_{0'0}(\omega)S^{a}(\omega) C^{a^{\free}\phi^{\free}}(\omega)\right] + O\lrbk{\frac{1}{N}}\\
    &=|S^{a}(\omega)|^2C^{\eta^{\free}}_{00'}(\omega)+\frac{1}{\sqrt{N}}\left[ (F_{00'}(\omega)S^{a}(\omega))^*C^{\phi a}(\omega)+F_{0'0}(\omega)S^{a}(\omega) C^{a\phi}(\omega)\right] + O\lrbk{\frac{1}{N}}\\
    &= |S^{a}(\omega)|^2C^{\eta^{\free}}_{00'}(\omega)+\frac{1}{\sqrt{N}}\left[ (F_{00'}(\omega)S^{a}(\omega))^*+F_{0'0}(\omega)S^{a}(\omega)\right]C^{\phi a}(\omega) + O\lrbk{\frac{1}{N}}.
\end{aligned}
    \label{eq:two-point_function_a_omega}
\end{equation}
In \cref{eq:two-point_function_a_omega}, note that $C^{\phi x}(\omega)=\phiprime\Cxf=C^{\phi x}(-\omega)$ using the Stein's lemma and $\Cxf$, $\Cphif$ are even functions. So for $a\in\{x,\phi\}$, we have $C^{a\phi}(\omega)=C^{\phi a}(\omega)$. Plugging \cref{eq:two-point_function_a_omega} into the definition of $\psi^{a}(\vec{\omega})$,
\begin{equation}
    \begin{aligned}
        &\frac{1}{N}\psi^{a}(\vec{\omega})=\mean{ C^{a}_{00'}(\omega_1)C^{a}_{00'}(\omega_2) } _{\MJ}=\mean{|S^{a}(\omega_1)S^{a}(\omega_2)|^2C^{\eta^{\free}}_{00'}(\omega_1)C^{\eta^{\free}}_{00'}(\omega_2)} _{\MJ}\\
        &+ \frac{1}{\sqrt{N}}\mean{|S^{a}(\omega_1)|^2[(F_{00'}(\omega_2)S^{a}(\omega_2))^*+(F_{0'0}(\omega_2)S^{a}(\omega_2))]C^{\eta^{\free}}_{00'}(\omega_1)C^{\phi a}(\omega_2)} _{\MJ}\\
        &+ \frac{1}{\sqrt{N}}\mean{|S^{a}(\omega_2)|^2[(F_{00'}(\omega_1)S^{a}(\omega_1))^*+(F_{0'0}(\omega_1)S^{a}(\omega_1))]C^{\eta^{\free}}_{00'}(\omega_2)C^{\phi a}(\omega_1) } _{\MJ}\\
        &+ \frac{1}{N}\left< F_{00'}^*(\omega_1)F_{0'0}(\omega_2)(S^{a}(\omega_1))^*S^{a}(\omega_2)+(F_{00'}(\omega_1)F_{00'}(\omega_2)S^{a}(\omega_1)S^{a}(\omega_2) )^*\right.\\
        &+\left.F_{0'0}(\omega_1)F_{0'0}(\omega_2)S^{a}(\omega_1)S^{a}(\omega_2)+F_{0'0}(\omega_1)F_{00'}^*(\omega_2)S^{a}(\omega_1)(S^{a}(\omega_2))^*\right> _{\MJ}C^{\phi a}(\omega_1)C^{\phi a}(\omega_2)+O\lrbk{\frac{1}{N^{3/2}}}.
    \end{aligned}
    \label{eq:psi_a_omega_full_version}
\end{equation}
For $\MJ$ with zero-mean i.i.d. entries, only self-averaging terms in the above are at leading order in $O\lrbk{\frac{1}{N}}$ as $N\to\infty$: $\mean{ C^{\eta^{\free}}_{00'}(\omega_1)C^{\eta^{\free}}_{00'}(\omega_2)}_{\MJ}$, $\frac{1}{N}\mean{ F_{00'}(\omega_1)F_{00'}(\omega_2)}_{\MJ}$, and $\frac{1}{N}\mean{F_{0'0}(\omega_1)F_{0'0}(\omega_2)}_{\MJ}$. Those mixed terms, e.g. $\frac{1}{\sqrt{N}}\mean{F_{00'}C^{\eta^{\free}}_{00'}}_{\MJ}$, are not at the leading order $O\lrbk{\frac{1}{N}}$ \cite{Clark2023}.
Hence the \cref{eq:psi_a_omega_full_version} simplifies to
\begin{equation}
\begin{aligned}
    \psi^a(\vec{\omega}) &=N |S^a(\omega_1)S^a(\omega_2)|^2\mean{C^{\eta^{\free}}_{00'}(\omega_1)C^{\eta^{\free}}_{00'}(\omega_2)}_{\MJ} \\
    &+\mean{F_{00'}(\omega_1)F_{00'}(\omega_2)} _{\MJ}S^a(\omega_1) S^a(\omega_2)C^{\phi a}(\omega_1)C^{\phi a}(\omega_2) \\
    &+\left(\mean{F_{00'}(\omega_1)F_{00'}(\omega_2)} _{\MJ}S^a(\omega_1) S^a(\omega_2)\right)^*C^{\phi a}(\omega_1)C^{\phi a}(\omega_2)+O\lrbk{\frac{1}{\sqrt{N}}}.
\end{aligned}
\label{eq:raw_psi_a_omega}
\end{equation}

There are two quenched-order averages in \cref{eq:raw_psi_a_omega}, i.e. $\mean{C^{\eta^{\free}}_{00'}(\omega_1)C^{\eta^{\free}}_{00'}(\omega_2)}_{\MJ}$ and $\mean{F_{00'}(\omega_1)F_{00'}(\omega_2)} _{\MJ}$, need to compute. To proceed, we use the definition of the recurrent input $\eta_{\mu}^{\free}$ from the unperturbed reservoir, note that $J_{\mu i}$ is independent of reservoir activity $a^{\free}_{i}(t)$.
\begin{equation}
\begin{aligned}
    \mean{ C^{\eta^{\free}}_{00'}(\omega_1)C^{\eta^{\free}}_{00'}(\omega_2)}_{\MJ}&=\mean{ \sum_{i,j=1}^NJ_{0i}^2J_{0'j}^2C^{\phi^{\free}}_{ij}(\omega_1)C^{\phi^{\free}}_{ij}(\omega_2)+\sum_{k\neq i, l\neq j}^NJ_{0i}J_{0k}J_{0'j}J_{0'l}C^{\phi^{\free}}_{ij}(\omega_1)C^{\phi^{\free}}_{kl}(\omega_2)
    }_{\MJ}\\
    &=\mean{\sum_{i=1}^{N}J_{0i}^2J_{0'i}^2C^{\phi^{\free}}_{ii}(\omega_1)C^{\phi^{\free}}_{ii}(\omega_2) + \sum_{i\neq j}^{N}J_{0i}^2J_{0'j}^2C^{\phi^{\free}}_{ij}(\omega_1)C^{\phi^{\free}}_{ij}(\omega_2)}_{\MJ} + 0\\
    &=\mean{\sum_{i=1}^{N}J_{0i}^2J_{0'i}^2}_{\MJ}C^{\phi^{\free}}(\omega_1)C^{\phi^{\free}}(\omega_2) + \mean{\sum_{i\neq j}^{N}J_{0i}^2J_{0'j}^2}_{\MJ}\mean{C^{\phi^{\free}}_{ij}(\omega_1)C^{\phi^{\free}}_{ij}(\omega_2)}_{\MJ}\\
    &=\frac{g^4}{N}C^{\phi}(\omega_1)C^{\phi}(\omega_2)+\frac{g^4}{N}\psiphitf+0.
\end{aligned}
\label{eq:C_eta_off_diag_average}
\end{equation}
The independence is used to separate one expectation to the product of two expectations and use $C^{\phi^{\free}}(\omega)=\Cphif+O\lrbk{\frac{1}{\sqrt{N}}}$ and drop the $O\lrbk{\frac{1}{\sqrt{N}}}$ terms. Similarly,
\begin{equation}
\begin{aligned}
    \mean{ F_{00'}(\omega_1)F_{00'}(\omega_2)} _{\MJ}&=N\mean{\sum_{i,j,k,l=1}^NJ_{0i}J_{ik}S^{\phi}_{ij}(\omega_1)S^{\phi}_{kl}(\omega_2)J_{j0'}J_{l0'}+J_{00'}^2+J_{00'}\sum_{i,j=1}^NJ_{0i}\lrsq{S^{\phi}_{ij}(\omega_1)+S^{\phi}_{ij}(\omega_2)}J_{j0'}}_{\MJ}\\
    &=N\mean{\sum_{i,j=1}^NJ_{0i}^2J_{j0'}^2S^{\phi}_{ij}(\omega_1)S^{\phi}_{ij}(\omega_2) + J_{00'}^2 + \sum_{k\neq i, l\neq j}^NJ_{0i}J_{ik}S^{\phi}_{ij}(\omega_1)S^{\phi}_{kl}(\omega_2)J_{j0'}J_{l0'}}_{\MJ} + 0\\
    &=N\mean{\sum_{i=1}^NJ_{0i}^2J_{i0'}^2S^{\phi}_{ii}(\omega_1)S^{\phi}_{ii}(\omega_2)+ \sum_{i\neq j}^NJ_{0i}^2J_{j0'}^2S^{\phi}_{ij}(\omega_1)S^{\phi}_{ij}(\omega_2) + J_{00'}^2}_{\MJ} + 0 \\
    &=g^4S^{\phi}(\omega_1)S^{\phi}(\omega_2)+g^4N\mean{ S^{\phi}_{00'}(\omega_1)S^{\phi}_{00'}(\omega_2)} _{\MJ}+g^2\\
    &=g^4S^{\phi}(\omega_1)S^{\phi}(\omega_2)+g^4\lrsq{S^{\phi}(\omega_1)S^{\phi}(\omega_2)}^2\mean{ F_{00'}(\omega_1)F_{00'}(\omega_2)} _{\MJ}+g^2.
\end{aligned}
\label{eq:F_four-point_function}
\end{equation}
Note that under the cavity method assumption, $\mean{S^{\phi}_{ij}(\omega_1)S^{\phi}_{ij}(\omega_2)}_{\MJ}=\mean{S^{\phi}_{00'}(\omega_1)S^{\phi}_{00'}(\omega_2)}_{\MJ}$, and $S^{\phi}_{00'}(\omega)=N^{-1/2}S^{\phi}(\omega)^2F_{00'}(\omega)$ \citep{Clark2023}. The independence between $J_{\mu i}$ and $S^{\phi}_{ij}(\omega)$ is also used. We can therefore solve  
$\mean{ F_{00'}(\omega_1)F_{00'}(\omega_2)} _{\MJ}$ from the last equation of \cref{eq:F_four-point_function},
\begin{equation}
    \mean{ F_{00'}(\omega_1)F_{00'}(\omega_2)}_{\MJ}=\frac{g^2}{1-g^2S^{\phi}(\omega_1)S^{\phi}(\omega_2)}.
\label{eq:F_mu_nu_four_point_function}
\end{equation}
Next, we derive the explicit expression of $S^{a}(\omega)$ for $a\in\{x,\phi\}$ in \cref{eq:raw_psi_a_omega}. Using the Furutsu-Novikov theorem \citep{furutsu1963,Novikov1965}, one can show that $S^{\phi}(\omega)=C^{\phi\tilde{\eta}}(\omega)/C^{\tilde{\eta}}(\omega)$ \citep{Clark2023}. Then applying the Stein's lemma (because $(x,\tilde{\eta})$ are jointly temporally Gaussian), $C^{\tilde{\phi}\eta}(\omega)=\phiprime C^{x\tilde{\eta}}(\omega)$, results in 
\begin{equation}
    S^{\phi}(\omega)=\phiprime S^{x}(\omega)=\frac{\phiprime}{1+i\omega}.
\label{eq:linear_response_phi_omega}
\end{equation}

Finally, we have all the equations needed to derive the explicit expression of $\psi^{a}(\omega)$ in terms of $g$, $\phiprime$ and $\Caf$ for $a\in\{x,\phi\}$. Plugging \cref{eq:C_eta_off_diag_average,eq:F_mu_nu_four_point_function,eq:linear_response_phi_omega} into \cref{eq:raw_psi_a_omega}, and using $C^{\phi x}(\omega)=\phiprime \Cxf$ (Stein's lemma), we obtain the frequency-dependent four-point functions valid for $\sigma\ge0$,
\begin{equation}
        \psiphitf=\left[\frac{|X(\vec{\omega})|^2}{\left|X(\vec{\omega}) - g^2\phiprime^2\right|^2}-1 \right]C^{\phi}(\omega_1)C^{\phi}(\omega_2),
        \quad \text{where } 
        X(\vec{\omega})=(1+i\omega_1)(1+i\omega_2),
\label{eq:final_derivation_psi_phi_omega}
\end{equation}
\begin{equation}
        \psixtf =\frac{g^4}{\left|X(\vec{\omega}) - g^2\phiprime^2\right|^2}C^{\phi}(\omega_1)C^{\phi}(\omega_2) + \left[\frac{\left|X(\vec{\omega})\right|^2-g^4\phiprime^4}{\left|X(\vec{\omega}) - g^2\phiprime^2\right|^2} - 1\right]C^{x}(\omega_1)C^{x}(\omega_2).
\label{eq:final_derivation_psi_x_omega}
\end{equation}

The second moment of these frequency-dependent covariance matrices can then be obtained by setting $\omega_1=-\omega_2=\omega$, and write $\psi^a(\vec{\omega})=\psi^a(\omega,-\omega)$, note that $X(\vec{\omega})=X(\omega,-\omega)=1+\omega^2$
\begin{equation}
\begin{aligned}
    \varphi_2\lrsq{\MCphif}&=\lim_{N\to\infty}\frac{1}{N}\left(\sum_{i=1}^N\lrsq{C^{\phi}_{ii}(\omega)}^2+\sum_{i\neq j}^NC^{\phi}_{ij}(\omega)C^{\phi}_{ij}(-\omega)\right)\\
    &=\lrsq{\Cphif}^2+\psiphif=\frac{\left(1+\omega^2 \right)^2}{\left(1+\omega^2-g^2\phiprime^2\right)^2}\lrsq{\Cphif}^2,
\end{aligned}
\label{eq:second_moment_phi_from_cavity}
\end{equation}
\begin{equation}
    \varphi_2\lrsq{\MCxf}=\lrsq{\Cxf}^2+\psixf=\frac{g^4}{\left(1+\omega^2-g^2\phiprime^2\right)^2}\left[ \Cphif\right]^2 + \frac{(1+\omega^2)^2-g^4\phiprime^4}{\left(1+\omega^2-g^2\phiprime^2\right)^2}\left[\Cxf\right]^2.
\label{eq:second_moment_x_from_cavity}
\end{equation}
Using \cref{eq:second_moment_phi_from_cavity}, we can obtain the frequency-dependent dimension for $\MCphif$ (\cref{eq:D_phi_omega} in the main text),
\[
\Dphif=\frac{\varphi_1^2\lrsq{\MCphif}}{\varphi_2\lrsq{\MCphif}}=\left(1-\frac{g^2\phiprime^2}{1+\omega^2}\right)^2.
\]
Note that the second moment of $\MCxf$ \cref{eq:second_moment_x_from_cavity} derived here using the two-cavity method matches exactly with \cref{eq:second_moment_Cov_x_nonzero_frequency_sigma>0} obtained using random matrix theory in \cref{app:second_moment_and_dimension}.

In summary, when $\sigma>0$, the single-neuron mean-field quantities $\Cxf$, $\Cphif$ and $\phiprime$ are changed. But because $C^{\xi}_{00'}=C^{\eta^{\free}\xi}_{00'}=0$ in the unperturbed cavity neuron dynamic equations (\cref{eq:unperturbed_activity}), the form of \cref{eq:final_derivation_psi_phi_omega,eq:final_derivation_psi_x_omega} remains the same as in the $\sigma=0$ case, while the effect of $\sigma$ is implicit through $\Cphif$, $\Cxf$ and $\phiprime$ terms.

\section{Scalar factor in $\MCphif$ spectrum ansatz}
\label{app:derivation_of_ansatz}
In \cref{sec:moment_to_ansatz}, we assume that 
$\MCphif$ has a similar expression as in the case of the linear dynamics when replacing $g$ with the effective connection strength $\geff=g\phiprime$, that is,
\begin{equation}
    \MCphif = m(\omega)
    \left(\MI-\frac{\phiprime}{1+i\omega}\MJ\right)^{-1}\left(\MI-\frac{\phiprime}{1-i\omega}\MJ^T\right)^{-1}.
\label{eq:Cphi_unknown_scalar}
\end{equation}
Here $m(\omega)$ is a scalar function to be determined.
Taking the first moment (i.e., normalized trace) of the two sides of \cref{eq:Cphi_unknown_scalar}, 
\[
\varphi\lrsq{\MCphif} = \frac{m(\omega)}{1- g^2\phiprime^2/(1+\omega^2)}.
\]
Here we have used the random matrix formula $\varphi\lrsq{\left(\MI-z\MJ_0\right)^{-1}\left(\MI-z^{\ast}\MJ_0^T\right)^{-1}} = (1-|z|^2)^{-1}$ for $\MJ_0$ with i.i.d. $\mathcal{N}(0,1/N)$ entries \citep{Hu2022} with $z=\frac{g\phiprime}{1+i\omega}$. 
Meanwhile, $\varphi\lrsq{\MCphif}=\Cphif$, so we have
\[
m(\omega) = \left(1-\frac{g^2\phiprime^2}{1+\omega^2}\right)\Cphif,
\]
which gives the complete $\MCphif$ ansatz \cref{eq:ansatz_of_Cov_phi_omega} in the main text.

\section{Random matrix derivation of $\MCxf$ moments and dimension}
\label{app:second_moment_and_dimension}
Here we use random matrix theory to derive the first and second moments of $\MCxf$, which is a different approach from the two-cavity method (\cite{Clark2023}, \cref{app:cavity_unit_four_point_function}). Note that covariance matrices are Hermitian positive semidefinite; intermediate random matrices such as $\MA(z)$ and $\MB(z)$ are generally non-Hermitian.

First, we elaborate some steps in deriving \cref{eq:Cov_x_omega_full}. 
In the main text (\cref{sec:cov_x}), we have derived the equations
$\MCxxif=\frac{1}{1+i\omega}\lrsq{\MJ\MCphixif+\sigma^2\MI}$
and 
$\MCphixif=\phiprime\MCxxif$, from which we can solve for $\MCphixif$,
\begin{equation}
\MCphixif = \frac{\phiprime\sigma^2}{1+i\omega}\left(\MI-\frac{\phiprime}{1+i\omega}\MJ\right)^{-1}.
\end{equation}
Plugging this and \cref{eq:ansatz_of_Cov_phi_omega} into \cref{eq:Cx_dynamic_eq}, with $\MC^{\xi\phi}(\omega)$ being the conjugate transpose of $\MC^{\phi\xi}(\omega)$, we get 
\begin{equation}
    \begin{aligned}
        \MCxf=&\left(1-\frac{g^2\phiprime^2}{1+\omega^2}\right)\frac{\Cphif}{1+\omega^2}\MJ\left(\MI-\frac{\phiprime}{1+i\omega}\MJ\right)^{-1}\left(\MI-\frac{\phiprime}{1-i\omega}\MJ^T\right)^{-1}\MJ^T\\
        &+\frac{\sigma^2}{1+\omega^2}\left[\frac{\phiprime}{1+i\omega}\MJ\left(\MI-\frac{\phiprime}{1+i\omega}\MJ\right)^{-1}+\frac{\phiprime}{1-i\omega}\left(\MI-\frac{\phiprime}{1-i\omega}\MJ^T\right)^{-1}\MJ^T +\MI\right].
    \end{aligned}
\label{eq:Cov_x_omega_form_2}
\end{equation}
To proceed, we can move a term $\frac{\sigma^2}{1+\omega^2} \frac{\phiprime^2}{1+\omega^2}\MJ\left(\MI-\frac{\phiprime}{1+i\omega}\MJ\right)^{-1}\left(\MI-\frac{\phiprime}{1-i\omega}\MJ^T\right)^{-1}\MJ^T$ from first term to the second term in \cref{eq:Cov_x_omega_form_2}  to complete a product. 
This shows that 
\[
\begin{aligned}
    \MCxf=&\lrsq{\left(1-\frac{g^2\phiprime^2}{1+\omega^2}\right)\frac{\Cphif}{1+\omega^2}-\frac{\sigma^2\phiprime^2}{(1+\omega^2)^2}}
    \MJ\left(\MI-\frac{\phiprime}{1+i\omega}\MJ\right)^{-1}\left(\MI-\frac{\phiprime}{1-i\omega}\MJ^T\right)^{-1}\MJ^T\\
    &+\frac{\sigma^2}{1+\omega^2}
    \left[\MI+\frac{\phiprime}{1+i\omega}\MJ\left(\MI-\frac{\phiprime}{1+i\omega}\MJ\right)^{-1}\right]
    \left[\MI+\frac{\phiprime}{1-i\omega}\left(\MI-\frac{\phiprime}{1-i\omega}\MJ^T\right)^{-1}\MJ^T\right].
\end{aligned}
\]
Note that 
\[
\MI+\frac{\phiprime}{1+i\omega}\MJ\left(\MI-\frac{\phiprime}{1+i\omega}\MJ\right)^{-1}
 = \left(\MI-\frac{\phiprime}{1+i\omega}\MJ\right)^{-1},
\]
and using the single neuron mean-field equation \cref{eq:single_neuron_MFT},
\[
\left(1-\frac{g^2\phiprime^2}{1+\omega^2}\right)\frac{\Cphif}{1+\omega^2}-\frac{\sigma^2\phiprime^2}{(1+\omega^2)^2} = \frac{\Cphif-\phiprime^2\Cxf}{1+\omega^2},
\]
we have (\cref{eq:Cov_x_omega_full} in the main text)
\begin{equation*}
\begin{split}
\MCxf &=
\frac{\Cphif-\phiprime^2 \Cxf}{1+\omega^2}
 \MJ \left(\MI-\frac{\phiprime}{1+i\omega}\MJ\right)^{-1}\left(\MI-\frac{\phiprime}{1-i\omega}\MJ^T\right)^{-1} \MJ^T\\
&+\frac{\sigma^2}{1+\omega^2} 
\left(\MI-\frac{\phiprime}{1+i\omega}\MJ\right)^{-1}\left(\MI-\frac{\phiprime}{1-i\omega}\MJ^T\right)^{-1}.
\end{split}
\end{equation*}

In the following, we first describe some general random matrix results, then apply them to computing $\MCxf$ moments.
Let $\MJ_0=\MJ/g$, whose entries are drawn i.i.d. from $\mathcal{N}(0,1/N)$. Let $z$ to be a complex number, also a local auxiliary variable in this section, with $|z|<1$, and denote 
$\MB(z)=\left(\MI-z\MJ_0\right)^{-1}$ and $\MA(z)=\MB(z)-\MI=z\MJ_0\left(\MI-z\MJ_0\right)^{-1}$. 
We have the following identities in the limit of $N\to \infty$ (recall that  $\varphi [\cdot]\coloneqq\lim_{N\to\infty}\frac{1}{N}\mathrm{Tr}\mean{\cdot}_{\MJ_0}$, $\varphi\lrsq{\MJ_0^{n}}=0$ for $n\geq1$ and $\varphi\lrsq{\MJ_0^n\lrbk{\MJ_0^{\dagger}}^m}=\delta_{nm}$):
\begin{equation}
\begin{aligned}
    \varphi\lrsq{z\MJ_0}=&z\varphi\lrsq{\MJ_0}=0,\\
    \varphi\lrsq{\MB(z)}=&\varphi\lrsq{\sum_{n=0}^{\infty}\left(z\MJ_0\right)^n}=1,\\
    \varphi\lrsq{\MA(z)}=&\varphi\lrsq{\sum_{n=1}^{\infty}\left(z\MJ_0\right)^n}=\varphi\lrsq{\MB(z)-\MI}=0,\\
    \varphi\lrsq{\MA^{2}(z)}=&\varphi\lrsq{\sum_{n=1}^{\infty}nz^{n+1}\MJ_0^{n+1}}=\sum_{n=1}^{\infty}nz^{n+1}\varphi\lrsq{\MJ_0^{n+1}}=0,\\
    \varphi\lrsq{\MB^{2}(z)}=&\varphi\lrsq{\left(\MA(z)+\MI\right)^2}=\varphi\lrsq{\MA^2(z)+2\MA(z)+\MI}=1,\\
    \varphi\lrsq{\MB(z)\MB^{\dagger}(z)}=&\frac{1}{1-|z|^2}\qquad\text{\citep{Hu2022}},\\
    \varphi\lrsq{\MA(z)\MA^{\dagger}(z)}=&\varphi\lrsq{\MB(z)\MB^{\dagger}(z)}-\varphi\lrsq{\MB(z)+\MB^{\dagger}(z)}+\varphi\lrsq{\MI}=\frac{1}{1-|z|^2}-1=\frac{|z|^2}{1-|z|^2},\\
    \varphi\lrsq{\MA^2(z)\MA^{\dagger}(z)}=&\varphi\lrsq{\left(\sum_{n=1}^{\infty}\left(z\MJ_0\right)^n\right)^2\left(\sum_{n=1}^{\infty}\left(z\MJ_0\right)^n\right)^{\dagger}}
    =\varphi\lrsq{\sum_{n=1}^{\infty}\sum_{m=1}^{\infty}nz^{n+1} \MJ_0^{n+1}\left(z^m\MJ_0^m\right)^{\dagger}}\\
    =&\sum_{n=1}^{\infty}n\varphi\lrsq{z^{n+1}\MJ_0^{n+1}\left(z^{n+1}\MJ_0^{n+1}\right)^{\dagger}}
    =\sum_{n=1}^{\infty}n|z|^{2n+2}=\frac{|z|^4}{\left(1-|z|^2\right)^2},\\
    \varphi\lrsq{\MB^2(z)\MB^{\dagger}(z)}=&\varphi\lrsq{\left(\MA(z)+\MI\right)^{2}\left(\MA(z)+\MI\right)^{\dagger}}=\varphi\lrsq{\MA^2(z)\MA^{\dagger}(z)}+\varphi\lrsq{2\MA(z)\MA^{\dagger}(z)+\MA^2(z)}\\
    &+\varphi\lrsq{2\MA(z)+\MA^{\dagger}(z)+\MI}=\frac{|z|^4}{\left(1-|z|^2\right)^2}+\frac{2|z|^2}{1-|z|^2}+1=\frac{1}{\left(1-|z|^2\right)^2},\\
    \varphi_2\lrsq{\MB(z)\MB^{\dagger}(z)}=&\varphi\lrsq{\MB(z)\MB^{\dagger}(z)\MB(z)\MB^{\dagger}(z)}=\frac{1}{\left(1-|z|^2\right)^4}\qquad\text{\citep{Hu2022}}.
\end{aligned}
\label{eq:identity_array}
\end{equation}
All these moment results are real-valued. We can similarly show the conjugate transpose versions, such as  $\varphi\lrsq{\MA^{\dagger}(z)}$, $\varphi\lrsq{(\MA^{\dagger}(z))^2 \MA(z)}$, which have the same values as their counterparts above.

To apply to $\MCxf$, note that $\MJeff = \frac{\phiprime}{1+i\omega}\MJ =\frac{g\phiprime}{1+i\omega}\MJ_0$, and let $z=\frac{g\phiprime}{1+i\omega}$. 
The norm $|z|=\geff(\omega) = \geff/\sqrt{1+\omega^2}<1$ (\cref{sec:proof_geff_bound}).
Let $m(\omega) = \Cphif\left(1-\frac{g^2\phiprime^2}{1+\omega^2}\right)
 = \Cphif [1-\geff^2(\omega)]$, we can then write \cref{eq:Cov_x_omega_form_2} as
\begin{equation}
    \MCxf = \frac{m(\omega)}{\phiprime^2} \MA\MA^{\dagger}+\frac{\sigma^2}{1+\omega^2}(\MA+\MA^{\dagger}+\MI).
\label{eq:Cov_x_omega_with_A}
\end{equation}

\subsection{First moment of $\MCxf$}
\label{app:first_moment_Cov_x}

Using identities in \cref{eq:identity_array}, the first moment of \cref{eq:Cov_x_omega_with_A} can be computed as (note $\varphi\lrsq{\MA}=\varphi\lrsq{\MA^{\dagger}}=0$),
\begin{equation}
    \varphi_1\lrsq{\MCxf} = \frac{m(\omega)}{\phiprime^2}\frac{\geff^2(\omega)}{1-\geff^2(\omega)}+0+0+\frac{\sigma^2}{1+\omega^2}=\frac{g^2 C^\phi(\omega)+\sigma^2}{1+\omega^2}.
\label{eq:first_moment_Cov_x_omega}
\end{equation}
Note that the left hand side is the homogeneous diagonal entry $\Cxf$. The above equation is exactly the same as the single neuron mean-field theory for the autocorrelation function \cref{eq:single_neuron_MFT}.

\subsection{Second moment of $\MCxf$}
\label{app:second_moment_of_Cov_x_omega}
We will first focus on the first term in \cref{eq:Cov_x_omega_with_A}, which is the only term when $\sigma=0$. So for $\sigma=0$, we will use $\MA=\MB-\MI$ and use the cyclic property of normalized trace, e.g. $\varphi\lrsq{\MB\MB^{\dagger}\MB}=\varphi\lrsq{\MB^2\MB^{\dagger}}$ etc., to compute $\varphi\lrsq{\MA(z)\MA^{\dagger}(z)\MA(z)\MA^{\dagger}(z)}$:
\begin{equation*}
    \begin{aligned}
    \varphi\left[\MA(z)\MA^{\dagger}(z)\MA(z)\MA^{\dagger}(z)\right]=&\varphi\left[(\MB-\MI)(\MB-\MI)^{\dagger}(\MB-\MI)(\MB-\MI)^{\dagger}\right]\\
    =&\left(\varphi\left[\MB\MB^{\dagger}\MB\MB^{\dagger}\right]-2\varphi\left[\MB^2\MB^{\dagger}\right]-2\varphi\left[\MB\left(\MB^{\dagger}\right)^2\right]+ \varphi\left[\MB^2\right] +\varphi\left[\left(\MB^{\dagger}\right)^2\right]\right.\\
    &\left.+4\varphi\left[\MB\MB^{\dagger}\right]- 2\varphi\left[\MB\right]-2\varphi\left[\MB^{\dagger}\right]+ \varphi(\MI)\right)\\
    =&\frac{|z|^4\lrbk{2-|z|^4}}{\lrbk{1-|z|^2}^4}.
    \end{aligned}
\end{equation*}
Next, we evaluate $|z|^2=\geff^2(\omega)$, and use \cref{eq:first_moment_Cov_x_omega},
\begin{equation}
\begin{aligned}
    \varphi_2\left[\MCxf\right]=&\frac{m^2(\omega)}{\phiprime^4}\frac{g^4\phiprime^4\left[2-\geff^4(\omega)\right]}{(1+\omega^2)^2\left[1-\geff^2(\omega)\right]^4}
    =\frac{2-\geff^4(\omega)}{\left[1-\geff^2(\omega)\right]^2}\left[\frac{g^2\Cphif}{1+\omega^2}\right]^2\\
    =&\frac{2-\geff^4(\omega)}{\left[1-\geff^2(\omega)\right]^2}\left[\Cxf\right]^2.
\end{aligned}
\label{eq:second_moment_Cov_x_nonzero_frequency_sigma0}
\end{equation}
Again, we used the random matrix identities in \cref{eq:identity_array} and the single-neuron mean-field identity \cref{eq:single_neuron_MFT} $g^2\Cphif=(1+\omega^2)\Cxf$ for $\sigma=0$ case.

Next, for the general case with $\sigma\ge 0$,  the second moment of $\MCxf$ involves a few additional terms that can be computed using \cref{eq:identity_array},
\begin{equation}
    \begin{aligned}
    \varphi_2\left[\MCxf\right] =& \frac{m^2(\omega)}{\phiprime^4} \varphi\left[\MA\MA^{\dagger}\MA\MA^{\dagger}\right]+\frac{m(\omega)\sigma^2}{\phiprime^2(1+\omega^2)}\varphi\left[2\MA^2\MA^{\dagger}+2\MA\left(\MA^{\dagger}\right)^2+2\MA\MA^{\dagger}\right]\\
    &+\frac{\sigma^4}{\left(1+\omega^2\right)^2} \varphi\left[\MA^2+\left(\MA^{\dagger} \right)^2+2\MA\MA^{\dagger}+2\MA+2\MA^{\dagger}+\MI\right]\\
    =&\frac{2-\geff^4(\omega)}{\left[1-\geff^2(\omega)\right]^2}\left[\frac{g^2\Cphif}{1+\omega^2}\right]^2+\frac{\sigma^2}{1+\omega^2}\frac{2\left[1+\geff^2(\omega) \right]}{1-\geff^2(\omega)}\frac{g^2\Cphif}{1+\omega^2}+\frac{\sigma^4}{\left(1+\omega^2\right)^2}\frac{1+\geff^2(\omega)}{1-\geff^2(\omega)}\\
    =&\frac{\frac{g^4}{\left(1+\omega^2\right)^2}}{\left[1-\geff^2(\omega)\right]^2}\left[\Cphif\right]^2+\frac{1-\geff^4(\omega)}{\left[1-\geff^2(\omega)\right]^2}\left[\Cxf\right]^2.
\end{aligned}
\label{eq:second_moment_Cov_x_nonzero_frequency_sigma>0}
\end{equation}
Again, we used the single-neuron mean-field identity $\frac{g^2\Cphif}{1+\omega^2}=\Cxf-\frac{\sigma^2}{1+\omega^2}$ (from \cref{eq:single_neuron_MFT}) in the last equality.
This result matches exactly with the one derived using the two-site cavity method (\cref{eq:second_moment_x_from_cavity}) and the result in \cite{Clark2023} when $\sigma=0$. 

Using the first and second moments of $\MCxf$ (\cref{eq:first_moment_Cov_x_omega,eq:second_moment_Cov_x_nonzero_frequency_sigma>0}), we can obtain the frequency dependent dimension of $\Dxf$ (\cref{eq:dim_x_w_sigma} in the main text),
\begin{equation*}
\begin{aligned}
    \Dxf=&\frac{\varphi_1^2\lrsq{\MCxf}}{\varphi_2\lrsq{\MCxf}}=\frac{\lrsq{\Cxf}^2}{\frac{1}{\lrsq{1-\geff^2(\omega)}^2}\lrsq{\frac{g^2\Cphif}{1+\omega^2}}^2+\frac{1-\geff^4(\omega)}{\lrsq{1-\geff^2(\omega)}^2}\lrsq{\Cxf}^2}\\
    =&\frac{\lrsq{1-\geff^2(\omega)}^2\lrsq{\Cxf}^2}{\lrsq{\Cxf-\frac{\sigma^2}{1+\omega^2}}^2 + [1-\geff^4(\omega)]\lrsq{\Cxf}^2}\\
    =&\frac{\lrsq{1-\geff^2(\omega)}^2}{1+\lrsq{1-\frac{\sigma^2}{(1+\omega^2)\Cxf}}^2-\geff^4(\omega)}.
\end{aligned}
\end{equation*}
In the special case when $\sigma=0$, similar to the cancellation when computing $\Dphif$, \cref{eq:dim_x_w_sigma} reduces to a simple expression without explicit dependence on $\Cxf$, 
\begin{equation}
\Dxf = \frac{\lrsq{1-\geff^2(\omega)}^2}{2 - \geff^4(\omega)}
 = \frac{\lrbk{1+\omega^2-g^2\phiprime^2}^2}{2\lrbk{1+\omega^2}^2 - g^4\phiprime^4}.
\label{eq:dim_x_w_sigma0}
\end{equation}

\section{Extending $\zeta$ properties to $\sigma \ge 0$ case}
\label{app:extend_zeta}
In this section, we extend the results on the relationship between $\Czetaf$, $\Cphif$, and $\Cxf$ in the main text \cref{sec:effective_dynamics} to the case with external noise input $\sigma\geq0$. When $\sigma=0$, $\zeta_i=\phi_i-\phiprimei x_i$ itself is the driving force and we call it effective noise since $\zeta$ in $\MCphif$ plays the role of $\xi$ in linear network covariance. For $\sigma>0$, the external noise $\xi_i$ will go into $x_i$, so we need to combine the effective recurrent input, i.e. $\zeta_i$, and external random input $\xi_i$ to be the overall effective noise $\tilde{\zeta}_i=\zeta_i+\frac{\phiprimei}{1+i\omega}\xi_i$.

Recall that the total effective noise $\tilde{\zeta}_i$ is
\begin{equation}
    \tilde{\zeta}_i(\omega)
    := \zeta_i(\omega)+\frac{\phiprimei}{1+i\omega}\xi_i(\omega)
    = \phi_i(\omega) - \phiprimei x_i(\omega) +\frac{\phiprimei}{1+i\omega}\xi_i(\omega)
    = \phi_i(\omega) -\frac{\phiprimei}{1+i\omega}\sum_{j=1}^NJ_{ij}\phi_j(\omega).
\label{eq:total_zeta_definition}
\end{equation}
Here we have used the Fourier transform of the network dynamics \cref{eq:network_model}, 
\[
x_i(\omega)=\frac{1}{1+i\omega}\lrbk{\sum_{j=1}^NJ_{ij}\phi_j(\omega)+\xi_i(\omega)}.
\]
From \cref{eq:total_zeta_definition}, the Fourier transform of firing rate $\phi_i(\omega)$ and $x_i(\omega)$ can be readily solved in terms of  $\tilde{\zeta}_j(\omega)$ in the vector form, similarly as \cref{eq:effective_dynamics_of_phi},
\begin{equation}
\begin{aligned}
    \vec{\phi}(\omega)&=\lrbk{\MI-\frac{1}{1+i\omega}\MPhiPrime\MJ}^{-1}\tilde{\vec{\zeta}}(\omega),\\
    (1+i\omega)\vec{x}(\omega)&=\MJ\lrbk{\MI-\frac{1}{1+i\omega}\MPhiPrime\MJ}^{-1}\tilde{\vec{\zeta}}(\omega)+\vec{\xi}\\
    &=\MJ\lrbk{\MI-\frac{1}{1+i\omega}\MPhiPrime\MJ}^{-1}\vec{\zeta}+\lrsq{\MJ\lrbk{\MI-\frac{1}{1+i\omega}\MPhiPrime\MJ}^{-1}\frac{\MPhiPrime}{1+i\omega}+\MI}\vec{\xi}
\end{aligned}
\label{eq:effective_dynamics_of_phi_sigma>0}
\end{equation}
From here on we invoke the homogeneity $\phiprimei\to\phiprime$ as $N\to\infty$, because without this step, $\MJ$ and $\MPhiPrime$ are not commutative, this could indicate that our expression of covariance matrices may not be general for heterogeneous network. Then we use $\MJeff\lrbk{\MI-\MJeff}^{-1}+\MI=\lrbk{\MI-\MJeff}^{-1}$ with $\MJeff=\frac{\phiprime}{1+i\omega}\MJ$ to get
\begin{equation}
\begin{aligned}
    \vec{\phi}(\omega) &= \lrbk{\MI-\MJeff}^{-1}\vec{\tilde{\zeta}}(\omega)\\
    (1+i\omega)\vec{x}(\omega) &= \MJ\lrbk{\MI-\MJeff}^{-1}\vec{\zeta}+\lrbk{\MI-\MJeff}^{-1}\vec{\xi}.
\end{aligned}
\label{eq:a_homogeneous_with_zeta}
\end{equation}

From above equations, we immediately obtain a general form of \cref{eq:Cov_phi_omega_from_zeta}
\begin{equation}
\MCphif=\lrbk{\MI-\MJeff}^{-1} \MC^{\tilde{\zeta}}(\omega) \lrbk{\MI-\MJeff^\dagger}^{-1}.
\label{eq:Cov_phi_omega_from_zeta_sigma}
\end{equation}

Accordingly, the condition (\cref{eq:ansatz_Cov_zeta}) sufficient to derive the $\MCphif$ ansatz (\cref{eq:ansatz_of_Cov_phi_omega}) is changed to 
\begin{equation}
\text{Ansatz:} \quad \MC^{\tilde{\zeta}}(\omega)  = \left(1-\frac{g^2\phiprime^2}{1+\omega^2}\right)\Cphif \cdot \MI \;\;\text{ as } N\to\infty.
\label{eq:ansatz_Cov_zeta_sigma}
\end{equation}

To motivate \cref{eq:ansatz_Cov_zeta_sigma}, note that $\tilde{\zeta}_i=\zeta_i+\frac{\phiprimei}{1+i\omega}\xi_i$, so $\MC^{\tilde{\zeta}}(\omega)$ expands to 4 terms,
\begin{equation}
\MC^{\tilde{\zeta}}(\omega) = \MCzetaf + \frac{\phiprime}{1-i\omega}\MC^{\zeta\xi}(\omega) + \frac{\phiprime}{1+i\omega}\MC^{\xi\zeta}(\omega) + \frac{\phiprime^2}{1+\omega^2}\MC^{\xi}(\omega).
\label{eq:Czeta_tilde_expansion}
\end{equation}
We assume $(\zeta_i, \xi_j)$ are uncorrelated, i.e. $\MC^{\zeta\xi}_{ij}(\omega)=0$, which means it has no effect in remaining part of this section.
For the last term in \cref{eq:Czeta_tilde_expansion}, note that $\MC^{\xi}(\omega) = \sigma^2 \cdot \MI$ since the external noise is i.i.d.
Finally, we use the ansatz for $\zeta$ (\cref{eq:ansatz_Cov_zeta}) and the expression of its diagonal entries (\cref{eq:Czeta_diag}). In particular, \emph{the first equality} of \cref{eq:Czeta_diag}, $\Czetaf = \Cphif-\phiprime^2\Cxf$, still holds since its derivation does not involve whether $\sigma$ is zero or positive. Therefore \cref{eq:Czeta_tilde_expansion} becomes (invoking the homogeneity of neurons to write $\phiprimei=\phiprime$ as $N\to \infty$),
\[
\MC^{\tilde{\zeta}}(\omega)=\left[\Cphif -\phiprime^2 \Cxf\right] \cdot \MI + 0 + 0 + \frac{\sigma^2 \phiprime^2} {1+\omega^2}\cdot \MI
= \left(1-\frac{g^2\phiprime^2}{1+\omega^2}\right)\Cphif \cdot \MI.
\]
We used \cref{eq:single_neuron_MFT} in the last equality.

To obtain \cref{eq:Cx_with_zeta}, from \cref{eq:a_homogeneous_with_zeta} and set $\MC^{\zeta\xi}(\omega)=0$, we have
\begin{equation}
(1+\omega^2)\MCxf=\MJ\lrbk{\MI-\MJeff}^{-1} \MC^{\zeta}(\omega) \lrbk{\MI-\MJeff^\dagger}^{-1}\MJ^T+\lrbk{\MI-\MJeff}^{-1} \MC^{\xi}(\omega) \lrbk{\MI-\MJeff^\dagger}^{-1}.
\label{eq:Cov_x_omega_from_zeta_sigma}
\end{equation}
When $N\to\infty$, we use $\MCzetaf=\lrbk{\Cphif-\phiprime^2\Cxf}\MI$ and $\MC^{\xi}(\omega)=\sigma^2\MI$, then we arrive at \cref{eq:Cx_with_zeta}.

\section{Asymptotic power-law tail of covariance spectra $p^{a,\omega}(\lambda)$}
\label{app:power_law_tail}
To derive the power-law tail approximation, we follow the method in Supplementary Materials B of \cite{Hu2022}. Briefly, the method assumes that as the parameter $\geff\rightarrow 1^-$, the probability density function of the spectrum $ p_{\geff}(\lambda)$ coverages to a finite, positive value for any fixed $\lambda> \lambda_-^0\ge 0$ ($\lambda_-^0$ is the limit of the left edge of the spectrum support), and its large value tail follows a power law 
\[
\lim_{\lambda \rightarrow \infty} \lim_{\geff\rightarrow 1^-} \lambda^{\beta} p_{\geff}(\lambda) = 1,
\] 
and the right edge of the spectrum's support diverges as $\lambda_{+} \propto (1-\geff^2)^{-\alpha}$.
Then the moments of the spectrum, as orders of $(1-\geff^2)$, are determined by the exponents $\alpha$, $\beta$:
\begin{equation}
    \mathbb{E}\lrsq{\lambda^n}\propto \delta^{-\alpha(-\beta+n+1)},
\quad \delta = (1-\geff^2),
\label{eq:Cov_spectrum_moment_n}
\end{equation}
which also means that we can solve for the exponent $\alpha, \beta$ from two moments of the spectrum.

Define
\begin{equation}
m(\omega)= \Cphif\left(1-\frac{g^2\phiprime^2}{1+\omega^2}\right), \;\;\text{and } n(\omega) = \frac{\sigma^2 \phiprime^2}{1+\omega^2}.
\label{eq:coef_m_n_define}
\end{equation}
The expression of $\MCxf$ (\cref{eq:Cx_with_zeta}) can be written as
\begin{equation}
\begin{split}
\phiprime^2 \MCxf &= [m(\omega) - n(\omega)]\;
\MJeff\left(\MI-\MJeff\right)^{-1}\left(\MI-\MJeff^{\dagger}\right)^{-1}\MJeff^{\dagger}
+ n(\omega)
\left(\MI-\MJeff\right)^{-1}\left(\MI-\MJeff^{\dagger}\right)^{-1}.
\end{split}
\end{equation}
This also allows us to consider the asymptotics of the frequency dependent spectrum by considering the limit of
\[
\geff(\omega) = \frac{g\phiprime}{\sqrt{1+\omega^2}} \rightarrow 1^-,\quad \text{and the orders of moments in } \delta = (1-\geff^2(\omega)),
\]
instead of $\geff\rightarrow 1^-$ and $\delta= 1-\geff^2(\omega)$. Note that from \cref{sec:proof_geff_bound}, when $\sigma>0$ the $g\phiprime<1$, thus $\geff(\omega)<\frac{1}{\sqrt{1+\omega^2}}$. We need to take $\omega\to0$ to achieve above limit.

As a corollary of the diagonal entries of $\MCzetaf$ (\cref{eq:Czeta_diag}, note its first equality is general and valid for $\sigma\ge 0$),  
\[
0\le C^{\zeta}(\omega) 
= \Cphif - \phiprime^2 \Cxf
= \Cphif - \frac{\phiprime^2 (g^2 \Cphif+\sigma^2)}{1+\omega^2}.
\]
This leads to an inequality,
\begin{equation}
m(\omega)\ge n(\omega)\ge 0.
\label{eq:coef_m_n_inequality}
\end{equation}
Note that the power-law property $p(\lambda)\propto \lambda^{-\beta}$ is invariant when multiplying the covariance matrix/eigenvalues with a scalar. Therefore, we can equivalently focus on deriving a power law for $\tilde{\MC}^x(\omega):= \phiprime^2 \MCxf/ m(\omega)$, which is
\begin{equation}
\tilde{\MC}^x(\omega) = \left(1-\frac{n(\omega)}{m(\omega)}\right) \M{P} + \frac{n(\omega)}{m(\omega)}\M{Q}, 
\label{eq:rescaled_Cov_x_omega}
\end{equation}
where
\[
\M{P} = \MJeff\left(\MI-\MJeff\right)^{-1}\left(\MI-\MJeff^{\dagger}\right)^{-1}\MJeff^{\dagger},
\quad \M{Q}=\left(\MI-\MJeff\right)^{-1}\left(\MI-\MJeff^{\dagger}\right)^{-1}.
\]
From the analytical expression for linear dynamics covariance spectrum \citep{Hu2022}, we know that the spectrum of $\M{Q}$ as $\geff(\omega)\rightarrow 1^-$ has a probability density function $p_{\M{Q}}(\lambda)$ that converge to a  finite value for any fixed $\lambda$. For the matrix $\M{P}$, we hypothesize the same is true.
The scalar coefficients for $\M{P}$, $\M{Q}$ in \cref{eq:rescaled_Cov_x_omega} are both between 0 and 1 and they sum to 1. Therefore, $\tilde{\M{C}}^x(\omega)$ as the linear combination also satisfies this finite limit property of the spectrum probability density function, and we can thus apply \cref{eq:Cov_spectrum_moment_n} to determine the power-law exponent for $\tilde{\M{C}}^x(\omega)$ spectrum.

Using the results from \cref{app:second_moment_and_dimension}
\begin{equation}
\varphi_1\lrsq{\tilde{\M{C}}^{x}(\omega)}= 
\frac{\geff^2(\omega)}{1-\geff^2(\omega)} + \frac{n(\omega)}{m(\omega)}.
\label{eq:first_moment_tilde_Cov_x_omega}
\end{equation}
Since the last term in the above is between $0$ and $1$, we see that  $\varphi_1\lrsq{\tilde{\M{C}}^x(\omega)} \propto \delta ^{-1}$ as $\geff(\omega)\rightarrow 1^-$.
For the second moment (\cref{eq:second_moment_Cov_x_nonzero_frequency_sigma>0}), 
\begin{equation}
\varphi_2\lrsq{\tilde{\MC}^{x}(\omega)} 
=\frac{\geff^4(\omega)}{\left[1-\geff^2(\omega)\right]^{4}} 
+\left( \frac{\geff^2(\omega)}{1-\geff^2(\omega)} + \frac{n(\omega)}{m(\omega)}
\right)^2\cdot \frac{1+\geff^2(\omega)}{1-\geff^2(\omega)}.
\label{eq:second_moment_tilde_Cov_x_omega}
\end{equation}
As $\geff(\omega)\rightarrow 1^-$, the first term above dominates and hence $\varphi_2\lrsq{\tilde{\MC}^{x}(\omega)} \propto \delta^{-4}$.
From these moments results, we can solve for the exponents as $\alpha =3$ and $\beta=5/3$, which means $\MCxf$ spectrum has a power-law tail as $\Pxf\propto\lambda^{-5/3}$.

\section{The left edge of $\Pxf$'s support when $\sigma=0$}
\label{app:left_support_of_x_spectrum}
Starting from the expression of the frequency-dependent covariance matrix of currents (\cref{eq:Cov_x_omega_full}, for any fixed $\omega$) at $\sigma=0$,
\[
\MCxf \propto  
\MJeff(\MI-\MJeff)^{-1} \left(\MI-\MJeff^{\dagger}\right)^{-1} \MJeff^{\dagger} 
=:\tilde{\MC}^{x}(\omega).
\]
Recall that $\MJeff = \frac{\phiprime}{1+i\omega}\MJ$ (\cref{eq:Cov_phi_omega_from_zeta}) and we introduced $\tilde{\MC}^{x}(\omega)$, differing with $\MCxf$ by a scalar, for ease of notation (it is the same as \cref{eq:rescaled_Cov_x_omega} while setting $\sigma=0$). 
Note that it is sufficient to show that the left edge of spectrum support of $\tilde{\MC}^{x}(\omega)$ being 0.

Since $\tilde{\MC}^{x}(\omega)$ is Hermitian, its minimal eigenvalue is given by
\begin{equation}
\lambda_{\min}\lrsq{\tilde{\MC}^{x}(\omega)} = \min_{\Vert \vec{v}\Vert =1} \vec{v}^{\dagger} \tilde{\MC}^{x}(\omega) \vec{v} = \min_{\Vert v\Vert =1}  \left\Vert \left(\MI-\MJeff^{\dagger}\right)^{-1} \MJeff^{\dagger} \vec{v} \right\Vert^2.
\label{eq:min_eig_Cx}
\end{equation}
Here the minimization is taken over all vectors $\vec{v}\in \mathbb{C}^N$, and $\Vert v \Vert = \sqrt{\sum_{i=1}^N |v_i|^2}$ is the vector 2-norm.
Note that 
\[
\min_{\Vert \vec{v}\Vert =1}  \left\Vert \MJeff^\dagger \vec{v} \right\Vert
= \frac{\phiprime}{\sqrt{1+\omega^2}}\cdot 
 \min_{\Vert \vec{v}\Vert =1}  \Vert \MJ^T \vec{v} \Vert,
\]
and $\min_{\Vert \vec{v}\Vert =1}  \left\Vert \MJ^T \vec{v} \right\Vert$ is the minimal singular value of $\MJ^T$. Since $J_{ij}$ are i.i.d. Gaussian distributed (\cref{sec:network_model,eq:network_model}), from the well-known Marchenko-Pastur law, the minimal singular value of $\MJ^T$ is 0 as $N\to\infty$. This means that in large networks $N\rightarrow \infty$, for any $\delta>0$, there exists a vector $\vec{v}_0$ (the minimal singular vector of $\MJ^T$) such that
\[
\Vert\MJ^T\vec{v}_0\Vert\le \delta.
\]
Consider this special $\vec{v}_0$ in \cref{eq:min_eig_Cx}, we have
\begin{equation}
\begin{aligned}
    \lambda_{\min}\lrsq{\tilde{\MC}^{x}(\omega)}
&\le \frac{\phiprime^2}{1+\omega^2}\cdot 
\left\Vert \lrbk{\MI-\MJeff^{\dagger}}^{-1} \MJ^T \vec{v}_0 \right\Vert^2
\le \frac{\phiprime^2}{1+\omega^2} \cdot \left\Vert\lrbk{\MI-\MJeff^{\dagger}}^{-1}\right\Vert^2 \left\Vert\MJ^T \vec{v}_0\right\Vert^2\\
&\le \frac{\delta^2 \phiprime^2}{1+\omega^2}\cdot 
\left\Vert\lrbk{\MI-\MJeff^{\dagger}}^{-1}\right\Vert^2 .
\end{aligned}
\label{eq:min_eig_bound_Cx}
\end{equation}
Note that $\left\Vert(\MI-\MJeff^\dagger)^{-1}\right\Vert^2$ (the operator norm or matrix 2-norm) is the square of the largest singular value of $(\MI-\MJeff^\dagger)^{-1}$, which is also the largest eigenvalue of $\lrbk{\MI-\frac{\phiprime}{1+i\omega}\MJ}^{-1}\lrbk{\MI-\frac{\phiprime}{1-i\omega}\MJ^{T}}^{-1}$. 
Using the covariance spectrum expression for networks with linear dynamics \citep{Hu2022}, in particular its right edge of support, we know that
\begin{equation}
\begin{aligned}
    &\lambda_{\max} \left[\lrbk{\MI-\frac{\phiprime}{1+i\omega}\MJ}^{-1}\lrbk{\MI-\frac{\phiprime}{1-i\omega}\MJ^{T}}^{-1}\right]\\
    &=\frac{2+5\geff^2(\omega)-\frac{\geff^4(\omega)}{4}+\frac{\geff(\omega)}{4}\lrsq{8+\geff^2(\omega)}^{3/2}}{2\lrsq{1-\geff^2(\omega)}^3} 
    \le \frac{7}{\lrsq{1-\geff^2(\omega)}^3}.
\end{aligned}
\label{eq:min_norm_bound_I_J}
\end{equation}
Here
\[
\geff(\omega) = \frac{\phiprime}{\sqrt{1+\omega^2}} \cdot g
\le \geff <1.
\]
In particular, this means $\left\Vert(\MI-\MJeff^\dagger)^{-1}\right\Vert^2$ is bounded by a fixed number depending only on $\geff(\omega)$ in large networks $N\rightarrow \infty$.
Going back to \cref{eq:min_eig_bound_Cx}, we have
\[
\lambda_{\min}\lrsq{\tilde{\MC}^{x}(\omega)} \le 
\frac{7\delta^2 \phiprime^2}{(1+\omega^2) \lrsq{1-\geff^2(\omega)}^3}
\le \frac{7\delta^2 }{\lrsq{1-\geff^2(\omega)}^3}.
\]
Since $\delta$ can be any positive number and $\tilde{\MC}^{x}(\omega)$ is positive semi-definite so all its eigenvalues are non-negative, this means that $\lambda_{\min}=0$, that is, the left edge of the support of $\tilde{\MC}^{x}(\omega)$'s spectrum is 0, when there is no external noise $\sigma=0$.

\section{Proof of $\geff < 1$ for $\sigma>0$}
\label{sec:proof_geff_bound}
Here we prove that the effective connection strength $\geff<1$ for all values of $g$ and $\sigma>0$.
Recall that (\cref{eq:Czeta_diag}, whose first equality holds generally for $\sigma\ge0$)
\[
\Czetaf = \Cphif-\phiprime^2\Cxf \ge 0.
\]
Multiplying both sides by $g^2$,
\begin{equation}
\geff^2 = g^2\phiprime^2 \le \frac{g^2\Cphif}{\Cxf}
= \frac{(1+\omega^2) \Cxf -\sigma^2}{\Cxf}
\le 1+\omega^2 - \frac{\sigma^2}{\Cxf}.
\label{eq:geff_inequality}
\end{equation}
We have used \cref{eq:single_neuron_MFT} in the second equality above. The division by $\Cxf$ is justified because $\Cxf\ge \sigma^2/(1+\omega^2)>0$ (\cref{eq:single_neuron_MFT}).
Since \cref{eq:geff_inequality} holds for any $\omega$, setting $\omega=0$ gives $\geff<1$.\\

\begin{figure}[H]
    \centering
    \includegraphics[width=0.65\linewidth]{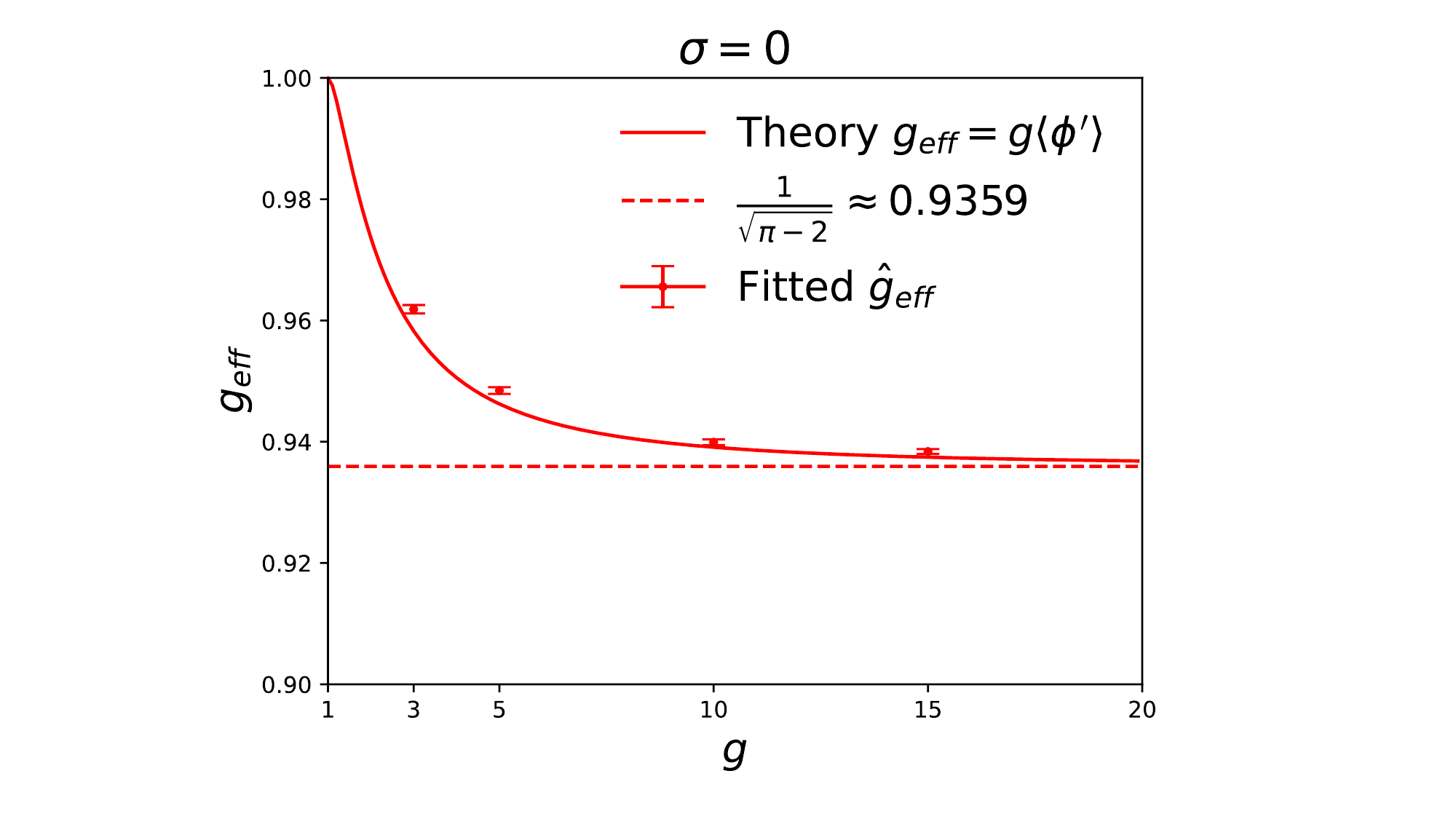}
    \caption{Numerical verification of the strict inequality $\geff<1$ for the special case of no noise $\sigma=0$.
    The $\geff=g\phiprime$ decreases from 1 monotonically with $g$ in the chaotic regime $g>1$. The solid curve is from the  mean-field theory (\cref{eq:single_neuron_MFT} and \cite{Schuecker2018}) and the dots (mean) and error bars (standard error across five $\MJ$ realizations) show the fitted $\hat{g}_{\text{eff}}$ from network simulations (\cref{eq:network_model}) 
    }
    \label{fig:geff_vs_g_sigma=0}
\end{figure}

For the special case without noise $\sigma=0$ and $g>1$ (so that the dynamics is not at a fixed point), the above proof can adapted to prove a weaker result $\geff \le 1$. 
In particular, we need to justify $C^x(\omega=0)>0$ for the step of dividing it in \cref{eq:geff_inequality}. 
As shown in \cite{Sompolinsky1988}, $C^x(\tau)$ decreases monotonically to zero with the time lag $\tau$, and hence $C^x(\omega=0) = \int_{-\infty}^\infty C^x(\tau)d\tau$ has no cancellation in the integral and is therefore positive. 
Lastly, although our proof only shows $\geff \le 1$ for $\sigma=0$, one can directly verify that $\geff<1$ and decreases monotonically with $g$ (\cref{fig:geff_vs_g_sigma=0}) by numerically solving the single-neuron mean-field theory \cref{eq:single_neuron_MFT}.

\section{Proof of non-negative dimension gap $\Delta D(\omega)=D^{\phi}(\omega)-D^{x}(\omega)$}
\label{app:non-negative_dimension_gap}
Here we provide a proof that for the frequency-dependent dimensions, the dimension gap $\Delta D(\omega)=D^{\phi}(\omega)-D^{x}(\omega)$ is always non-negative for all frequencies and is valid for all parameters $g$ and $\sigma$, as long as the dynamics is not at a fixed point (when $\sigma=0$ and $g<1$). 

We start from \cref{eq:dim_x_w_sigma},
\[
\Dxf = \frac{\left(1+\omega^2-g^2\phiprime^2 \right)^2}{\left(1+\omega^2\right)^2+\left(1+\omega^2-\frac{\sigma^2}{\Cxf}\right)^2-g^4\phiprime^4}.
\]
Using \cref{eq:single_neuron_MFT}, the second term in the denominator can be written as
\[
\left(\frac{(1+\omega^2) \Cxf -\sigma^2}{\Cxf}\right)^2
 = \left(\frac{g^2\Cphif}{\Cxf}\right)^2
 \ge  g^4\phiprime^4.
\]
The last inequality is due to $\Czetaf = \Cphif-\phiprime^2\Cxf \ge 0$ (\cref{eq:Czeta_diag} in the main text), since $\Czetaf$ is the Fourier transform of the autocorrelation function of $\zeta_i$.
Using this inequality in the $\Dxf$ expression,
\[
\Dxf  \le \frac{\left(1+\omega^2-g^2\phiprime^2 \right)^2}{\left(1+\omega^2\right)^2} = \left(1 -\frac{g^2\phiprime^2}{1+\omega^2} \right)^2 = \Dphif,
\]
which proves the desired dimension gap. The last equality above is according to \cref{eq:D_phi_omega}.

\section{Additional details on inferring network parameters}
\label{app:data_analysis_method}
Before describing the details of the inference method (\cref{sec:infer_parameter_main_text}), we first need to explain a fundamental degeneracy in model parametrization when we observe only the firing rates and not the currents, which is the most typical situation in experiments.  
Let $\{\phi_i(t)\}$, $i\in\{1,\cdots,N\}$, be the observed firing rate data for $N$ neurons, and let $t=j\Delta t$ for $j\in\{1,\cdots,N_t\}$, where $N_t$ is the number of recording frames and $\Delta t$ is the recording time step.  
If we scale and shift the currents as $x_i(t) \rightarrow k(x_i(t)-b)$, and simultaneously substitute the network parameters as $J\rightarrow kJ$ (meaning $g \rightarrow kg$, $\sigma \rightarrow k\sigma$, and the neuronal firing-rate function $\phi(\cdot) \rightarrow \phi(\cdot/k+b)$), then it is easy to see that the dynamics of the new network is identical to that of the original one for any $k>0$ and $b$. Therefore, there is a degeneracy in the network parametrization, and given the firing rates, one can only identify the equivalence class of parameters under the transformations $(g,\sigma,\phi(\cdot)) \rightarrow (kg, k\sigma,\phi(\cdot/k+b))$.

The first step is to estimate the neuronal firing-rate function $\phi(x)$ using the approach in \cite{Lim2015}. Briefly, their approach is based on two assumptions: (i) $\phi(x)$ is monotonically increasing, and (ii) the distribution of currents $\{x_i(t)\}$ across time is Gaussian. The second assumption is justified in large randomly connected neural network models, including ours, similarly as in the classic mean-field theory, where the neuron receives a large number of weakly correlated inputs. 
Under these assumptions, without loss of generality, we can assume that $\{x_i(t)\}$ follows a standard normal distribution $N(0,1)$. This is related to the degeneracy discussed in \cref{sec:infer_parameter_main_text} and will uniquely select a representative set of parameters from the equivalence class. With these assumptions, the firing-rate function $\phi(x)$ can be inferred from the activity data by comparing the cumulative distributions \citep{Lim2015}.
This allows us to estimate the currents $\{x_i(t)\}$ via $\hat{\phi}^{-1}$.

To obtain the dynamic gain $\phiprime$ within $\geff$ (\cref{eq:g_eff_definition}), one can either numerically compute the derivative using the estimated $\hat{\phi}(\cdot)$, or, for better numerical accuracy, compute $\phiprime$ as
$
\hat{\phiprimei}= \frac{1}{N_t}\sum_{j=1}^{N_t}\hat{x}_i(j\Delta t) \phi_i(j\Delta t),
$
which is based on Stein's lemma: $\mean{x_i(t)\phi_i(t)}=\mean{\phi_i^{\prime}(t)}\mean{x_i^2(t)}=\phiprimei$, under the distribution assumption of $x_i(t)$. 
We then estimate the network-wide $\phiprime$ by averaging the dynamic gain across neurons. 
The effective recurrent connection strength $\geff$ can be estimated by fitting the firing rate covariance spectrum \citep{Hu2022}. Then, according to \cref{eq:g_eff_definition}, $g$ can be estimated as $\hat{g} =\hat{g}_{\text{eff}}/\hat{\phiprime}$. Therefore, we have individually estimated the two factors contributing to the effective recurrent connection strength.

Lastly, we need to estimate the external input magnitude $\sigma$. In the main text, we describe an intersection method (\cref{fig:contour_geff0.95}AB). Here we introduce an alternative temporal difference method (\cref{fig:contour_geff0.95}C), which is more accurate when the temporal resolution of the recorded neural activity is sufficiently high. In our network simulations, this method works well when $\Delta t / \tau \le 0.1$.
Consider the temporal discretization of the dynamic equation \cref{eq:network_model},
\begin{equation}
    x_i(t+\Delta t) - x_i(t)
+ x_i(t) \Delta t  = \sum_j J_{ij} \phi_j \Delta t  + \Delta \xi_i(t) + o(\Delta t).
\label{eq:discretized_dynamics}
\end{equation}
Now consider the variance of the above equation, and use the mean-field theory to substitute the random recurrent input term. The variance of the right-hand side is
\[
\Delta t ^2 g^2 \mean{\phi^2} + \Delta t \sigma^2.
\]
Note that the variance of the left-hand side of \cref{eq:discretized_dynamics} can be computed from the inferred currents $\hat{x}_i(t)$, and $\mean{\phi_i^2}$ can be computed from the firing rate data. Then we can solve for $\sigma$ from the above using the previously estimated $\hat{g}$.

The effectiveness of the method is illustrated in \cref{fig:contour_geff0.95}.
To facilitate comparison with the ground-truth parameters, here we choose another convention to represent the equivalence class of inferred network parameters by requiring that the maximum slope of $\phi(\cdot)$ occurs at 0 and equals 1 (which is consistent with the true firing-rate function in our network model \cref{eq:network_model} in \cref{sec:network_model}).

\section{Justification for $d\geff/d\sigma < 0$}
\label{sec:par_geff_par_sigma}
In the main text, we use the property of $\geff$ decreasing with $\sigma$ when fixing $g$ to estimate $\sigma$ in \cref{sec:infer_parameter_main_text}. 
Here we provide the justification of this property. According to the definition of $\geff=g\phiprime$ and use the chain rule
\begin{equation}
     \frac{d\geff}{d\sigma} = g\frac{d\phiprime}{d\Czero}\frac{d\Czero}{d\sigma},
     \quad \Czero = \mean{x_i^2(t)}_{t}.
\label{eq:dgeff_dsigma_chain_rule}
\end{equation}

For the first factor $\frac{d\phiprime}{d\Czero}$ in \cref{eq:dgeff_dsigma_chain_rule}, 
\begin{equation}
    \phiprime = \mathbb{E}_{t}\lrsq{\phi^\prime(x_i(t))} = \mathbb{E}_Z\lrsq{\phi^\prime\lrbk{\sqrt{\Czero} Z}}
= 2\int_0^\infty \phi^\prime \lrbk{\sqrt{\Czero} z}\frac{1}{\sqrt{2\pi}}e^{-z^2/2} \,dz.
\label{eq:phi_prime_Gaussian_integral}
\end{equation}
Where $Z$ is a standard normal random variable.
Note that $\phi^\prime (x) = 1/\cosh^2(x) =4/\lrbk{e^x+e^{-x}}^2$ is an even function and is decreasing for $x>0$; therefore the integral \cref{eq:phi_prime_Gaussian_integral} shows $\frac{d\phiprime}{d\Czero}<0$.

For the second factor in \cref{eq:dgeff_dsigma_chain_rule}, we claim that $\frac{dc_0}{d\sigma}>0$ based on the intuition that when the variance of the external input $\xi_i$ to the neurons (i.e., $\sigma^2$) increases, the resulting variance of the neurons' current $x_i(t)$ (i.e., $\Czero$) will increase. A direct proof of this property is left for future work.
Combining the results on the two factors in \cref{eq:dgeff_dsigma_chain_rule}, we conclude that $d\geff/d\sigma <0$.

\section{Extension of the theory for partially symmetric networks}
\label{app:extension_to_partially_symmetric}
This section includes additional details of the theoretical derivations needed to generalize the covariance matrix ansatz to partially symmetric networks (\cref{app:dynamic_gain_part_symm}), a hybrid approach of obtaining the theoretical values of $\beta$, $g_{c,\text{eff}}$, and $C^{\zeta}$ using large-network simulations (\cref{app:estimate_beta_and_C_zeta}), and a theoretical cross-checking of the ansatz with previous results on the dimensionality (\cref{app:random_matrix_derivation_part_symm,app:first_second_moment_part_symm}), where \cref{app:random_matrix_derivation_part_symm} includes the necessary random matrix results.

\subsection{Derivation of the dynamic gain $\beta$}
\label{app:dynamic_gain_part_symm}
Motivated by the fact that the generalized anstaz \cref{eq:ansatz_of_Cov_phi_omega_symmetric}, when choosing an appropriate $\beta$ value, matches closely with network simulation (\cref{fig:summary_symm_J}BCEF), we describe here a heuristic derivation for theoretical value of $\beta$ (\cref{eq:beta_equation}), under which the $\MCzetaf$ matrix of the effective recurrent noise becomes a scalar matrix $\Czetaf\MI$ and leads to the $\phi$ and $x$ covariance matrix ansatz \cref{eq:ansatz_of_Cov_phi_omega_symmetric}.

We start from the unperturbed reservoir neurons' activity and cavity neurons' activity (similar to \cref{eq:unperturbed_activity}) in the case of $\sigma=0$ without external input for simplicity.
The single-neuron dynamics equation can be written as
\begin{equation}
\left\{
\begin{aligned}
    (1+\partial_t)x_{i}^{\free}(t)&=\sum_{j=1}^NJ_{ij}\phi_j^{\free}(t)=:\eta_i(t)+g^2\kappa\int_{-\infty}^{t}S^{\phi}_{ii}(t-t')\phi_i^{\free}(t')dt',\\
    (1+\partial_t)x_{\mu}^{\free}(t)&=\sum_{i=1}^{N}J_{\mu i}\phi_{i}^{\free}(t)=:\eta_{\mu}(t).
\end{aligned}
\right.
\label{eq:unperturbed_activity_part_symm}
\end{equation}
Note that for partially symmetric network, the total recurrent input $\sum_{j=1}^NJ_{ij}\phi^{\free}_j(t)$ is non-Gaussian, but it can be separated to a Gaussian part and a non-Gaussian part \citep{Marti2018,Clark2023}. The Gaussian component of recurrent input $\eta$ is generated by the same mechanism and plays a similar role as $\eta^{\free}$ in \cref{eq:unperturbed_activity}. For ease of notation we suppress the superscript ``free'' for $\eta$ throughout this section. 

Following the same steps as in \cref{app:cavity_unit_four_point_function} to introduce perturbation $\delta\phi_i(t')$, we get the following equation written in the same form as \cref{eq:a_t_first_order_expansion} (but note that $F_{\mu\mu}(t,t^{\dprime})$ is $O\lrbk{\sqrt{N}}$ here), the current and firing rate of the cavity neurons when the reservoir neurons are perturbed with $J_{i\mu}, J_{\mu i}, J_{\mu\nu}\neq0$ follows 
\[
a_{\mu}(t)=a_{\mu}^{\free}(t)+\frac{1}{\sqrt{N}}\int^tdt'S^a_{\mu\mu}(t,t')\int^{t'}dt''\sum_{\nu\in\{0,0'\}}F_{\mu\nu}(t',t'')\phi^{\free}_{\nu}(t'')+O\lrbk{\frac{1}{N}}.
\]
In the frequency domain
\begin{equation}
    a_{\mu}(\omega) = a^{\free}_{\mu}(\omega)+\frac{1}{\sqrt{N}}S^a_{\mu\mu}(\omega)\sum_{\nu\in\{0,0'\}}F_{\mu\nu}(\omega)\phi_{\nu}^{\free}(\omega)+O\lrbk{\frac{1}{N}}.
\label{eq:perturbation_freq}
\end{equation}

Using the definition $\zeta_{\mu}(\omega):=\phi_{\mu}(\omega)-\beta(\omega)x_{\mu}(\omega)$, we have (we used the homogeneity $S^a(\omega)=\lim_{N\to\infty}\mean{S^{a}_{ii}(\omega)}_{i}$ to drop the subscript $i$)
\begin{equation}
    \zeta_{\mu}(\omega) = \phi_{\mu}^{\free}(\omega) -\beta(\omega)x_{\mu}^{\free}(\omega) 
    + \frac{1}{\sqrt{N}} \lrsq{S^{\phi}(\omega) -\beta(\omega)S^x(\omega)} \sum_{\nu\in\{0,0'\}} F_{\mu\nu}(\omega)\phi_{\nu}^{\free}(\omega) + O\lrbk{\frac{1}{N}}.
\label{eq:zeta_perturbed}
\end{equation}
Intuitively, if we set 
\begin{equation}
\beta(\omega)=S^{\phi}(\omega)/S^x(\omega),
\label{eq:beta_equation1}
\end{equation}
it will cancel the second term in the above and make the effective noise $\zeta$ insensitive to the recurrent perturbation transmitted through $F_{\mu\nu}$. 

Next, we show that under this choice of $\beta(\omega)$ (\cref{eq:beta_equation1}), the magnitude of the off-diagonal entries of $\MCzetaf$ is reduced to a higher order at $O(1/N)$. 
After we plug \cref{eq:beta_equation1} into \cref{eq:zeta_perturbed}, $\zeta_{\mu}(\omega)=\phi_{\mu}^{\free}(\omega)-\beta(\omega)x_{\mu}^{\free}(\omega)+O(1/N)$, and the off-diagonal covariance
\begin{equation}
C^{\zeta}_{00'}(\omega) = C^{\phi^{\free}}_{00'}(\omega) -\beta(\omega)C^{x^{\free}\phi^{\free}}_{00'}(\omega) -\beta^{\ast}(\omega)C^{\phi^{\free}x^{\free}}_{00'}(\omega) + \left|\beta(\omega)\right|^2C^{x^{\free}}_{00'}(\omega)+O\lrbk{\frac{1}{N^2}}.
\label{eq:zeta_off_diag_eq1}
\end{equation}
There are four terms at $O\lrbk{\frac{1}{\sqrt{N}}}$. To further simplify \cref{eq:zeta_off_diag_eq1}, we need to generalize the covariances $C^{a^{\free}}_{00'}(\omega)$ of i.i.d. connectivity network, $a\in\{x,\phi\}$, \cref{eq:two-point_function_of_a}, to the case of partially symmetric networks. 
This generalization is valid because $\lrbk{\eta_{\mu}(t), \eta_{\nu}(t')}$ are jointly Gaussian for $\mu,\nu\in\{0,0'\}$ \citep{Clark2023}, so the requirement of the Price's theorem is met.
In the unperturbed network, $J_{i\mu}=J_{\mu\nu}=0$, and there is no reciprocal loop $J_{\mu i}J_{i\mu}=0$, hence $a^{\free}_{\mu}(t)=\int_{-\infty}^{t}S^{a}_{\mu\mu}(t,s)\eta_{\mu}(s)ds$ is driven by the Gaussian component $\eta_{\mu}$ only (\cref{eq:unperturbed_activity_part_symm}). 
We can write $a^{\free}_{\mu}(t)=a^{\free}_{\mu}\lrsq{\eta_{\mu}(s)}(t)$, and the covariance
\[
C^{a^{\free}b^{\free}}_{00'}(\tau)=\mean{a^{\free}_0(t)b^{\free}_{0'}(t+\tau)}_t=\mean{a^{\free}_0\lrsq{\eta_0(s)}(t)b^{\free}_{0'}\lrsq{\eta_{0'}(s')}(t+\tau)}_{t,s,s'}.
\]
 Here $\mean{\cdot}_s$ represents the integral over $s$. 
 Next, we expand $C^{a^{\free}b^{\free}}_{00'}(\tau)$ near $0$ to the leading order $O\lrbk{\frac{1}{\sqrt{N}}}$ and apply the Price's theorem (eq.~(B3) in \cite{Clark2023}),
\begin{equation}
    \begin{aligned}
        \mean{a^{\free}_0(t)b^{\free}_{0'}(t+\tau)}_t&=\mean{\frac{\partial\mean{a^{\free}_0(t)b^{\free}_{0'}(t+\tau)}_t}{\partial C^{\eta}_{00'}(s'-s)}C^{\eta}_{00'}(s'-s)}_{s,s'}+O\lrbk{\frac{1}{N}}\\
        &=\int ds\int ds'\mean{\frac{\delta a^{\free}_{0}(t)}{\delta\eta_0(s)}\frac{\delta b^{\free}_{0'}(t+\tau)}{\delta\eta_{0'}(s')}}_tC^{\eta}_{00'}(s'-s)+O\lrbk{\frac{1}{N}}\\
        &=\int ds\int ds'S^{a}(-s)S^{b}(\tau-s')C^{\eta}_{00'}(s'-s)+O\lrbk{\frac{1}{N}}.
    \end{aligned}
\end{equation}
Its Fourier transform is 
\begin{equation}
    C^{a^{\free}b^{\free}}_{00'} (\omega)=S^a(\omega)\lrbk{S^b(\omega)}^{\ast}C^{\eta}_{00'}(\omega)+O\lrbk{\frac{1}{N}}. 
\label{eq:C_afree_bfree_freq}
\end{equation}
Plugging \cref{eq:C_afree_bfree_freq} into \cref{eq:zeta_off_diag_eq1}, we have
\begin{equation}
\begin{aligned}
    C^{\zeta}_{00'}(\omega) 
    &= \lrsq{\left|S^{\phi}\right|^2-\beta S^x\lrbk{S^{\phi}}^{\ast}-\beta^{\ast}S^{\phi}\lrbk{S^x}^{\ast}+|\beta|^2\left|S^x\right|^2}C^{\eta}_{00'}+O\lrbk{\frac{1}{N}}\\
    &= O\lrbk{\frac{1}{N}}.
\end{aligned}
\label{eq:C_zeta_off_diag_derivation}
\end{equation}
Note that in the last equality, the leading order terms cancel due to the choice of $\beta(\omega)=S^{\phi}(\omega)/S^x(\omega)$.
Therefore, the off-diagonal entries of $\MCzetaf$ are $O\lrbk{\frac{1}{N}}$, which is of a higher order comparing to the firing rates and currents counterpart $C^{a}_{ij}(\omega)=O\lrbk{\frac{1}{\sqrt{N}}}$, $i\neq j$, $a\in \{x,\phi\}$.

As usual, the cavity neurons are the same as reservoir neurons when the perturbations are included in large networks. Therefore, to the leading order,
\begin{equation}
    \MCzetaf\rightarrow\Czetaf\MI, \text{ as } N\rightarrow \infty.
\label{eq:Czeta_mat_limit_reciprocal}
\end{equation} 
Here, the diagonal entries of $\MCzetaf$, according to the definition of $\zeta$, are
\begin{equation}
   \Czetaf = C^{\phi}(\omega)-\frac{S^{\phi}(\omega)}{S^{x}(\omega)}C^{x\phi}(\omega)-\lrbk{\frac{S^{\phi}(\omega)}{S^{x}(\omega)}}^{\ast}C^{\phi x}(\omega) + \left|\frac{S^{\phi}(\omega)}{S^{x}(\omega)}\right|^2C^x(\omega).
\label{eq:C_zeta_def_part_symm}
\end{equation}

Lastly, to derive the firing-rate and current covariance matrix ansatz, note that using the definition of $\zeta$ and the neuron dynamics equation (\cref{eq:network_model}), it is straightforward to show  (similar to in \cref{app:extend_zeta} but with $\sigma=0$) that the following algebraic identities always hold:
\begin{equation}
    \MCphif=\lrbk{\MI-\MJeff}^{-1}\MCzetaf\lrbk{\MI-\MJeff}^{-\dagger},
\label{eq:part_symm_cov_phi}
\end{equation}
\begin{equation}
    \MCxf=\frac{1}{1+\omega^2}\MJ\lrbk{\MI-\MJeff}^{-1}\MCzetaf\lrbk{\MI-\MJeff}^{-\dagger}\MJ^T,
\label{eq:part_symm_cov_x}
\end{equation}
where $\MJeff=\frac{\beta(\omega)}{1+i\omega}\MJ$. 
Then the generalized covariance matrix ansatz (\cref{eq:ansatz_of_Cov_phi_omega_symmetric}) follows directly from \cref{eq:Czeta_mat_limit_reciprocal,eq:C_zeta_def_part_symm}.

For the case of our original i.i.d. random connectivity model, $\kappa=0$, $S^{\phi}(\omega)=\frac{\phiprime}{1+i\omega}$ and $S^x(\omega)=\frac{1}{1+i\omega}$ \citep{Clark2023},
hence $\beta(\omega) = S^{\phi}(\omega) /S^x(\omega)= \phiprime$, and the generalized ansatz reduces to the original case (\cref{eq:ansatz_of_Cov_phi_omega}).

\subsection{Mean-field theory expression for $\Czetaf$}
\label{app:C_zeta_f_part_symm}
Here we use single-neuron mean-field theory equations to derive another expression for the scalar factor $\Czetaf$ (\cref{eq:C_zeta_def_part_symm}), which we will use in \cref{app:random_matrix_derivation_part_symm}. 
Since \cref{eq:C_zeta_def_part_symm} involves the correlations $C^{x\phi}(\omega)$ and $C^{\phi x}(\omega)$, as well as $\Cxf$, we derive these quantities from the single-neuron mean-field theory equations.
When $\sigma=0$, the single-neuron dynamics in the frequency domain are given by \citep{Clark2023,Marti2018}
\begin{equation}
    (1+i\omega)x(\omega)=\eta(\omega)+g^2\kappa S^{\phi}(\omega)\phi(\omega).
    \label{eq:single_neuron_eq_part_symm}
\end{equation}
Here, $\eta(t)$ is the Gaussian component of the recurrent input with autocorrelation function $C^{\eta}(\tau)=\mean{\eta(t)\eta(t+\tau)}=g^2C^{\phi}(\tau)$, or equivalently $C^{\eta}(\omega)=g^2\Cphif$ in the frequency domain \citep{Clark2023,Marti2018}. The second term on the RHS of \cref{eq:single_neuron_eq_part_symm} is the non-Gaussian recurrent input contribution arising from partial symmetry in $\MJ$. To evaluate the terms in \cref{eq:C_zeta_def_part_symm}, we proceed with the following steps. 

\paragraph{Relation between $S^{\phi}(\omega)$ and $S^x(\omega)$}
We can simplify \cref{eq:C_zeta_def_part_symm} by eliminating one of $S^{\phi}(\omega)$ and $S^x(\omega)$.
Multiplying both sides of \cref{eq:single_neuron_eq_part_symm} by $\eta^{\ast}$ and taking the expectation, 
then using $C^{a\eta}(\omega)=S^a(\omega)C^{\eta}(\omega)$, $a\in \{x,\phi\}$ (which follows from the Furutsu-Novikov theorem; \cite{Clark2023}, Appendix A), and canceling $C^{\eta}(\omega)$ from both sides, we obtain:
\begin{equation}
\begin{aligned}
    (1+i\omega)C^{x\eta}(\omega) &= \Cetaf + g^2\kappa S^{\phi}(\omega)C^{\phi\eta}(\omega),\\
  \Rightarrow
  (1+i\omega)S^x(\omega) &= 1+g^2\kappa\lrsq{S^{\phi}(\omega)}^2.
\end{aligned}
\label{eq:S_x_part_symm}
\end{equation}

\paragraph{Expressions for $C^{x\phi}(\omega)$ and $\Cxf$} 
To derive the expression for $C^{x\phi}(\omega)$ (and its complex conjugate $C^{\phi x}(\omega)$), we multiply both sides of \cref{eq:single_neuron_eq_part_symm} by $\phi^{\ast}$  and take the expectation. Using
$C^{\eta\phi}(\omega)=\lrbk{S^{\phi}(\omega)}^{\ast}C^{\eta}(\omega)=g^2\lrbk{S^{\phi}(\omega)}^{\ast}C^{\phi}(\omega)$ \citep{Clark2023,Marti2018}, we obtain:
\begin{equation}
\begin{aligned}
    (1+i\omega)C^{x\phi}(\omega) &= C^{\eta\phi}(\omega) + g^2\kappa S^{\phi}(\omega)\Cphif,\\
    &= g^2\lrsq{\lrbk{S^{\phi}(\omega)}^{\ast}+\kappa S^{\phi}(\omega)}\Cphif.
\end{aligned}
\label{eq:C_x_phi_part_symm}
\end{equation}

To derive the expression for $\Cxf$, we multiply each side of \cref{eq:single_neuron_eq_part_symm} by its complex conjugate and take the expectation. Applying the Furutsu-Novikov theorem to $C^{\phi\eta}(\omega)$ and $C^{\eta\phi}(\omega)$, and using $\Cetaf=g^2\Cphif$, this gives:
\begin{equation}
\begin{aligned}
    \Cxf &= \frac{g^2}{1+\omega^2}\left\{1+g^2\kappa\lrsq{\lrbk{S^{\phi}(\omega)}^2+\lrbk{\lrbk{S^{\phi}(\omega)}^{\ast}}^2} + g^2\kappa^2\left|S^{\phi}(\omega)\right|^2\right\}\Cphif.
\end{aligned}
\label{eq:C_x_part_symm}
\end{equation}

\paragraph{Expression for $\Czetaf$} We now have all the quantities needed in \cref{eq:C_zeta_def_part_symm} to compute  $\Czetaf$.
Plugging \cref{eq:S_x_part_symm,eq:C_x_phi_part_symm,eq:C_x_part_symm} into \cref{eq:C_zeta_def_part_symm} and after some straightforward algebra, we obtain:
\begin{equation}
    C^{\zeta}(\omega) 
    = \frac{1-g^2\lrabs{S^{\phi}(\omega)}^2}{\left|1+g^2\kappa \lrbk{S^{\phi}(\omega)}^2\right|^2}\Cphif.
\label{eq:C_chi_part_symm}
\end{equation}
\cref{eq:C_chi_part_symm} gives the desired scalar factor $\Czetaf$ in terms of $\Cphif$ and the linear response function $S^{\phi}(\omega)$. Note that $\lrbk{S^{\phi}}^2$ in the denominator is an ordinary square rather than a modulus square, since it comes directly from \cref{eq:S_x_part_symm}. 
Note that since $\Czetaf$ and $\Cphif$ are Fourier Transforms of autocorrelation functions, they are all positive which implies $g^2\lrabs{S^{\phi}}^2<1$. 
Using the $S^x(\omega)$ $S^{\phi}(\omega)$ relation in \cref{eq:S_x_part_symm}, \cref{eq:C_chi_part_symm} can be alternatively written as,
\begin{equation}
    C^{\zeta}(\omega)
   = \frac{1-g^2\lrabs{S^{\phi}(\omega)}^2}{(1+\omega^2)\lrabs{S^x(\omega)}^2}C^{\phi}(\omega)
   = \lrbk{\frac{1}{(1+\omega^2)\lrabs{S^x(\omega)}^2}-\frac{g^2\lrabs{\beta(\omega)}^2}{1+\omega^2}}C^{\phi}(\omega).
\label{eq:coeff_temp}
\end{equation}

\subsection{large-network estimation of theoretical $\beta$, $g_{c,\mathrm{eff}}=g(1+\kappa)\beta$ and $\Czetaomega$}
\label{app:estimate_beta_and_C_zeta}
Here we describe methods to compute the theoretical $\beta$, $g_{c,\text{eff}}$ and $\Czetaf$ in partially symmetric networks by combining the mean-field theory and large-network simulations, because a complete mean-field theory to determine those quantities in partially symmetric networks is rather involved. 
We first estimate the zero-frequency linear response function $S^{\phi}$ and $S^{x}$, then determine $\beta=S^{\phi}/S^x$ and $g_{c,\text{eff}}=g(1+\kappa)\beta$. Lastly, we compute the zero-frequency version of $C^{\zeta}$ using the estimated $\beta$ and then reconstruct and the theoretical covariance matrix.

Denote $C^{ab,\omega}_{sim}=\frac{1}{N}\sum_{i=1}^{N}C^{ab,\omega}_{sim,ii}$ to be the zero-frequency correlation function between $a,b\in\{x,\phi\}$ (note that $C^{ab,\omega}$ and $S^{a,\omega}$ represents the $\omega=0$ version, defined in our notation manner) computed from simulating a large network (e.g., $N=10000$). We use $\hat{C}^{\zeta,\omega}$, $\hat{S}^{a}$ and $\hat{\beta}$ to represent the value estimated from large-network simulation.
For $\sigma\geq0$, the mean-field equation \cref{eq:single_neuron_eq_part_symm} is modified as
\begin{equation}
    (1+i\omega)x(\omega)=\eta(\omega)+g^2\kappa S^{\phi}(\omega)\phi(\omega)+\xi(\omega)=\tilde{\eta}(\omega)+g^2\kappa S^{\phi}(\omega)\phi(\omega),
\label{eq:single_neuron_eq_part_symm_general}
\end{equation}
where $\tilde{\eta}=\eta+\xi$ is the overall Gaussian input, note that $\eta$ and $\xi$ are uncorrelated\citep{Marti2018} and thus $C^{\tilde{\eta}}(\omega)=C^{\eta}(\omega)+\sigma^2$.

\paragraph{Estimating $S^{\phi}$} Next, we multiply $\phi^{\ast}$ on both sides of \cref{eq:single_neuron_eq_part_symm_general} and take the expectation, using Furutsu-Novikov theorem $C^{\tilde{\eta}\phi}=\lrbk{S^{\phi}}^{\ast}C^{\tilde{\eta}}$ and the mean-field equation $C^{\eta}(\omega)=g^2C^{\phi}(\omega)$ \citep{Marti2018,Clark2023}, we have
\begin{equation}
    \begin{aligned}
        (1+i\omega)C^{x\phi}(\omega)&=C^{\tilde{\eta}\phi}(\omega)+g^2\kappa S^{\phi}(\omega)C^{\phi}(\omega)\\
        &= \lrbk{S^{\phi}}^{\ast}(\omega)C^{\tilde{\eta}}(\omega)+g^2\kappa S^{\phi}(\omega)C^{\phi}(\omega)\\
        &= \lrbk{S^{\phi}(\omega)}^{\ast}\lrsq{g^2\Cphif(\omega)+\sigma^2}+g^2\kappa S^{\phi}(\omega)C^{\phi}(\omega)\\
        &= \lrsq{\lrbk{S^{\phi}(\omega)}^{\ast}+\kappa S^{\phi}(\omega)}g^2\Cphif+\lrbk{S^{\phi}(\omega)}^{\ast}\sigma^2
    \end{aligned}
\label{eq:C_x_phi_general_part_symm}
\end{equation}

By setting $\omega=0$ in \cref{eq:C_x_phi_general_part_symm} and approximating those correlation functions by with one estimated from the large-network simulation, we have the following equation to determine $\hat{S}^{\phi,\omega}$
\begin{equation}
    \hat{S}^{\phi,\omega} = \frac{C^{x\phi,\omega}_{sim}}{g^2(1+\kappa) C^{\phi,\omega}_{sim}+\sigma^2}.
\label{eq:S_phi_estimator}
\end{equation}

\paragraph{Estimating $S^{x}$ and $\beta$} \cref{eq:S_x_part_symm} is still valid when $\sigma>0$ because $C^{\eta\xi}(\omega)=0$. Evaluate \cref{eq:S_x_part_symm} at $\omega=0$ gives
\begin{equation}
    \hat{S}^{x,\omega} = 1 + g^2\kappa\lrbk{\hat{S}^{\phi,\omega}}^2.
\label{eq:S_x_estimation}
\end{equation}
The theoretical zero-frequency $\beta$ is then estimated according to \cref{eq:beta_equation},
\begin{equation}
\hat{\beta} = \hat{S}^{\phi,\omega} / \hat{S}^{x,\omega}.
\label{eq:beta_estimation}
\end{equation}
We use the above equations compute the $g_{c,\text{eff}}=g(1+\kappa)\hat{\beta}$ curves in \cref{fig:g_eff_curve_for_diff_kappa}A-E.

\paragraph{Estimating $C^{\zeta}$} For the scalar factor, $\hat{C}^{\zeta,\omega}$, in the generalized covariance matrix ansatz \cref{eq:ansatz_of_Cov_phi_omega_symmetric}, it can be computed following the definition of $\zeta=\phi-\beta x$,
\begin{equation}
    \hat{C}^{\zeta,\omega} = C^{\phi,\omega}_{sim} - \hat{\beta} (C^{x\phi,\omega}_{sim} + C^{\phi x,\omega}_{sim}) + \hat{\beta}^2C^{x,\omega}_{sim}.
\label{eq:C_zeta_estimation}
\end{equation}

The theoretical covariance matrix (when $\sigma=0$), is then computed as 
\begin{equation}
\begin{aligned}
    \hat{\MC}^{\phi,\omega} &= \hat{C}^{\zeta,\omega}\lrsq{\MI-\hat{\beta}\MJ}^{-1} \lrsq{\MI-\hat{\beta}\MJ}^{-T},\\
    \hat{\MC}^{x,\omega} &= \hat{C}^{\zeta,\omega}\MJ\lrsq{\MI-\hat{\beta}\MJ}^{-1} \lrsq{\MI-\hat{\beta}\MJ}^{-T}\MJ^T,
\end{aligned}
\label{eq:Cov_matrix_estimation}
\end{equation}

\subsection{Random matrix results}
\label{app:random_matrix_derivation_part_symm}
Let $\MP(\gamma)=\lrbk{\gamma\MI-\MJ_0}^{\dagger}\lrbk{\gamma\MI-\MJ_0}$ with $\MJ_0=\MJ/g$ and 
\begin{equation}
    \MC(\gamma)=\MP^{-1}(\gamma)=\lrabs{\gamma}^{-2}\lrbk{\MI-\gamma^{-1}\MJ_0}^{-1}\lrbk{\MI-\gamma^{-1}\MJ_0}^{-\dagger},
\label{eq:MC_part_symm}
\end{equation}
where $\gamma=a+ib \in \mathbb{c}$ lies outside the limiting spectral support of $\MJ_0$, the eigenvalue distribution of $\MP$, denoted as $p_{\MP}(\lambda;\gamma)$, while its Stieltjes transform $\Delta_{\MP}(z;\gamma)$ (here $z$ is a local transform variable only in the section about partially symmetric connection) follows:
\begin{equation}
    z=-\frac{1}{\Delta_{\MP}(\Delta_{\MP}+1)}+\frac{a^2}{\lrsq{\Delta_{\MP}(1+\kappa)+1}^2}+\frac{b^2}{\lrsq{\Delta_{\MP}(1-\kappa)+1}^2}.
    \label{eq:Stieltjes_MP_equation}
\end{equation}
In later derivation, $\gamma$ will be specialized as $\gamma=\frac{1+i\omega}{g\beta}$ corresponding to the covariance matrix ansatz \cref{eq:ansatz_of_Cov_phi_omega_symmetric}. Using the following two equations, we can rewrite \cref{eq:Stieltjes_MP_equation} of $\MP(\gamma)$ to $\MC(\gamma)$:
\begin{equation}
    -\frac{1}{z^2}\Delta_{\MC}\lrbk{\frac{1}{z};\gamma}-\frac{1}{z}=\Delta_{\MP}(z;\gamma),
    \label{eq:Stiel_inver}
\end{equation}
\begin{equation}
    M_{\MC}(z;\gamma)=\sum_{n=1}^{\infty}\varphi_n\lrsq{\MC(\gamma)}z^n=-\Delta_{\MC}\lrbk{\frac{1}{z};\gamma}\frac{1}{z}-1.
    \label{eq:moment_generation_function}
\end{equation}
Here $\varphi_n\lrsq{\MC(\gamma)}$ is the n-th moment of $\MC(\gamma)$ and $M_{\MC}$ is the moment generating function. Finally, the equation of $M_{\MC}$ follows:
\begin{equation}
    \frac{1}{z}=-\frac{1}{M_{\MC}(M_{\MC}+z)}+\frac{a^2}{\lrsq{M_{\MC}(1+\kappa)+z}^2}+\frac{b^2}{\lrsq{M_{\MC}(1-\kappa)+z}^2}.
\label{eq:equation_of_MC}
\end{equation}
On both sides, we can multiply $zM_{\MC}(M_{\MC}+z)\lrsq{M_{\MC}(1+\kappa)+z}^2\lrsq{M_{\MC}(1-\kappa)+z}^2$, to get
\begin{equation}
\begin{aligned}
    &M_{\MC}(M_{\MC}+z)\lrsq{M_{\MC}(1+\kappa)+z}^2\lrsq{M_{\MC}(1-\kappa)+z}^2\\
    =& z\left\{-\lrsq{M_{\MC}(1+\kappa)+z}^2\lrsq{M_{\MC}(1-\kappa)+z}^2+M_{\MC}(M_{\MC}+z)\lrsq{a^2\lrbk{M_{\MC}(1-\kappa)+z}^2+b^2\lrbk{M_{\MC}(1+\kappa)+z}^2}\right\}
\end{aligned}
\label{eq:eq_of_MC_full}
\end{equation}
and replace $M_{\MC}$ by its expansion \cref{eq:moment_generation_function}.

\paragraph{Derivation of $\varphi_1$} 
For the rest of this section, we use the shorthand $\varphi_n=\varphi_n\lrsq{\MC(\gamma)}$ for $n=1,2$. 
One can solve for $\varphi_1$ by comparing the coefficients of $z^n$ in \cref{eq:eq_of_MC_full}. 
Note that the lowest order of $z$ in the left-hand side (LHS) is $z^6$, while the apparent lowest order in the  right-hand side (RHS) is $z^5$, so its coefficient of $z^5$ must be 0:
\begin{equation}
    -\lrsq{\varphi_1(1+\kappa)+1}^2\lrsq{\varphi_1\lrbk{1-\kappa}+1}^2+\varphi_1\lrbk{\varphi_1+1}\lrsq{a^2\lrbk{\varphi_1(1-\kappa)+1}^2+b^2\lrbk{\varphi_1(1+\kappa)+1}^2}=0.
\label{eq:varphi_1_factorization}
\end{equation}
This is a quartic equation in $\varphi_1$, but turns out can be factorized as we simplify and combine terms.
First, let $gS^{\phi}(\omega) =: r= u+iv\in\mathbb{C}$ and $\theta=|r|^2=u^2+v^2$, 
\begin{equation}
    \begin{aligned}
        a&=\mathrm{Re} \lrsq{\frac{1+g^2\kappa\lrbk{S^{\phi}}^2}{gS^{\phi}}}
        =\frac{u\lrbk{1+\kappa\theta}}{\theta},\\
        b&=\mathrm{Im}\lrsq{\frac{1+g^2\kappa\lrbk{S^{\phi}}^2}{gS^{\phi}}}
        =\frac{-v\lrbk{1-\kappa\theta}}{\theta}.
    \end{aligned}
\label{eq:a_b_expression}
\end{equation}
Substituting \cref{eq:a_b_expression} into \cref{eq:varphi_1_factorization}, we obtain the following expression for the LHS of \cref{eq:varphi_1_factorization}
\begin{equation}
\scalebox{1.2}{$\frac{-\theta^2\lrsq{\varphi_1(1+\kappa)+1}^2\lrsq{\varphi_1(1-\kappa)+1}^2+\varphi_1\lrbk{\varphi_1+1}\lrsq{u^2\lrbk{\varphi_1(1-\kappa)+1}^2(1+\kappa\theta)^2 +v^2\lrbk{\varphi_1(1+\kappa)+1}^2(1-\kappa\theta)^2}}{\theta^2}.$}
\label{eq:varphi_1_step1}
\end{equation}
Define $m_1=u^2\lrbk{\varphi_1(1-\kappa)+1}^2$ and $m_2=v^2\lrbk{\varphi_1(1+\kappa)+1}^2$. Then, \cref{eq:varphi_1_step1} can be written as
\begin{equation}
\begin{split}
&\frac{-\theta^2\lrsq{\varphi_1(1+\kappa)+1}^2\lrsq{\varphi_1(1-\kappa)+1}^2+\varphi_1\lrbk{\varphi_1+1}\lrsq{m_1\lrbk{1+\kappa^2\theta^2+2\kappa\theta} +m_2\lrbk{1+\kappa^2\theta^2-2\kappa\theta}}}{\theta^2}\\
&=
\frac{-\theta\lrsq{\varphi_1(1+\kappa)+1}^2\lrsq{\varphi_1(1-\kappa)+1}^2+\varphi_1\lrbk{\varphi_1+1}2\kappa (m_1-m_2)}{\theta}+\frac{\varphi_1(\varphi_1+1)(1+\kappa^2\theta^2)(m_1+m_2)}{\theta^2}.
\end{split}
\label{eq:varphi_1_step3}
\end{equation}
For the first fraction in the above, substitute $\theta=u^2+v^2$, and then replace $u^2\lrbk{\varphi_1(1-\kappa)+1}^2$ and $v^2\lrbk{\varphi_1(1+\kappa)+1}^2$ by $m_1$ and $m_2$,
\begin{equation}
\begin{aligned}
    &\frac{-(u^2+v^2)\lrsq{\varphi_1(1+\kappa)+1}^2\lrsq{\varphi_1(1-\kappa)+1}^2+ \varphi_1\lrbk{\varphi_1+1}2\kappa(m_1-m_2)}{\theta}\\
    &=\frac{-m_1\lrbk{\varphi_1(1+\kappa)+1}^2 - m_2\lrbk{\varphi_1(1-\kappa)+1}^2+ \varphi_1\lrbk{\varphi_1+1}2\kappa(m_1-m_2)}{\theta}  \\
    &=\frac{(m_1+m_2)\lrsq{-(1+\kappa^2)\varphi_1^2-2\varphi_1-1}}{\theta}.
\end{aligned}
\label{eq:varphi_1_step4}
\end{equation}
Plugging this into \cref{eq:varphi_1_step3} and note the common factor of $m_1+m_2$, we can rewrite \cref{eq:varphi_1_factorization} as
\begin{equation}
\frac{(m_1+m_2)\lrsq{\varphi_1\lrbk{\varphi_1+1}\lrbk{1+\kappa^2\theta^2}-\theta\lrbk{(1+\kappa^2)\varphi_1^2+2\varphi_1+1}}}{\theta^2}=0.
\label{eq:varphi_1_step5}
\end{equation}

Note that $\theta=|r|^2 = \lrabs{gS^{\phi}(\omega)}^2>0$,
and if
\begin{equation}
    m_1+m_2 = u^2\lrsq{\varphi_1(1-\kappa)+1}^2+v^2\lrsq{\varphi_1(1+\kappa)+1}^2= 0,
\label{eq:m1_plus_m2}
\end{equation}
it implies $u=v=0$, and hence $r=0$, which contradicts with $ \lrabs{gS^{\phi}(\omega)}>0$.
Therefore, nontrivial solutions of $\varphi_1$ according to \cref{eq:varphi_1_step5} come from   
\begin{equation}
    \varphi_1\lrbk{\varphi_1+1}\lrbk{1+\kappa^2\theta^2}-\theta\lrsq{(1+\kappa^2)\varphi_1^2+2\varphi_1+1}=0
\label{eq:phi_1_2_factor_equals_0}
\end{equation}
This is a quadratic equation in $\varphi_1$, and can be further factorized as
\begin{equation}
\lrsq{(-1+\theta)\varphi_1+\theta}\lrsq{-1+(-1+\kappa^2\theta)\varphi_1}=\lrsq{(-1+|r|^2)\varphi_1+|r|^2}\lrsq{-1+(-1+\kappa^2|r|^2)\varphi_1}.
\label{eq:varphi_1_two_factors}
\end{equation}
This leads to two possible solutions of $\varphi_1$.
A solution corresponding to the covariance matrix ansatz should satisfy $\varphi_1>0$. Moreover, from the mean-field theory of $\Czetaf$ (\cref{eq:C_chi_part_symm}),  $|r| = \lrabs{gS^{\phi}(\omega)}< 1$.
These show that the only proper solution of $\varphi_1$ corresponds to the first factor of \cref{eq:varphi_1_two_factors},
\begin{equation}
    \varphi_1=\frac{|r|^2}{1-|r|^2}=\frac{g^2\lrabs{S^{\phi}}^2}{1-g^2\lrabs{S^{\phi}}^2}.
\label{eq:first_moment_root2_part_symm}
\end{equation}

\paragraph{Derivation of $\varphi_2$} Similar to the derivation of $\varphi_1$, the second moment $\varphi_2$ can be derived from matching the $z^6$ coefficient in \cref{eq:eq_of_MC_full}. 
To extract the $z^6$ terms on RHS, we can make the replacements $M_{\MC}\rightarrow\varphi_1z$ and $M_{\MC}^n\rightarrow n\varphi_1^{n-1}\varphi_2z^{n+1}$ for $n\geq2$ on RHS of \cref{eq:eq_of_MC_full}, the latter replacement can equivalently be written as $\varphi_2z^{n+1}\frac{d\varphi_1^n}{d\varphi_1}$.
This gives a convenient way of computing the $z^6$ term on RHS using the derivative.
Now comparing the $z^6$ terms from both sides and dividing by $z^6$,
\begin{equation}
\begin{aligned}
    &\varphi_1(\varphi_1+1) \lrsq{\varphi_1(1+\kappa)+1}^2\lrsq{\varphi_1(1-\kappa)+1}^2\\
    =& \varphi_2\frac{d}{d\varphi_1}\left\{-\lrsq{\varphi_1(1+\kappa)+1}^2\lrsq{\varphi_1(1-\kappa)+1}^2+\varphi_1(\varphi_1+1)\lrsq{a^2\lrbk{\varphi_1(1-\kappa)+1}^2+b^2\lrbk{\varphi_1(1+\kappa)+1}^2}\right\}.
\end{aligned}
\label{eq:eq_of_varphi2_full}
\end{equation}
The two sides of \cref{eq:eq_of_varphi2_full} can be evaluated by plugging in the $\varphi_1$ and $a,b$ expressions \cref{eq:first_moment_root2_part_symm,eq:a_b_expression}:
\begin{equation}
    \begin{aligned}
        \varphi_1&=\frac{g^2\lrabs{S^{\phi}}^2}{1-g^2\lrabs{S^{\phi}}^2}=\frac{|r|^2}{1-|r|^2},\quad
        \varphi_1+1=\frac{|r|^2}{1-|r|^2}+1=\frac{1}{1-|r|^2}, \quad
        \varphi_1(1\pm\kappa)+1 = \frac{1\pm\kappa|r|^2}{1-|r|^2},\\
        \text{LHS}
        &=\frac{|r|^2\lrbk{1-\kappa^2|r|^4}^2}{\lrbk{1-|r|^2}^6},\\
        \frac{\text{RHS}}{\varphi_2}
        &= \frac{\lrbk{1-\kappa^2|r|^4}\lrbk{1-2\kappa(u^2-v^2)+\kappa^2\lrbk{u^2+v^2}^2}}{|r|^2\lrbk{1-|r|^2}^2}
         = \frac{\lrbk{1-\kappa^2|r|^4}\lrabs{1-\kappa r^2}^2}{|r|^2\lrbk{1-|r|^2}^2}.\\
    \end{aligned}
\label{eq:second_moment_equations_all}
\end{equation}
Lastly, to compute $\varphi_2$, we plug in the particular value of $r=gS^{\phi}(\omega)$ (when $\gamma=\frac{1+i\omega}{g\beta}$) in the above,
\begin{equation}
\begin{aligned}
    \left. \varphi_2\lrsq{\MC(\gamma)} \right|_{\gamma=\frac{1+i\omega}{g\beta}} 
    = \frac{|r|^4\lrbk{1-\kappa^2|r|^4}}{\lrbk{1-|r|^2}^4\lrabs{1-\kappa r^2}^2} 
    = \frac{g^4\lrabs{S^{\phi}}^4\lrbk{1-g^4\kappa^2\lrabs{S^{\phi}}^4}}{\lrbk{1-g^2\lrabs{S^{\phi}}^2}^4\lrabs{1-g^2\kappa\lrbk{S^{\phi}}^2}^2}.
\end{aligned}
\label{eq:second_moment_MP}
\end{equation}

\subsection{Cross-checking the first two moments and dimension with the cavity method}
\label{app:first_second_moment_part_symm}
\cref{app:random_matrix_derivation_part_symm} computed the first two moments of $\MC(\gamma)$ evaluated at $\gamma=\frac{1+i\omega}{g\beta(\omega)}$ using random matrix theory. We now apply these formulas to the generalized covariance matrix ansatz (\cref{eq:ansatz_of_Cov_phi_omega_symmetric}) to obtain the first two moments of covariance eigenvalues and hence the dimension, and then compare them with the previous results derived using the two-site cavity method \citep{Clark2023}.

\paragraph{First moment} The ansatz of $\MCphif$ in \cref{eq:ansatz_of_Cov_phi_omega_symmetric} can be expressed using $\MC(\gamma)$ as:
\begin{equation}
    \MCphif=C^{\zeta}(\omega)\lrabs{\gamma}^2\MC(\gamma).
\label{eq:Cov_phi_with_general_expression}
\end{equation}
To compute its first moment, plugging in $\Czetaf$ from \cref{eq:C_chi_part_symm}, $\gamma=\frac{(1+i\omega)S^x(\omega)}{gS^{\phi}(\omega)}$, and $\varphi_1\lrsq{\MC(\gamma)}$ from \cref{eq:first_moment_root2_part_symm}:
\begin{equation}
\begin{aligned}
    &\varphi_1\lrsq{\MCphif}=C^{\zeta}(\omega)\left.\lrabs{\gamma}^2\varphi_1\lrsq{\MC(\gamma)}\right|_{\gamma=\frac{1+i\omega}{g\beta(\omega)}}\\
    =& \frac{1-g^2\lrabs{S^{\phi}}^2}{(1+\omega^2)\lrabs{S^x}^2}\Cphif\frac{(1+\omega^2)\lrabs{S^x}^2}{g^2\lrabs{S^{\phi}}^2}\frac{g^2\lrabs{S^{\phi}}^2}{1-g^2\lrabs{S^{\phi}}^2}=\Cphif.
\end{aligned}
\label{eq:ansatz_first_moment_part_symm}
\end{equation}

This reproduces exactly the mean-field result for the diagonal entries of $\MCphif$, that is, $\varphi\lrsq{\MCphif}=\Cphif$.

\paragraph{Second moment}
The second moment of $\MCphif$ using \cref{eq:Cov_phi_with_general_expression} is
\begin{equation}
\begin{aligned}
&\varphi_2\lrsq{\MCphif}
=\lrsq{\Czetaf}^2\left.\lrabs{\gamma}^4\varphi_2\lrsq{\MC(\gamma)}\right|_{\gamma=\frac{1+i\omega}{g\beta(\omega)}}\\
    &=\frac{\lrbk{1-g^2\lrabs{S^{\phi}(\omega)}^2}^2}{\lrabs{1+g^2\kappa \lrbk{S^{\phi}(\omega)}^2}^4}\lrsq{\Cphif}^2\frac{\lrabs{1+g^2\kappa\lrbk{S^{\phi}(\omega)}^2}^4}{g^4\lrabs{S^{\phi}(\omega)}^4}\frac{g^4\lrabs{S^{\phi}(\omega)}^4\lrbk{1-g^4\kappa^2\lrabs{S^{\phi}(\omega)}^4}}{\lrbk{1-g^2\lrabs{S^{\phi}(\omega)}^2}^4\lrabs{1-g^2\kappa\lrbk{S^{\phi}(\omega)}^2}^2}\\
    &=\frac{1-g^4\kappa^2\lrabs{S^{\phi}(\omega)}^4}{\lrbk{1-g^2\lrabs{S^{\phi}(\omega)}^2}^2\lrabs{1-g^2\kappa\lrbk{S^{\phi}(\omega)}^2}^2}\lrsq{\Cphif}^2.
\end{aligned}
\label{eq:ansatz_second_moment_part_symm}
\end{equation}
In \cite{Clark2023}, the authors computed $\varphi_2\lrsq{\MCphif}$ using the two-site cavity method.
To compare the two results, we restate below the results in \cite{Clark2023} using our notations.
The second moment contains terms from both diagonal and off-diagonal entries. 
The off-diagonal terms are described by the four-point function $\psi^{\phi}(\omega_1,\omega_2):=N\mean{C^{\phi}_{ij}(\omega_1)C^{\phi}_{ij}(\omega_2)}$, which describes the products of off-diagonal entries at two frequencies $\vec{\omega}=(\omega_1,\omega_2)$. 
For partially symmetric networks, the four-point function $\psi^{\phi}$ from \cite{Clark2023} is
\begin{equation}
    \begin{aligned}
        \psi^{\phi}(\vec{\omega}) &= \lrbk{M(\vec{\omega})-1}C^{\phi}(\omega_1)C^{\phi}(\omega_2),\\
        \text{where } M(\vec{\omega}) &= \frac{1}{\lrabs{1-g^2S^{\phi}(\omega_1)S^{\phi}(\omega_2)}^2}\frac{1-g^4\kappa^2\lrabs{\lrbk{S^{\phi}(\omega_1)}^{\ast}S^{\phi}(\omega_2)}^2}{\lrabs{1-g^2\kappa\lrbk{S^{\phi}(\omega_1)}^{\ast}S^{\phi}(\omega_2)}^2}.
    \end{aligned}
\end{equation}
The second moment is computed by setting $(\omega_1,\omega_2)=(-\omega,\omega)$,
\begin{equation*}
\begin{aligned}
    \varphi_2\lrsq{\MCphif}
    &=\psi^{\phi}(-\omega,\omega)+C^{\phi}(-\omega)C^{\phi}(\omega) 
    =M(-\omega,\omega)\lrsq{\Cphif}^2\\
    &=\frac{1}{\lrabs{1-g^2S^{\phi}(-\omega)S^{\phi}(\omega)}^2}\frac{1-g^4\kappa^2\lrabs{\lrbk{S^{\phi}(-\omega)}^{\ast}S^{\phi}(\omega)}^2}{\lrabs{1-g^2\kappa\lrbk{S^{\phi}(-\omega)}^{\ast}S^{\phi}(\omega)}^2}\lrsq{\Cphif}^2\\
    &=\frac{1}{\lrbk{1-g^2\lrabs{S^{\phi}(\omega)}^2}^2}\frac{1-g^4\kappa^2\lrabs{S^{\phi}(\omega)S^{\phi}(\omega)}^2}{\lrabs{1-g^2\kappa S^{\phi}(\omega)S^{\phi}(\omega)}^2}\lrsq{\Cphif}^2\\
    &=\frac{1-g^4\kappa^2\lrabs{S^{\phi}(\omega)}^4}{\lrbk{1-g^2\lrabs{S^{\phi}(\omega)}^2}^2\lrabs{1-g^2\kappa \lrsq{S^{\phi}(\omega)}^2}^2}\lrsq{\Cphif}^2.
\end{aligned}
\end{equation*}
This is exactly the same as our random matrix theory result \cref{eq:ansatz_second_moment_part_symm}.

\paragraph{Dimension}
As an immediate corollary of the first two moments, the frequency-dependent dimension of $\MCphif$
for partially symmetric networks is given by the following equation and agrees with the two-site cavity method result \citep{Clark2023},
\begin{equation}
    \Dphif=\frac{\varphi_1^2\lrsq{\MCphif}}{\varphi_2\lrsq{\MCphif}}=\frac{\lrbk{1-g^2\lrabs{S^{\phi}(\omega)}^2}^2\lrabs{1-g^2\kappa\lrsq{S^{\phi}(\omega)}^2}^2}{1-g^4\kappa^2\lrabs{S^{\phi}(\omega)}^4}.
    \label{eq:D_phi_Clark}
\end{equation}

\section{Zero-timelag dimension and spectrum}
\label{sec:time-lag_dimension}
The zero-timelag dimensions $\Datau$, $a\in\{x,\phi\}$ can be computed numerically from the frequency-dependent four point functions derived in \cref{app:cavity_unit_four_point_function} similarly as the noiseless case $\sigma=0$ in \cite{Clark2023}. Here we describe the key steps for completeness.

By definition using the matrix trace, $\Datau$ can be written by separating into diagonal and off-diagonal terms,
\begin{equation}
\Datau=\frac{\left(C^{a,\tau}\right)^2}{\left(C^{a,\tau}\right)^2+\psi^{a,\tau}}.
\label{eq:zero_time_dim}    
\end{equation}
For the diagonal terms, $\Cxtau=\mean{x(t)x(t)}=:c_0$,
$\Cphitau=\mean{\phi(x(t))\phi(x(t))}=\int_{-\infty}^{\infty} \frac{1}{\sqrt{2\pi}}\lrsq{\phi\left(\sqrt{c_0}z\right)}^2e^{-\frac{z^2}{2}}dz$ (recall that $x(t)$ is a Gaussian process as $N\to\infty$), and $c_0$, $\Cphit$, $\Cxt$ can be determined by numerically solving the single-neuron mean-field theory \cref{eq:single_neuron_MFT} as in \cite{Schuecker2018}.

The off-diagonal term in \cref{eq:zero_time_dim} is given by the time-lag four point function
\[
\psi^{a,\tau} =  N\left.\mean{ C^{a}_{ij}(\tau=0)C^{a}_{ij}(\tau=0)} _{\MJ}\right|_{i\neq j}.
\]
It can be obtained by (numerically) taking the 2D inverse Fourier Transform of the frequency-dependent four point function $\psi^{a}(\omega_1,\omega_2)$, which has been derived for the general case of $\sigma\ge 0$ in \cref{app:cavity_unit_four_point_function}, as \cref{eq:final_derivation_psi_phi_omega,eq:final_derivation_psi_x_omega}.

The asymptotic values of $\Datau$ as $g\to\infty$ can be derived based on an analysis of the asymptotic version of the single-neuron mean-field theory \cref{eq:single_neuron_MFT} discussed in \cite{Crisanti2018} and computed in similarly as in \cite{Clark2023}. 
In particular, $\Cxt = O(g^2)$ and $\Cphit=O(1)$, so we can drop the $\sigma^2$ term in \cref{eq:single_neuron_MFT}, when $g\rightarrow \infty$ while fixing $\sigma$. 
This means that we can solve for the diagonal term quantities,  $c_0$, $\Cphit$, and $\Cxt$ in \cref{eq:zero_time_dim}, from the asymptotic version of \cref{eq:single_neuron_MFT} in \cite{Crisanti2018}, their results will be exactly the same as the case when $\sigma=0$. 
For the off-diagonal terms, we can use the asymptotic version of \cref{eq:single_neuron_MFT}, $g^2\Cphif=(1+\omega^2)\Cxf$ (same as setting $\sigma=0$), to combine the two terms in \cref{eq:final_derivation_psi_x_omega}, which gives
\[
\psixtf=\lrsq{\frac{2\left|X(\vec{\omega})\right|^2-g^4\phiprime^4}{\left|X(\vec{\omega})-g^2\phiprime^2 \right|^2}-1}C^{x}(\omega_1)C^{x}(\omega_2).
\]
Observe that the above four-point function for currents and the four-point function for rates \cref{eq:final_derivation_psi_phi_omega} only depend on $\geff=g\phiprime$, $\Cphif$, and $\Cxf$, where their asymptotic values do not dependent on $\sigma$.
Therefore, the asymptotic dimensions are found to be 
$\lim_{g\to\infty}\Dxtau\approx0.0602$, $\lim_{g\to\infty}\Dphitau\approx0.126$, the same as those in \citep{Clark2023} for any fixed $\sigma\ge 0$.

The zero-timelag covariance spectra $p^{a,\tau}(\lambda)$, $a\in\{x,\phi\}$ is computed from the frequency-dependent covariance matrix theory using a Monte Carlo method.
First, we compute $\MCphitau$, $\MCxtau$ using an entry-wise inverse Fourier transform of the frequency-dependent covariance matrices, \cref{eq:ansatz_of_Cov_phi_omega,eq:Cov_x_omega_full}, for a large $N$ and a given realization of the random $\MJ$. 
More specifically, $C^{a,\tau}_{ij} = C^{a}_{ij} (\tau=0) =\frac{1}{2\pi}\int_{-\infty}^{\infty}C^{a}_{ij}(\omega)d\omega$, and we computed it simply by a summation $\MCatau=\frac{\Delta\omega}{2\pi}\sum_{k=-K}^K\M{C}^{a}(k \Delta \omega )$.
After getting $\MCphitau$, $\MCxtau$, we compute their eigenvalues to get the zero-timelag spectrum $p^{a,\tau}(\lambda)$. To improve the accuracy of the spectrum, the process is repeated for multiple $\MJ$ realizations and the spectra are averaged.
One numerical consideration to note is that the network size $N$ needs to be set larger when $\geff$ is closer to 1, especially to resolve the large right edge of the spectrum support in such cases.

\section{Time-sampled theoretical spectrum}
\label{sec:finite-sample_theory}
As noted in \cite{Hu2022}, when the number of time samples $N_T$ or recording length of neural activity is not substantially larger than the number of neurons $N$ in the network, the effect due to finite time samples on the spectrum and dimension cannot be neglected. To address this, we can extend the theory for the covariance spectrum to the time-sampled case following the method in \cite{Hu2022}.

The method in \citep{Hu2022} assumes that the neural activity $\vec{x}(t)$ follow a multivariate Gaussian distribution,
the sampled covariance matrix $\hat{\MC}^{x}$ has the same eigenvalues as $\MC^x \M{Z}\M{Z}^T$, where $\MC^x$ is the exact covariance matrix and $Z_{ij}$ is drawn i.i.d. from $\mathcal{N}(0,1/N_T)$.
For the firing rates $\vec{\phi}$, since its covariance (\cref{eq:ansatz_of_Cov_phi_omega}) has the same form as case with linear dynamics, the analytical time-sampled spectrum result in \cite{Hu2022} can be directly used to obtain the time-sampled $\Pphif$ by replacing the $g$ with $\geff(\omega)=\frac{g\phiprime}{\sqrt{1+\omega^2}}$ (\cref{eq:geff_w}). 
This time-sampled spectrum $\Pphif$ was used in \cref{fig:Eigen_spectrum_total}C-F and \cref{fig:special_spectrum_autocorrelation}A, and for fitting $\geff$ in \cref{fig:g_eff_phase_diagram}B. 
Note that although the multivariate Gaussian distribution assumption may not hold for nonlinear dynamics, numerically, this correction remains accurate, consistent with universality expectations in random matrix theory.

For dimension, the time-sampled theory in \cite{Hu2022} is general and applies to any covariance matrix. In particular, the dimension of the sample covariance matrix $\hat{D}$ is adjusted from the exact covariance dimension $D$ by $\hat{D}=\frac{D}{1+\gamma D}$ with $\gamma=N/N_T<1$.
We used this time-sampled dimension in \cref{fig:Dimension_gap}BD and \cref{fig:zero-frequency_dim_change_with_N}.

\section{Network simulation and additional numerical details}
\label{sec:methods}
The network model \cref{eq:network_model} is simulated with first order finite difference method (i.e., the Euler–Maruyama method),
\begin{equation}
    x_i(t+\Delta t)= \left(1-\Delta t\right)x_i(t)+\Delta t \sum_{j=1}^NJ_{ij}\phi_j(t)+ \sqrt{\Delta t}\xi_i(t).
\label{eq:first_order_finite_difference_simulation}
\end{equation}
The external input, $\xi_i(t)$, modeled as white noise is drawn i.i.d. from the Gaussian distribution $\mathcal{N}\left(0, \sigma^2\right)$. 
The random connectivity $\MJ$ is drawn and fixed at the beginning of the simulation. 
To match exactly with the mean-field theory, one would need $N\to\infty$, $\Delta t\to0$, and the total simulation time $T\to\infty$. We have chosen the simulation parameters balancing computational cost as listed in \cref{tab:table_of_parameters} for all figures. In general, $N \sim 10^3$, $T\sim 100N$, and time step $\Delta t=0.1$ is sufficient to achieve high accuracy.

The zero-timelag covariance matrix $\MCphitau$ is computed from the simulated activity as 
\begin{equation}
    C^{\phi,\tau}_{ij}=\frac{\Delta t}{T}\sum_{k=1}^{\frac{T}{\Delta t}}\phi_{i}(t_k)\phi_j(t_k).
\end{equation}

To compute the frequency-dependent covariance matrix, $C^a(\omega)$, we first choose a large time bin width $\tbin$, where the correlation function has decayed to be sufficiently small, usually $\tbin=100$. 
The total simulation time $T$ is divided into $N_\text{bin}=T/t_\text{bin}$ intervals.
We compute the Fourier transform on the $n$-th interval as $a_i^{(n)}(\omega)=\frac{1}{\sqrt{\tbin}}\int_{(n-1)\tbin}^{n\tbin}a_i(t)dt$ with $a\in\{x,\phi\}$. 
Then, $C^{a}_{ij}(\omega)=\frac{1}{N_\text{bin}}\sum_{n=1}^{N_\text{bin}}a_i^{(n)}(\omega)a_j^{(n)}(-\omega)$ (Note that $a_j^{(n)}(-\omega)$ is the complex conjugate of $a_j^{(n)}(\omega)$). 
The cross-covariance matrices such as $\MC^{\phi x}(\omega)=\frac{1}{N_{\text{bin}}}\sum_{n=1}^{N_{\text{bin}}} \phi_i^{(n)}(\omega)x_j^{(n)}(-\omega)$ are computed similarly.

\newpage

\begin{longtable}{|p{2cm}|p{3cm}|p{1cm}|p{3.5cm}|p{1cm}|p{0.8cm}|p{1.2cm}|p{0.5cm}|}
    \caption{Additional parameters used in the figures. Those that do not apply are marked with ``--''.}
    \label{tab:table_of_parameters} \\
    \hline
    Figure & $g$ & $\sigma$ & $N$ & $T$ & $\tbin$ & $\Delta t$ & $\omega$\\
    \hline
    \endfirsthead
    
    \multicolumn{8}{c}{{\bfseries Table \thetable{} continued from previous page}} \\
    \hline
    Figure & $g$ & $\sigma$ & $N$ & $T$ & $\tbin$ & $\Delta t$ & $\omega$\\
    \hline
    \endhead
    
    \hline
    \multicolumn{8}{r}{{Continued on next page...}} \\
    \endfoot
    
    \hline
    \endlastfoot

    \cref{fig:example_covariance_and_relative_error}ABEF & 5 & 0.495 & 10000 & 200N & 100 & 0.1 & 0\\
    \hline
    \cref{fig:example_covariance_and_relative_error}CD & 5 & 0.495 & 500, 1000, 2000, 3000, 5000, 6000, 8000, 10000 & 200N & 100 & 0.1 & 0\\
    \hline
    \cref{fig:Eigen_spectrum_total}BE & 5 & 0.495 & 4000 & 200N & 100 & 0.01 & 0\\
    \hline
    \cref{fig:Eigen_spectrum_total}C & 1.7 & 1 & 4000 & 200N & 100 & 0.01 & 0\\
    \hline
    \cref{fig:Eigen_spectrum_total}D & 0.4 & 0.495 & 3000 & 200N & 100 & 0.1 & 0\\
    \hline
    \cref{fig:Eigen_spectrum_total}F & 10 & 0 & 4000 & 200N & 100 & 0.01 & 0\\
    \hline
    \cref{fig:g_eff_phase_diagram}B & 0.4, 0.9, 1.2, 1.37, 1.5, 2, 3, 5 & 0.495 & 3000 & 200N & 100 & 0.1 & 0\\
    \hline
    \cref{fig:Dimension_gap}A & 2, 4, 6, 8, 10, 12, 14, 16, 18, 20 & 0 & 4000(5 realizations) & 50000  & -- & 0.1 & --\\
    \hline
    \cref{fig:Dimension_gap}B & 3, 5, 10, 15 & 0 & 2000(30 realizations) & 200N  & 100 & 0.1 & 0\\
    \hline
    \cref{fig:Dimension_gap}C & 0.4, 0.6, 0.9, 1.1, 1.37, 1.5, 2, 5 & 0.495 & 2000(5 realizations) & 200N & -- & 0.1 & --\\
    \hline
    \cref{fig:Dimension_gap}D & 0.4, 0.6, 0.9, 1.1, 1.37, 1.5, 2, 5 & 0.495 & 2000(5 realizations) & 200N & 100 & 0.1 & 0\\
    \hline
    \cref{fig:contour_geff0.95}ABC & 1.1, 1.5, 2.2 & 0.3324, 0.6742, 1.0173 & 1000 & 10000 & 100 & $\Delta t_{\text{sim}}=0.01$, $\Delta t_{\text{obs}}=0.1$ & 0 \\
    \hline
    \cref{fig:special_spectrum_autocorrelation}A & 5 & 0.495 & 2000 & 200N & 100 & 0.01 & 0.2\\
    \hline
    \cref{fig:special_spectrum_autocorrelation}B & 5 & 0.495 & 10000 & 200N & -- & 0.1 & --\\
    \hline
    \cref{fig:special_spectrum_autocorrelation}C & 0.4 & 0.495 & 2000 & 200N & -- & 0.02 & --\\
    \hline
    \cref{fig:different_combination_geff=0.7}A & 0.8 & 0.495 & 4000(50 realizations for MC $x$ spectrum) & -- & -- & -- & 0\\
    \hline
    \cref{fig:different_combination_geff=0.7}B & 5 & 0.495 & 4000(50 realizations for MC $x$ spectrum) & -- & -- & -- & 0.9\\
    \hline
    \cref{fig:different_combination_geff=0.7}C & 5 & 0 & 4000(50 realizations for MC $x$ spectrum) & -- & -- & -- & 0.9\\
    \hline
    \cref{fig:dimension_gap_and_derivative}A & 0.4, 0.9, 1.2, 1.37, 1.5 & 0.495 & 3000 & 200N & -- & 0.1 & -- \\
    \hline
    \cref{fig:dimension_gap_and_derivative}B & 0.4, 0.9, 1.2, 1.37, 1.5 & 0.495 & 3000 & 200N & 100 & 0.1 & 0 \\
    \hline
    \cref{fig:dimension_gap_and_derivative}C & 0.4, 0.9, 1.5, 2, 3 & 1 & 3000 & 200N & -- & 0.1 & -- \\
    \hline
    \cref{fig:dimension_gap_and_derivative}D & 0.4, 0.9, 1.5, 2, 3 & 1 & 3000 & 200N & 100 & 0.1 & 0 \\
    \hline
    \cref{fig:zero-frequency_dim_change_with_N}A-D & 3, 15 & 0 & 100, 200, 500, 1000, 2000 & 200N & 100 & 0.1 & 0 \\
    \hline
    \makecell{\cref{fig:summary_Plin}CD,\\ \cref{fig:summary_sparse_J}CD,\\ \cref{fig:summary_EI_J}CD} & 0.4 & 0.495 &\makecell{1000 (simulation)\\2000 (Monte-Carlo,\\100 realizations.)} & 2000N & 100 & 0.1 & 0 \\
    \hline
    \makecell{\cref{fig:summary_Plin}EF, \\ \cref{fig:summary_sparse_J}EF, \\ \cref{fig:summary_EI_J}EF, \\
    \cref{fig:general_theory_main_text}BC, \\
    \cref{fig:general_theory_main_text}EF} & 5 & 0.495 &\makecell{4000 (simulation)\\8000 (Monte-Carlo,\\100 realizations.)} & 200N & 100 & 0.01 & 0 \\
    \hline
    \cref{fig:summary_symm_J}ABC, \cref{fig:part_symm_spectrum_supplement}AB,
    \cref{fig:general_theory_main_text}HI & 10, $\kappa=0.4$ & 0 & 4000 & 200N & 150 & 0.02 & 0 \\
    \hline
    \cref{fig:summary_symm_J}DEF, \cref{fig:part_symm_spectrum_supplement}C & 10, $\kappa=-0.4$ & 0 & 4000 & 200N & 100 & 0.01 & 0 \\
    \hline
    \cref{fig:g_eff_curve_for_diff_kappa}A & $g(1+\kappa)$= 0.4, 0.8, 0.9, 1.0, 1.1, 1.2, 1.3, 1.4, 1.5, 1.8, 2, 2.5, 3, 4, 5, 6, 7, 8, 9, 10 & 0.495 & 10000 (5 realizations) & $500\tbin$ & 150 & 0.033 & 0 \\
    \hline
    \cref{fig:g_eff_curve_for_diff_kappa}C & $g(1+\kappa)$= 0.4, 0.8, 0.9, 1.0, 1.1, 1.2, 1.3, 1.4, 1.5, 1.8, 2, 2.5, 3, 4, 5, 6, 7, 8, 9, 10 & 0.495 & 10000 (15 realizations) & $500\tbin$ & 40 & 0.01 & 0 \\
    \hline
    \cref{fig:g_eff_curve_for_diff_kappa}B & $g(1+\kappa)$= 0.4, 0.8, 0.9, 1.0, 1.1, 1.2, 1.3, 1.4, 1.5, 1.8, 2, 2.5, 3, 4, 5, 6, 7, 8, 9, 10 & 0.495 & 10000 (5 realizations) & $500\tbin$ & 100 & 0.02 & 0 \\
    \hline
    \cref{fig:g_eff_curve_for_diff_kappa}D & $g(1+\kappa)$= 0.4, 0.8, 0.9, 1.0, 1.1, 1.2, 1.3, 1.4, 1.5, 1.8, 2, 2.5, 3, 4, 5, 6, 7, 8, 9, 10 & 0.495 & 10000 (5 realizations) & $500\tbin$ & 50 & 0.01 & 0 \\
    \hline
    \cref{fig:geff_vs_g_sigma=0} & $g$=3, 5, 10, 15 & 0 & 2000 (30 realizations) & 200N & 100 & 0.1 & 0 \\
    \hline
\end{longtable}

\end{appendix}

\end{document}